\documentclass[apj,tighten,iop,twocolumn]{emulateapj}
\usepackage{graphicx,lscape,sidecap,subfig}
\newcommand{\funits}{erg~cm$^{-2}$~s$^{-1}$}

\newcommand{\kmps}{km s$^{-1}$}

\newcommand{\lunits}{erg~s$^{-1}$}
\newcommand{\lnuunits}{erg~s$^{-1}$~Hz$^{-1}$}

\newcommand{\mstar}{$M_{\ast}$}
\newcommand{\msun}{$M_{\odot}$}
\newcommand{\msuny}{\msun yr$^{-1}$}

\newcommand{\tmone}{$^{-1}$}
\newcommand{\tmtwo}{$^{-2}$}

\newcommand{\na}{$^{\ast}$}

\newcommand{\airac}{$\alpha_{\rm IRAC}$}

\newcommand{\aox}{$\alpha_{\rm OX}$}

\newcommand{\defhone}{{\it Def}$_{\rm H I}$}

\newcommand{\flims}{$f_{\rm lim,S}$}
\newcommand{\flimh}{$f_{\rm lim,H}$}
\newcommand{\fnu}{$f_{\nu}$}

\newcommand{\fofe}{$f_{\nu 1.4, {\rm est}}$}
\newcommand{\fofo}{$f_{\nu 1.4, {\rm obs}}$}

\newcommand{\ftwp}{$f_{20{\rm,p}}$}
\newcommand{\ftwi}{$f_{20{\rm,i}}$}
\newcommand{\ftsi}{$f_{36{\rm,i}}$}
\newcommand{\ftftp}{$f_{352{\rm,p}}$}
\newcommand{\ftfti}{$f_{352{\rm,i}}$}

\newcommand{\lha}{$L_{{\rm H}\alpha}$}

\newcommand{\lqq}{\lq\lq}

\newcommand{\lof}{$L_{\nu 1.4}$}

\newcommand{\lx}{$L_X$}
\newcommand{\lxlims}{$L_{X, {\rm lim,S}}$}
\newcommand{\lxlimh}{$L_{X, {\rm lim,H}}$}

\newcommand{\lxte}{$L_{X,{\rm 0.5-8.0 keV}}$}

\newcommand{\lxhe}{$L_{X,{\rm 2.0-8.0 keV}}$}
\newcommand{\lxht}{$L_{X,{\rm 2.0-10.0 keV}}$}

\newcommand{\nh}{$N_{\rm H}$}

\newcommand{\nexcs}{$N_{\rm exc,S}$}
\newcommand{\nexch}{$N_{\rm exc,H}$}
\newcommand{\nhgal}{$N_{\rm H}^{\rm gal}$} 
\newcommand{\nhint}{$N_{\rm H}^{\rm int}$} 

\newcommand{\nobss}{$N_{\rm obs,S}$}
\newcommand{\nobsh}{$N_{\rm obs,H}$}
\newcommand{\nulnu}{$\nu L_{\nu}$}
\newcommand{\rqq}{\rq\rq}

\newcommand{\x}{X-ray}

\newcommand{\zmed}{$z_{\rm med}$}

\newcommand{\ha}{H$\alpha$}

\newcommand{\hone}{H{\sc \,i}}

\newcommand{\ohb}{[O{\sc \,iii}]/H$\beta$}
\newcommand{\niha}{[N{\sc \,ii}]/H$\alpha$}
\newcommand{\siha}{[S{\sc \,ii}]/H$\alpha$}
\newcommand{\oha}{[O{\sc \,i}]/H$\alpha$}

\newcommand{\chandra}{{\it Chandra}}

\newcommand{\hst}{{\it HST}}

\newcommand{\rosat}{{\it ROSAT}}
\newcommand{\spitzer}{{\it Spitzer}}

\newcommand{\swift}{{\it Swift}}

\newcommand{\xmm}{{\it XMM-Newton}}
\newcommand{\wone}{{\it uvw1}}
\newcommand{\wtwo}{{\it uvw2}}

\newcommand{\mtwo}{{\it uvm2}}
\newcommand{\uvot}{{UVOT}}

\newcommand{\galfit}{{\sc galfit}}

\newcommand{\xspec}{{\sc xspec}}
\newcommand{\wav}{{\sc wavdetect}}
\newcommand{\ace}{{\sc ae}}

\newcommand{\er}{Equation~\ref}
\newcommand{\fr}{Fig.~\ref}
\newcommand{\scr}{Sec.~\ref}
\newcommand{\tr}{Table~\ref}

\newcommand{\exi}{\begin{equation}}
\newcommand{\exo}{\end{equation}}
\newcommand{\tm}{\tablenotemark}
\newcommand{\tx}{\tablenotetext}
\newcommand{\aer}[3]{$#1^{#2}_{#3}$} 
\newcommand{\scier}[4]{$#1^{#3}_{#4} \times 10^{#2}$} 
\newcommand{\ten}[2]{$#1\times 10^{#2}$} 

\newcommand{\tgt}{\hspace{-.1cm}}

\def\spose#1{\hbox to 0pt{#1\hss}} 
\def\approxlt{\mathrel{\spose{\lower 3pt\hbox{$\sim$}}
        \raise 2.0pt\hbox{$<$}}}
\def\approxgt{\mathrel{\spose{\lower 3pt\hbox{$\sim$}}
        \raise 2.0pt\hbox{$>$}}}

\slugcomment{Accepted by ApJS}

\shorttitle{\chandra-\swift\ View of Hickson Compact Groups}
\shortauthors{Tzanavaris et al.}

\begin{document}

\title{A Chandra-Swift View of Point Sources in Hickson Compact Groups:
High AGN fraction but a dearth of strong AGNs}

\author{
P.~Tzanavaris\altaffilmark{1,2,3},
S.~C.~Gallagher\altaffilmark{4}, 
A.~E.~Hornschemeier\altaffilmark{1}, 
K.~Fedotov\altaffilmark{4,5},
M.~Eracleous\altaffilmark{5,6},
W.~N.~Brandt\altaffilmark{5,6},
T.~D.~Desjardins\altaffilmark{4},
J.~C.~Charlton\altaffilmark{5}, 
C.~Gronwall\altaffilmark{5,6}
}

\altaffiltext{1}{Laboratory for X-ray Astrophysics, 
NASA/Goddard Spaceflight Center, Mail Code 662, Greenbelt, Maryland, 20771, USA}
\altaffiltext{2}{Department of Physics and Astronomy,
The Johns Hopkins University, Baltimore, MD 21218, USA}
\altaffiltext{3}{NPP Fellow}
\altaffiltext{4}{Department of Physics and Astronomy,
The University of Western Ontario, London, ON N6A 3K7, Canada}
\altaffiltext{5}{Department of Astronomy and Astrophysics, 
The Pennsylvania State University, University Park, PA 16802, USA}
\altaffiltext{6}{The Institute for Gravitation and the Cosmos,
The Pennsylvania State University, University Park, PA 16802, USA}
\altaffiltext{5}{Herzberg Institute Of Astrophysics, Victoria, BC, V9E 2E7, Canada}



\begin{abstract}
We present \chandra\ \x\ point source catalogs for
9 Hickson Compact Groups (HCGs, 37 galaxies) 
at distances $34 - 89$~Mpc.
We perform
detailed \x\ point source detection and photometry, and interpret the point source population by means of
simulated hardness ratios. We thus estimate \x\ luminosities (\lx)
for all sources, most of which are too weak for reliable
spectral fitting. For all sources, we provide catalogs with counts, count rates,
power-law indices ($\Gamma$), hardness ratios, and \lx, in the full
($0.5-8.0$ keV), soft ($0.5-2.0$ keV) and hard ($2.0-8.0$ keV) bands. 
We use
optical emission-line ratios from the literature to re-classify 24 galaxies
as star-forming, accreting onto a supermassive black hole (AGNs),
transition objects, or low-ionization nuclear emission regions
(LINERs).  Two-thirds of our galaxies have nuclear \x\ sources with \swift/UVOT
counterparts. Two nuclei have \lxte~$ >
10^{42}$~\lunits, are strong multi-wavelength AGNs and follow the known
\aox--\nulnu$_{({\rm near UV})}$ correlation for strong AGNs. Otherwise, most nuclei
are \x\ faint, consistent with either a low-luminosity AGN
or a nuclear \x\ binary population, and fall in the \lq\lq non-AGN locus\rq\rq\
in \aox--\nulnu$_{({\rm near UV})}$ space, which also hosts
other, normal, galaxies. Our results suggest that HCG \x\ nuclei in high specific star formation rate
spiral galaxies are likely dominated by star formation, while those with low
specific star formation rates in earlier types likely harbor a weak AGN.  
The AGN fraction in HCG galaxies with $M_R \le -20$ and \lxte~$\ge
10^{41}$~\lunits\ is \aer{0.08}{+0.35}{-0.01}, somewhat
higher than the $\sim 5\%$ fraction in galaxy clusters.
\end{abstract}

\vspace{1cm}
\keywords{galaxies: nuclei --- galaxies: active --- ultraviolet: galaxies --- X-rays: galaxies --- catalogs}

\section{Introduction}
By virtue of their selection criteria, Hickson Compact Groups (HCGs)
constitute a distinct class among small galaxy agglomerations.  The
Hickson catalog \citep{hickson1982,hickson1992} comprises 92
spectroscopically confirmed, nearby (median redshift \zmed~$=0.03$,
$\sim 130$ Mpc) compact groups with three or more members with
accordant redshifts (i.e., within 1000~\kmps\ of the group mean). The
characteristic physical properties of CGs \citep{hickson1992} 
include galaxy separations
of the order of a few galaxy radii (median projected separations $\sim
40h^{-1}$ kpc), low velocity dispersions (radial median $\sim 200$
\kmps) and high galaxy number densities (up to $10^8 h^2$
Mpc$^{-2}$). These conditions favor galaxy interactions, as
demonstrated by the spectacular examples of HCG 92 \citep[Stephan's
  Quintet, e.g.][]{fedotov2011} and HCG 31 \citep{gallagher2010}. It
is then natural to ask what influence this interaction-prone
environment has on processes related to star-formation or accretion
onto a nuclear supermassive black hole.

With regards to star formation, recent work suggests that, compared to
non-compact group environments, star formation is accelerated, leading
to rapid exhaustion of the gas supply sustaining star forming
activity.  This result follows from ultraviolet and infrared
star-formation estimates that show significant discontinuities in
mid-infrared colors and ultraviolet+infrared specific star formation
rates \citep[SSFRs,][]{johnson2007,tzanavaris2010,walker2010,walker2012}.  In
particular, the discontinuities indicate a bimodality between galaxies
with high levels of star formation and those with little star
formation.  The latter have also been found to exhibit high levels
of \lq\lq
\hone\ deficiency\rq\rq, \defhone, as defined by \citet{verdes2001}.
These authors predict an expected \hone\ mass for field galaxies of a
given morphological type and compare it to the \hone\ mass of compact
group galaxies, thus calculating \defhone. Taken together, the lack
of galaxies with intermediate mid-infrared colors and SSFRs, as well
as the high \defhone\ values are suggestive of accelerated and then
abruptly truncated
star formation.

\begin{deluxetable*}{cccc cccc}
\tablecolumns{8}
\tablewidth{0pc} 
\tablecaption{Chandra observation log for this HCG sample \label{tab-xdata}}
\tablehead{ 
\colhead{HCG ID}
&\colhead{Obs. ID}
&\colhead{Obs. Start Date}
&\colhead{Detector}
&\colhead{Obs. time (ks)}
&\colhead{Obs. type}
&\colhead{PI}
&\colhead{References}
\\
\colhead{(1)}
&\colhead{(2)}
&\colhead{(3)}
&\colhead{(4)}
&\colhead{(5)}
&\colhead{(6)}
&\colhead{(7)}
&\colhead{(8)}
}
\startdata
HCG 7    &    8171 & 2007-09-13 & ACIS-S     &     19.4 & GTO    & Garmire & \\
HCG 7    &    9588 & 2007-09-16 & ACIS-S     &     16.9 & GTO    & Garmire &  \\
&&&& {\bf 36.3} &&& \citet{konstantopoulos2010}  \\
HCG 16   &     923 & 2000-11-16 & ACIS-S     &     {\bf 12.7} & GO     & Mamon  & \citet{jeltema2008}    \\
HCG 22   &    8172 & 2006-11-23 & ACIS-S     &     {\bf 32.2} & GTO    & Garmire  & \citet{desjardins2013}  \\
HCG 31   &    9405 & 2007-11-15 & ACIS-S     &     {\bf 36.0} & GO     & Gallagher & \citet{smith2012} \\
HCG 42   &    3215 & 2002-03-26 & ACIS-S     &     {\bf 32.1} & GO     & Ponman  & \citet{jeltema2008}   \\
HCG 59   &    9406 & 2008-04-12 & ACIS-S     &     {\bf 38.9} & GO     & Gallagher & \citet{desjardins2013} \\
HCG 62   &     921 & 2000-01-25 & ACIS-S     &     49.1 & GO     & Vrtilek  &  \\
HCG 62   &   10462 & 2009-03-02 & ACIS-S     &     68.0 & GO     & Rafferty &  \\
HCG 62   &   10874 & 2009-03-03 & ACIS-S     &     52.0 & GO     & Rafferty &  \\
&&&& {\bf 169.2} &&& \citet{jeltema2008}\\
HCG 90   &     905 & 2000-07-02 & ACIS-I     &     {\bf 50.2} & GO     & Bothun  & \citet{jeltema2008}   \\
HCG 92   &    7924 & 2007-08-17 & ACIS-S     &     94.4 & GO     & Vrtilek    &\\
HCG 92   &     789 & 2000-07-09 & ACIS-S     &     20.0 & GO     & Trinchieri &\\
&&&& {\bf 114.4} && & \citet{osullivan2009}
\enddata
\tablecomments{Columns are: (1) HCG group name; (2) observation ID; (3) start date of observation; (4) detector;
(5) exposure time; (6) observation type (Guarranteed Time observing or General Observer proposal; (7) principal investigator; (8) references (first publication using these data).
Total exposure times for each group appear in bold. 
}
\end{deluxetable*}

The importance of accretion onto a nuclear supermassive black hole
(SMBH) in compact groups (\lq\lq AGN\rq\rq\footnote{In line with
  common usage in the literature we shall use the acronym \lq\lq
  AGN\rq\rq\ (active galactic nucleus) to refer to accretion onto a
  nuclear supermassive black hole. Strictly this is incorrect as
  nuclear activity can also be due to star formation.}) has not been
thoroughly investigated and is not well established.  In galaxy
clusters \citet{dressler1985} found fewer AGNs compared to the field
\citep[but see also][and below]{martini2006}. 
Compared to clusters, compact groups of galaxies
have lower 
velocity dispersions making prolonged close interactions
more likely. It is thus possible that the level of AGN activity
is different. On the theoretical and computational side, simulation work
\citep[e.g.][]{hopkins2010} suggests that major galaxy mergers are a
leading mechanism that can trigger inflow of rotationally supported
gas to feed a central SMBH. Note though that this would also provide fuel for
intense star formation and could trigger
nuclear starbursts \citep[e.g.][]{mihos1996}.  Other feeding
mechanisms
include supernova winds, minor interactions, and disk
instabilities. Several observational surveys
have provided insight on the connection between AGNs and galaxy interactions.
For instance
\citet{kartaltepe2010} find that AGNs are common in Ultraluminous and
Hyperluminous Infrared Galaxies (ULIRGs and HyLIRGs), which are known
to result from major mergers. In addition, the AGN fraction in this
population increases with
infrared luminosity.  Recently, \citet{silverman2011} find increased
AGN activity in pairs compared to isolated galaxies.  On the other
hand, several authors find minor interactions and secular evolution to
be most important in triggering AGN activity
\citep[e.g.][]{grogin2005,georgakakis2009,cisternas2011,deng2013}.






In the optical regime, \citet{coziol1998a,coziol1998b,coziol2004} used
emission-line ratios in several samples (up to 91 galaxies in
27 compact groups) to determine the type of
nuclear activity in compact group galaxies, consistently finding that
strong and low-luminosity (\lha~$\lesssim 10^{39}$~\lunits) AGNs
(LLAGNs) each make up no more than $\sim 10$\%\ of the total CG
populations \citep[see][Table 3]{coziol2004}.  Depending on the specific
sample, star-forming galaxies represent a fraction up to $\sim
34$\%\ of the population, with the remaining galaxies showing no
emission lines. Both LLAGNs and AGNs are found mainly in optically
luminous early type galaxies with little on-going star formation that are
in the centers of evolved groups. This finding was interpreted to
indicate that such group cores are old, collapsed systems
where star formation activity has ceased.  According to this
interpretation, high central densities of group cores induced
gravitational interactions, which accelerated star formation,
rapidly consuming all of the available fuel.
  
It is important to note that the fractions for \lq\lq
LLAGNs\rq\rq\ presented by these authors also include low-ionization
nuclear emission regions (LINERs), the nature of which is still a
matter of debate. LINERs are characterized by high ratios of narrow
optical low ionization oxygen emission lines \citep{heckman1980} and
are found in about half of all nearby galaxies
\citep{ho1997}. Candidate power sources for LINERs include (1) weak
AGNs \citep[e.g.][]{halpern1983,ferland1983}, (2) hot stars
\citep[e.g.][]{terlevich1985,filippenko1992,shields1992}, and (3)
shocks \citep[e.g.][]{heckman1980,dopita1996}.  Although weak AGNs
have been found in the majority ($\sim 75$\%) of LINERs
\citep[e.g.][]{barth1998,ho2001,filho2004,nagar2005,maoz2005,flohic2006,gonzalez-martin2009}, they
cannot account for the total LINER emission in the majority of cases
\citep{eracleous2010a}. In fact, for most LINERs
\citet{eracleous2010a} show that there is an energy deficit problem:
Star formation and AGN activity are not able to provide a sufficient
number of ionizing photons to account for the observed
emission lines.

In the most recent optical study \citet[][hereafter M10]{martinez2010}
compiled a large spectroscopic sample of 280 galaxies in 64 HCGs and
used emission-line ratios to classify the type of nuclear activity,
providing an estimate for the AGN fraction in HCGs.

They classified
23\%\ of galaxies as AGNs, 
10\%\ as transition objects (TO), 
and
14\%\ as star forming (SF), 
with the remainder of the galaxies showing no
emission lines.
According to this study, although AGNs appear to 
be the most numerous emission-line galaxy class
in CGs, they have characteristically low \ha\
luminosities (median \ten{7.1}{39}~\lunits) and
virtually no broad emission lines, suggestive
of LLAGNs. 
However, these authors use a restricted set
of line ratios that precludes distinguishing
between LINERs and AGN.  

In this paper
we use the \citet{kewley2006} method to
reclassify
the galaxies of M10. This allows us
to also identify LINER systems. To stress that this is an optically based
classification, we use the
designations optAGN, optTO, optSF, optLINER.

Work in different wavelength regimes can provide complementary insight
into these questions.  \citet{gallagher2008} used
$1-24$\micron\ 2MASS+\spitzer\ nuclear data to probe the nuclear
activity in 46 galaxies from 12 nearby HCGs.  They found that the
spectral index, \airac, of a power law fit to the
$4.5-8.0$\micron\ IRAC data cleanly separates MIR-active from
MIR-inactive HCG nuclei.  Unfortunately, the exact origin of activity
(whether AGN or star-formation) cannot be deduced by this method.
In particular, these authors show that hot dust emission
can be responsible for their results, and this can be due
either to hard ionizing AGN continua or AGB populations in
star forming galaxies.
On
the other hand, \citet{roche1991} have shown that MIR-inactivity
(\airac~$>0$) is associated with {\it low}-luminosity AGN activity.

Due to the high-energy emission generated by supermassive black hole
accretion, by far the best direct diagnostic for strong AGN activity
is nuclear \x\ emission. Compared to the optical, the \x\ regime
offers the advantage that the nuclear emission is not diluted by
starlight from the host galaxy, while dust obscuration is very
significantly mitigated due to the higher, penetrating power of
\x\ radiation.  
Unfortunately, this simple picture is complicated by 
the combined effect of
two factors.
First, \x\ starlight sometimes can actually
dilute AGN emission. This is because
\x\ binary (XRB) populations in circumnuclear star
clusters also emit in the \x\ regime, although 
{\it individual} XRBs typically have lower luminosities than
strong AGNs.
Second, as the name implies, LLAGNs emit at low
\x\ luminosities. 
Adopting a fiducial threshold of
\lxte~$=10^{41-42}$~\lunits,
it is only at higher \x\ luminosities
that
nuclear \x\ emission can be
attributed to an AGN with high probability.
Thus the situation becomes increasingly
ambiguous at progressively fainter luminosities, making
it challenging to distinguish between
\x\ emission due to unresolved populations of circumnuclear XRBs and
that of LLAGNs.
In this regime high angular resolution becomes
critical for distinguishing nuclear from circumnuclear emission.

Although earlier studies did detect \x\ emission in HCGs, they were
hampered by poor angular resolution and the lack of
hard \x\ sensitivity, making it difficult to disentangle
the contributions from point source (nuclear or extra-nuclear) and
diffuse emission, and essentially concentrated on studying the diffuse
component. Using \rosat\ data, \citet{ponman1996} detected a diffuse
IGM in $\sim 75$\%\ of a large HCG sample, while \citet{mulchaey2003},
using a low-redshift sample of 109 groups that included poor, compact
as well as rich, non-compact systems found diffuse, extended
\x\ emission in 61 groups (56\%).  In an effort to understand the relevance
of ram-pressure stripping and strangulation due to a hot IGM in the
most \hone-deficient HCGs, \citet{rasmussen2008} also examined the
level of nuclear activity in a sample of 8 HCGs, finding no
significant enhancement. However, they do not carry out a 
detailed high angular resolution study 
to provide more specific results
on the nature of nuclear activity in their systems.

The level of AGN activity in galaxy {\it clusters} has already
been systematically investigated in the \x\ regime, leading to
differing conclusions \citep[e.g. see][for a review]{ehlert2013}.
Using a multi-wavelength
approach that includes emission lines, \x\ spectral properties and
\x\ to visible-wavelength flux ratios in rich clusters,
\citet{martini2006} find that $\sim 5$\% of cluster galaxies more
luminous than $M_R = -20$ host AGNs with \lxte~$ > 10^{41}$~\lunits.
They notably also find a discrepancy between the AGN fraction
determined from optical spectroscopy and a higher fraction suggested by
\x\ luminosities. Interestingly, \citet{shen2007} compare the environments
of poor groups and clusters using a combined
optical and \x\ approach. They conclude that
poor groups host AGNs that are in an optically dominant phase, whereas
those in clusters are dominant in the \x s, leading to the
findings of \citet{martini2006}.
 
In compact groups there has been to-date no
systematic study of nuclear \x\ emission. 
In this paper we take advantage of the
superb angular resolution of the \chandra\ \x\ observatory to carry
out detailed point source detection in a sample of 9 compact groups
(37 galaxies).  This paper has two main goals: First, we make
available full \x\ source catalogs based on the \chandra\ observations
in 9 compact group fields with detailed information on counts, fluxes,
luminosities and hardness ratios.  Second, we focus on point sources
located in HCG galaxy nuclei. Using \chandra\ and
\swift/Ultra-Violet and Optical Telescope \citep[\uvot;][]{2005SSRv..120...95R}
data, we combine 
\x\ and ultraviolet (UV) nuclear photometry, and compare with radio and
optical diagnostics to assess the nature of nuclear
activity in compact group galaxies. In a separate paper, we
discuss the diffuse \x\ emission in the same sample of compact groups
\citep{desjardins2013}. Some of the \chandra\ data have first been presented
previously in a different context. We give appropriate references in
\tr{tab-xdata}.

The structure of the paper is as follows: Section 2 introduces our
sample.  Section 3 discusses X-ray data and analysis and point source
detections.  Section 4 presents UV nuclear data
and analysis.  Section 5 presents multiwavelength analyses,
including new optical emission-line ratio classifications, radio data and
a combined \x-UV analysis. Section 6 presents estimates on the
AGN fraction in HCGs and Section 7 discusses our findings.
We conclude with a summary in Section 8.


%

\section{Sample selection}
Our original multi-wavelength HCG sample comprises 11 groups compiled from
the original HCG catalog of 92 spectroscopically confirmed compact
groups \citep{hickson1992}. 
This has been the most widely used \citep[\lq\lq benchmark\rq\rq,][]{lee2004}
of all CG catalogs. Although valid concerns regarding selection biases
about this catalog have been raised \citep[e.g.][]{mamon1994,ribeiro1998},
comparisons with recent larger catalogs show that many HCG galaxy properties
such as surface brightness, or angular and linear diameter
are in fact close to median values for the corresponding distributions
\citep{lee2004}. 

\begin{figure*}
\epsscale{1.2}
\plotone{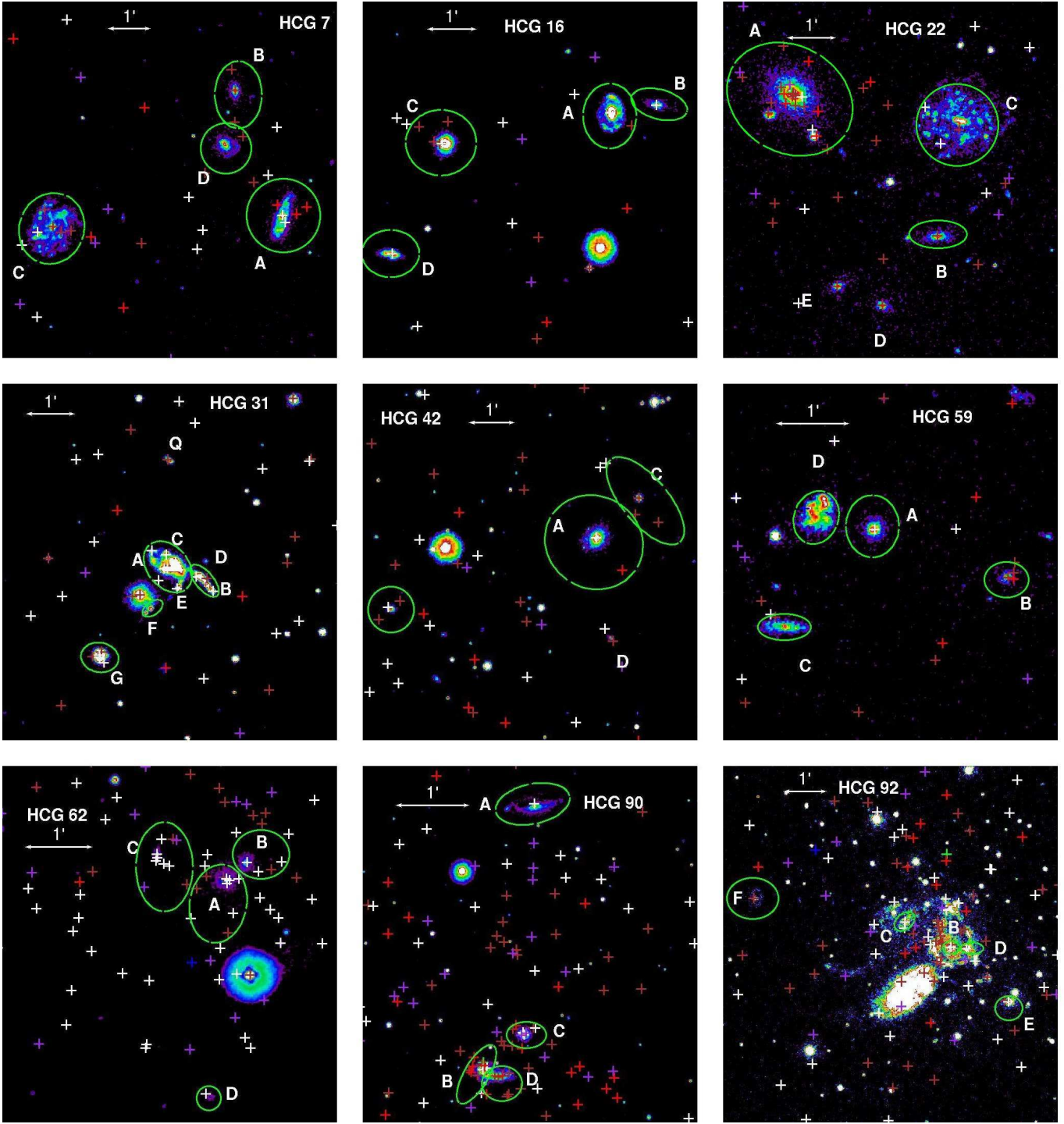}
\caption{\swift-UVOT false color \wone\ images with \chandra\ X-ray point
  sources overlaid. UV intensity increases from violet (lowest) to blue to
  green to red to white (highest).
  The color coding of source symbols
  indicates whether they are detected in the full band only (red), the
  soft band only (green), the hard band only (blue), the full and the
  hard band (purple), the full and the soft band (brown), the soft and
  the hard (blue-green) or the full, soft and hard bands (white). The
  green ellipses define galaxy regions from the mid-infrared work
  of 
  \citet{johnson2007}. The bright source right of center in
  the HCG 62 field is a foreground star.}
\label{fig:uvxims}
\end{figure*}

In order to ensure that
our sample would be observable with a range of ground- and space-based
instruments for our long-term multi-wavelength campaign, the selection
was based on membership (a minimum of three giant galaxies with
accordant redshifts, i.e., within 1000 \kmps\ of the group mean),
distance ($\lesssim 4500$ \kmps), and angular extent ($\lesssim
8$\arcmin\ in diameter). 

In this paper we present 9 of these groups, for which
both archival \chandra\ \x\ and \swift\ UVOT 
ultraviolet
data are available. 
\swift/UVOT false color images of the group fields, with
detected \chandra\ \x\ point sources overlaid are
shown in \fr{fig:uvxims}.
An observation log for the \chandra\ data is
presented in \tr{tab-xdata}.
The \chandra\ observations include
Guaranteed Time observing (HCGs 7 and 22, P.I. Garmire).
An observation log for the \swift\ UVOT data is
presented in \citet{tzanavaris2010}. In addition, note that in the
present work we have included UVOT data for HCGs 90 and 92 (see below).
The group and galaxy ID's, as well as morphological types, can be found in 
the first two columns of \tr{tab-multi}, which also provides an overview of our
multiwavelength results (\scr{sec:multiw}).

\begin{figure*}[ht]
\epsscale{1.4}
\centering
\newcommand{\wid}{10cm}
\subfloat
{
\includegraphics[width=\wid]{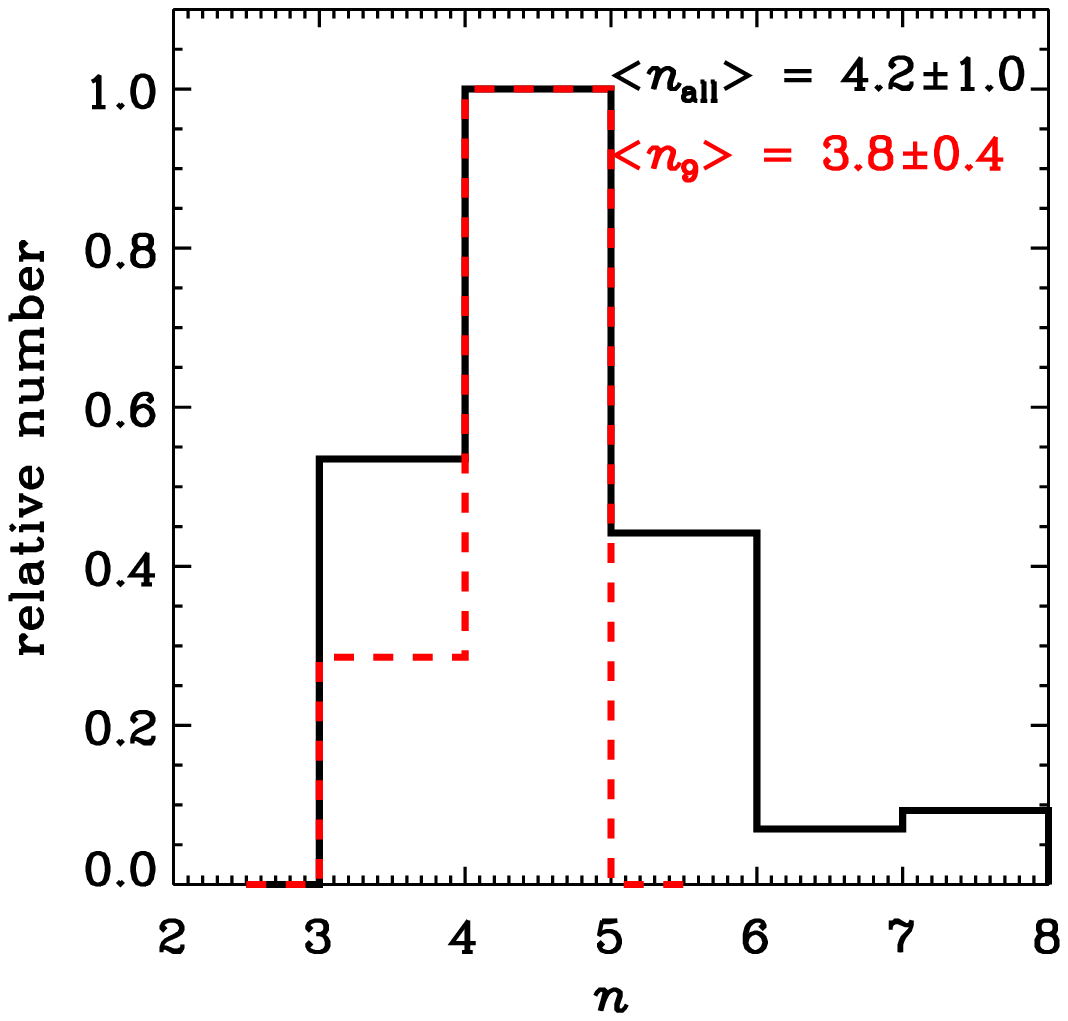}\hspace{-3cm}
\includegraphics[width=\wid]{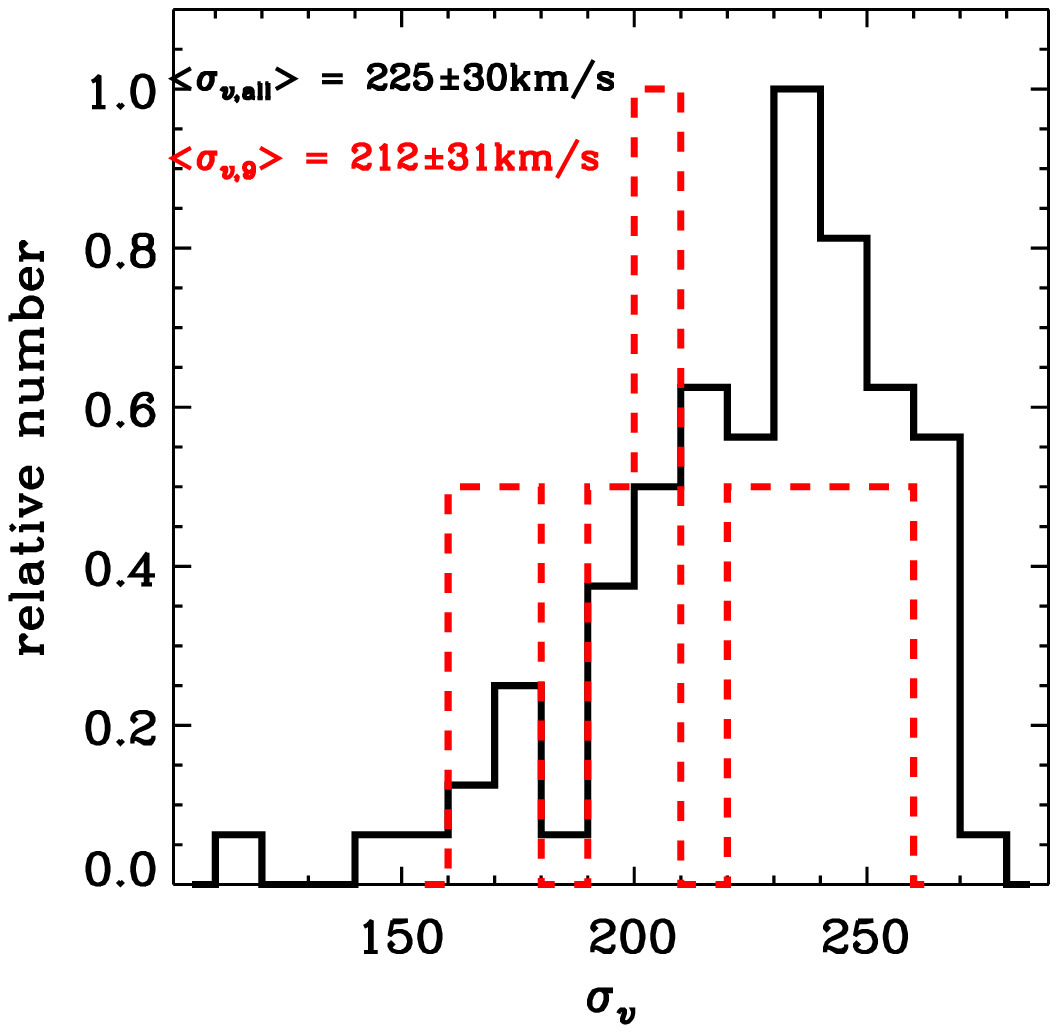}
}
{
\includegraphics[width=\wid]{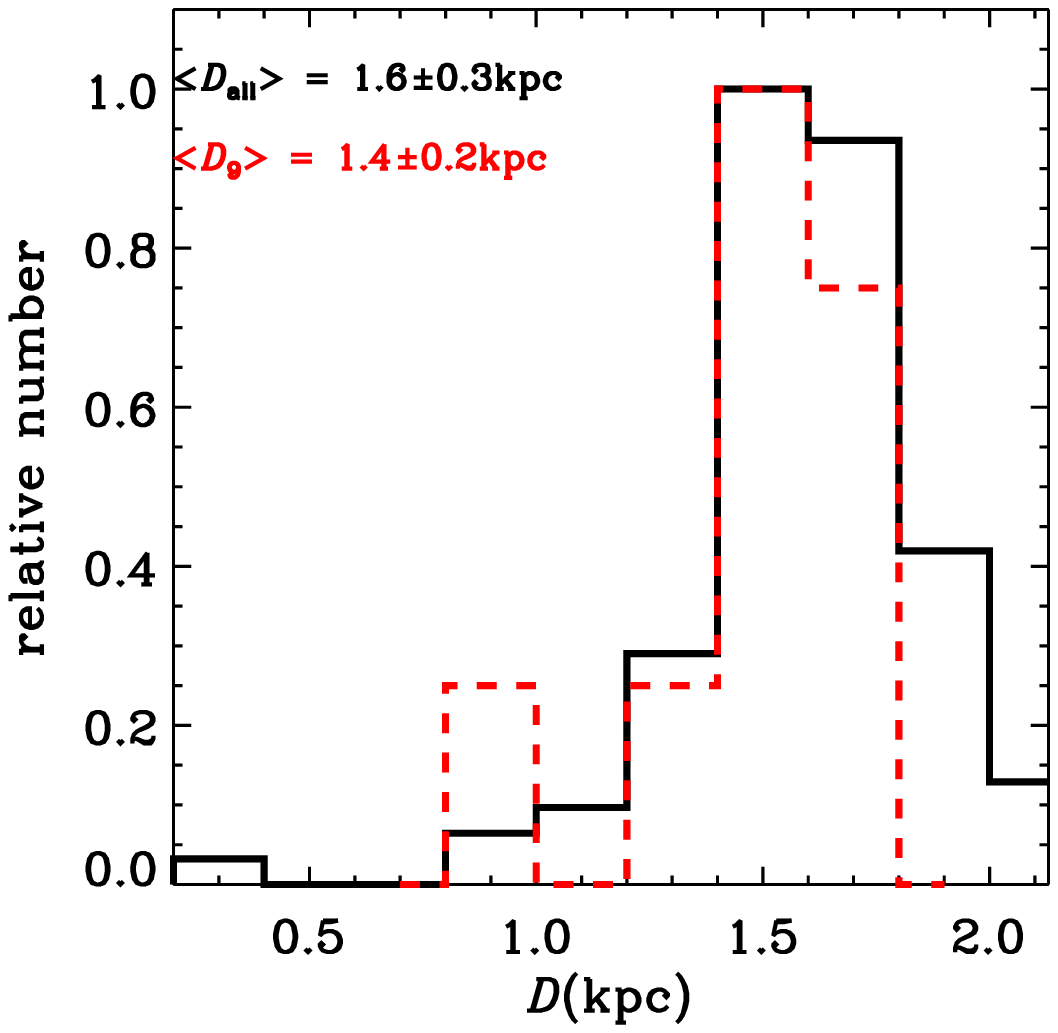}\hspace{-3cm}
\includegraphics[width=\wid]{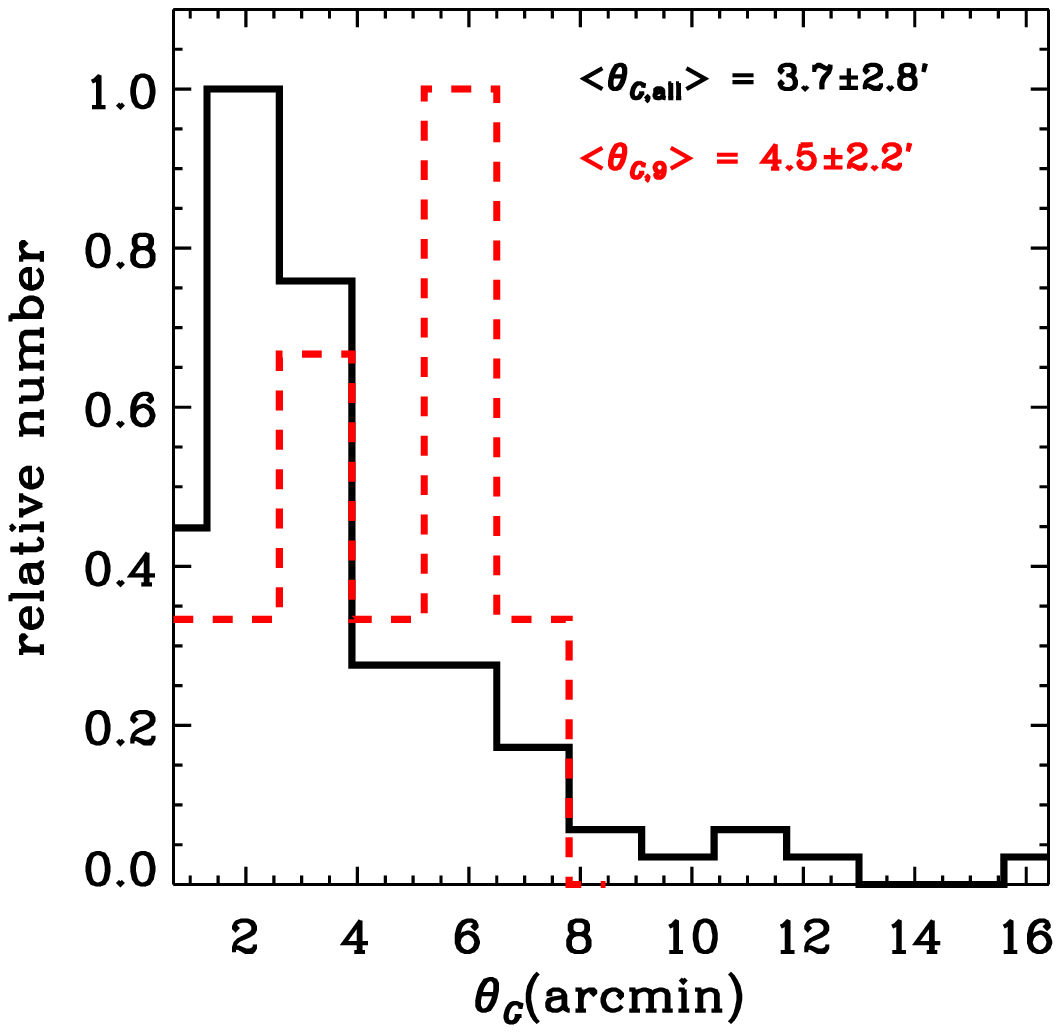}
}
\caption{Histograms for
physical properties of the nine galaxy groups in this paper 
({\it red, dashed lines}) and
the full sample of 92 Hickson Compact Groups \citep[][{\it black, continuous
lines}]{hickson1982,hickson1992}. 
Clockwise from top left, properties shown are 
mean number of galaxies 
per group with accordant redshifts,
mean 
radial velocity dispersion of galaxies per group,  
mean projected separation,
mean angular size of galaxies per group.
Mean values and standard deviations are shown in the upper right corner
of each panel.
}
\label{fig:comp}
\end{figure*}

As the nine groups used in this paper represent a small archival
sub-sample of the full set of 92 compact groups, 
we do not a priori expect them to be fully representative of
all HCGs. We further investigate this issue by comparing
the distributions of a number of characteristic properties,
established and tabulated by \citet{hickson1982} and \citet{hickson1992},
both for the full set and our sub-sample.
These include the number of galaxies per group, $n$,
the radial velocity dispersion, $\sigma_v$,
the median projected separation, $D$,
and the angular diameter, $\theta_G$. 
The distributions of these properties both for the full sample
and our nine-group subsample our shown in \fr{fig:comp}.
The distribution for $n$ is very similar for the two populations,
with means of $<n_{\rm 92}> = 4.2\pm1.0$ and
$<n_{\rm 9}> = 3.8\pm0.4$ for the 92 and nine HCG samples, respectively.
The Kolmogorov-Smirnov (KS) test gives a modestly high probability 
of 0.005 that
the two distributions come from the same parent population.
The situation for $D$ is similar
($<D_{\rm 92}> = (1.6\pm0.3)$~kpc and
$<D_{\rm 9}> = (1.4\pm0.2)$~kpc), with an even higher
KS probability (0.2). The $\sigma_v$
distributions are less similar in terms of their peaks but
their means are fully consistent within the uncertainties
($<\sigma_{v,\rm 92}> = (225\pm30)$~\kmps\ and
$<\sigma_{v,\rm 9}> = (212\pm31)$~\kmps). However, the KS probability
that the distributions are the same is relatively high (0.3).
Finally, the
distributions for $\theta_G$ are also less similar, and are consistent
within 2$\sigma$
($<\theta_{G,\rm 92}> = (3.7\pm2.8)$\arcmin\ and
$<\theta_{G,\rm 9}> = (4.5\pm2.2)$\arcmin). 
This is to be expected, as our selection criteria
are biased towards more nearby systems. Again though, the
KS probability is relatively high (0.3). Thus, overall, we find that
the sub-sample used in this paper is reasonably
representative of
HCGs as a class.

\section{X-ray data analysis}
\subsection{Point source detection and photometry}\label{sec:xsrc}
Each group was observed at the aim point of the back-illuminated S3 CCD
of \chandra 's Advanced CCD Imaging Spectrometer (ACIS), with the
exception of HCG 90, which was observed with the ACIS-I array.  The
data were processed using standard \chandra\ \x\ Center aspect
solution and grade filtering, from which the level 2 events file was
produced. Thanks to \chandra 's $\sim 1$\arcsec\ angular
resolution and the proximity of our galaxies, we can detect a
multitude of individual point sources in our fields. We have thus been able to
carry out a detailed point source detection and characterization
consisting of four stages as follows.

First, the CIAO 4.1.2\footnote{http://cxc.harvard.edu/ciao} wavelet
detection tool \wav\ \citep{freeman2002} was used in the soft
($S$, 0.5--2.0 keV), hard ($H$, 2.0--8.0 keV) and full ($F$, 
0.5--8.0 keV) bands to
detect candidate point sources in each band. The lower limit of 0.5
keV matches the well-calibrated part of the response, while above 8.0
keV the effective area of the \chandra\ mirrors is known to drop
considerably and the particle background increases significantly.  The chip
field (1024$\times$1024 pixels for S3 and 2048$\times$2048 pixels for
I0-I3) was searched with \wav\ at the $10^{-5}$ false-probability
threshold in all three energy bands. Although a lower
probability threshold ($10^{-6}$ per CCD) is often used to ensure low false
positive detections, the situation is more complicated for false
negatives, especially near an observation's detection limit \citep{kim2004}. We
thus chose to detect a greater number of spurious sources at this
stage, which were excluded at subsequent stages of the analysis as
explained below.  
For comparison, we show the numbers of sources detected with
\wav\ using these two probability thresholds
for each HCG field in the three X-ray bands in \fr{fig:w5w6}.
Wavelet scales used were 1, 1.414, 2, 2.828, 4,
5.657 and 8.0 pixels to cover a wide variety of source sizes, as well
as to take into account the variation of the point spread function (PSF)
size across the ACIS
CCD.  Source lists produced by \wav\ for each band were 
cross-correlated to produce a single list of
positions for candidate point sources in each field. 
This matching used each source's PSF ellipses, whose
size and orientation depend on the detector used
(ACIS-S or I), position on the detector and roll angle of observations.

The second stage of the point source analysis involved using the
software {\sc acis extract}
\citep[\ace,][\footnote{Package and User's Guide available at http://www.astro.psu.edu/xray/acis/acis\_analysis.html}]{broos2010}
to perform aperture photometry for sources in the \wav\ source list.
Some of the features of \ace\ that make it a
good choice for ACIS point source extraction include
\begin{enumerate}  \setlength{\itemsep}{-2pt}
\item Construction of PSF-shaped aperture extraction regions separately
  for each source and each observation. These regions encircle $\sim
  90 - 60$\% of the photon energy at 1.5 keV, depending on how crowded
  a given field region is.
\item Construction of background regions that exclude pixels from
  neighboring sources, and, where appropriate, a model of the wings of
  a neighboring source's contaminating emission.
\item Use of the \chandra\ Calibration Database for producing
  Ancillary Reference Files (ARFs) and Response Matrix Files (RMFs)
  for each source and observation, appropriately merging these for
  multiple observations.
\item Aperture corrections by means of calculation of
  the energy fraction falling inside the PSF
  region at 5 different energies.
\end{enumerate}

Our initial catalogs included {\it all} sources detected at the \ace\ stage.
We then flagged sources with negative net (source $-$ background) counts 
in a given band as
non-detections in that band. Such sources were assigned the detection flag 1
in that band. Further,  
we obtained Poisson $\pm1\sigma$ errors on net counts by using 
the method of \citet{gehrels1986}. If the measured net counts
minus the lower 
$2\sigma$ error thus calculated were $\le 0$, sources were also 
flagged as non-detections. To distinguish these from the previous
type of non-detections, these were assigned the detection flag 5 in that band.
For non-detections we estimated upper 
confidence limits\footnote{We use the term {\it upper confidence
limit} to stress that this is an estimate of the upper edge
of a confidence interval for the source intensity, regardless of
the detection procedure. The term {\it upper limit}
should be reserved to characterize the detection process
\citep{kashyap2010}}.
for fluxes and confidence levels CL~$= 0.90$
by following \citet[]{kraft1991}.
In this approach, the probability that
a source flux, $S$, lies between $S_{\rm min}$ and $S_{\rm max}$
is given
by 
\exi 
{\rm CL} = \int_{S_{\rm min}}^{S_{\rm max}} f_{N,B} (S) dS \ ,   
\exo
where the posterior probability function for parameter $S$
as a function of observed counts $N$ and mean background $B$
is given by 
\exi 
f_{N,B} (S) = C \frac {e^{-(S+B)}(S+B)^N} {N!}      
\exo
and $C$ is a normalization constant (equation (8) in
\citet{kraft1991}.)

\begin{figure*}\vspace{-5cm}
\epsscale{1.0}
\plotone{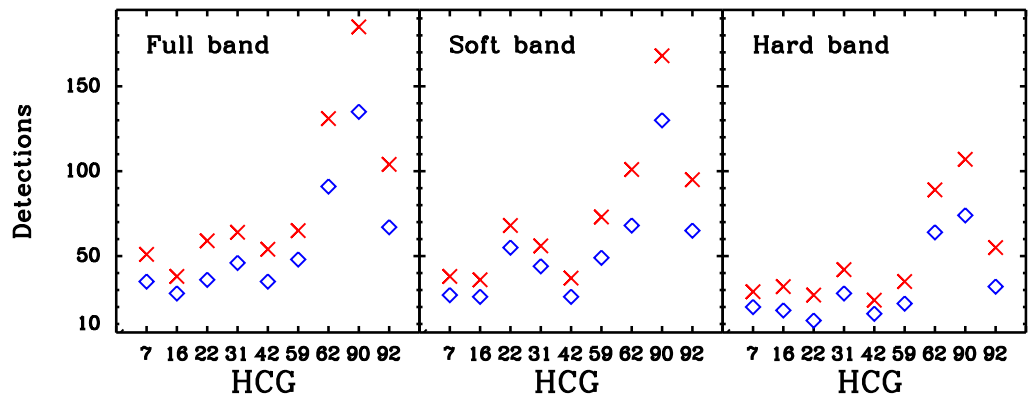}\vspace{-0.5cm}
\caption{Number of sources detected with \wav\ in three \chandra\
\x\ bands for each HCG field. Red crosses are for a significance threshold
of $10^{-5}$ and blue diamonds for a significance threshold of $10^{-6}$.}
\label{fig:w5w6}
\end{figure*}

In our catalogs, we also include alternative detection criteria
(see below).

Most of our sources
have few net counts, precluding
reliable spectral fitting. 
Thus at the third and final stage of the
point source analysis, we applied the method of
\citet{gallagher2005}, which makes use of hardness ratios to obtain rough
estimates of spectral shapes and, hence, fluxes and luminosities for
each source. The hardness ratio is defined as 
\exi
{\rm HR} \equiv (H-S) /
(H+S), \label{equ:hr}
\exo
where $H$ and $S$ represent net counts in the hard and soft
bands, respectively.  For each source this method compares the observed
hardness ratio to that obtained from simulated spectra, in order to
estimate the power-law index $\Gamma$ (where the photon flux is given by
$f_E \propto E^{-\Gamma}$
photon cm\tmtwo\ s\tmone\ keV\tmone) and associated \x\ flux and
luminosity. 
We used the \x\ spectral modeling tool {\sc xspec}
\citep{arnaud1996}, version 12.5.0, to construct grids of simulated
spectra. For each source, we used the corresponding Galactic column
density, \nhgal, as well as the ARFs and RMFs produced by \ace. We imposed a
simple, absorbed power law model ({\tt tbabs*po} in {\sc XSPEC}) and
varied $\Gamma$ in the range $-1 \rightarrow +4$ to obtain simulated
count rates in the full, soft, and hard band, and, thus, simulated HR
values. By comparing with the observed HR, we estimated best $\Gamma$
values and corresponding fluxes and luminosities. 
We illustrate this process in \fr{fig:HR}. Each panel corresponds to
a single X-ray HCG field. For each source detected in at least one
band in this field we plot the simulated HR values against
corresponding $\Gamma$ values. Thus, each grey curve in a panel 
is made up from a set of simulated ${\rm HR} - \Gamma$ pairs for
a single detected source.
Given an {\it observed} HR value and a simulated ${\rm HR} - \Gamma$ curve, 
there is then a unique $\Gamma$ value that provides the best {\it observed}
$\Gamma$ estimate, as indicated by the blue triangles\footnote{We stress 
that there is only one blue triangle per curve, although
the high density of curves in \fr{fig:HR} may suggest otherwise.}.  

\begin{figure*}
\hspace{-1.2cm}
\includegraphics[width=20cm,angle=0]{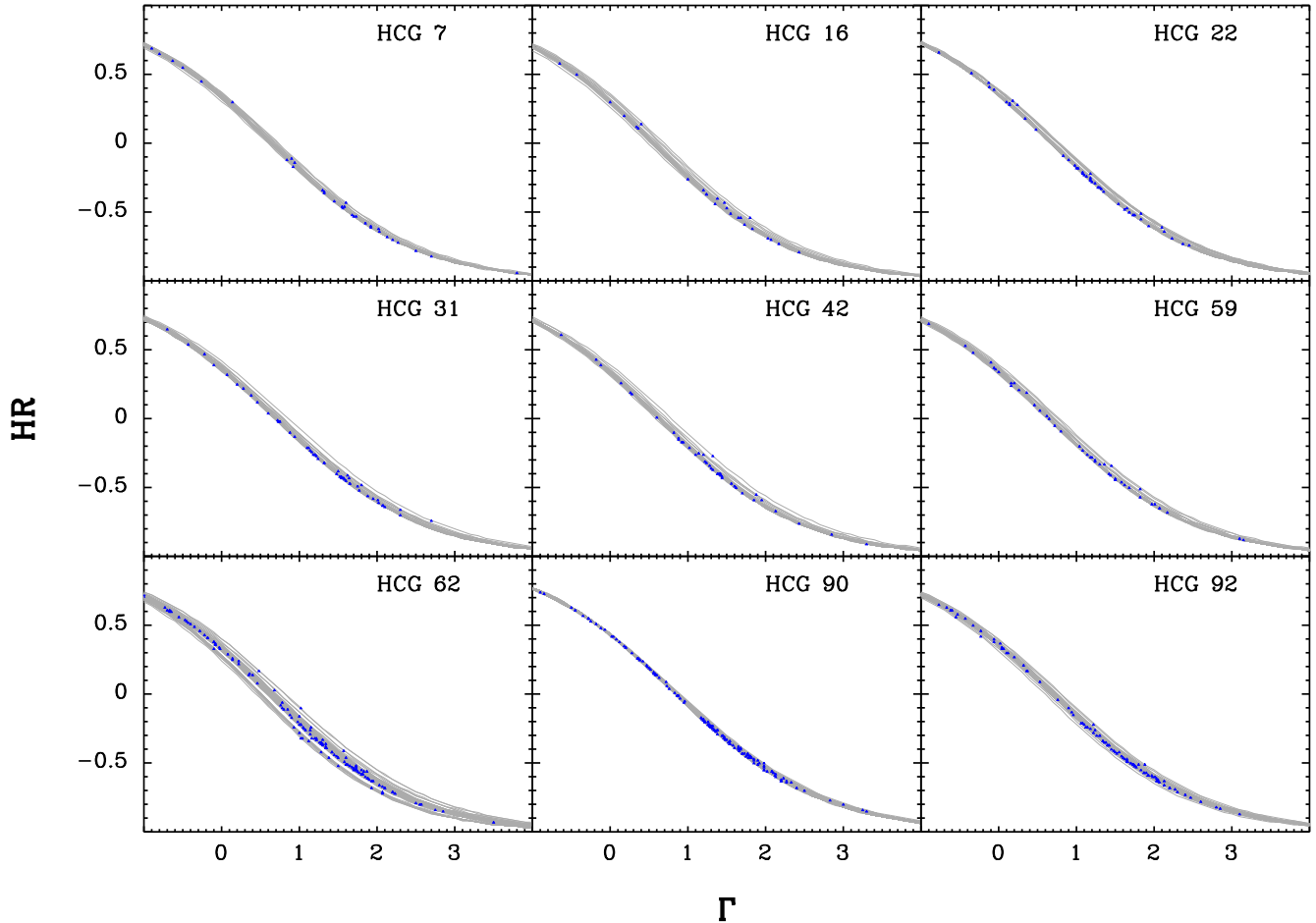}\vspace{-0.5cm}
\caption{Estimating \x\ spectral slopes for point sources in our fields.
Each grey curve is constructed from each detected source's
simulated HR -- $\Gamma$ grid, based on 
\xspec\ simulated spectra that
assume a simple absorbed power law, Galactic \nh\ at the
source right ascension and declination, as well
as ARFs and RMFs specific to the source's position on
the ACIS CCDs. A $\Gamma$ value that best matches
the observed HR can then be obtained for each curve
and corresponding source, as indicated by the blue triangles.     
}
\label{fig:HR}
\end{figure*}

Due to the
simplicity of our model, which assumes no intrinsic absorption (\nhint~=~0), 
it is likely that some $\Gamma$ values are incorrect. However, $\Gamma$ and
\nh\ are degenerate, so that this should have a minimal effect on
luminosity, which is the quantity we are most interested in.
As a comparison, for the nucleus of HCG 62~A our method yields luminosities
\lxhe~=~\ten{1.0}{39}~\lunits\ and
$L_{X,{\rm 0.5-8.0 keV}} = $~\ten{3.2}{39}~\lunits.
These are in a good agreement with
the results reported by \citet{rafferty2013}. 
Using the same \chandra\ data, these authors
carry out more detailed spectral fitting that includes
both intrinsic and Galactic absorption, as well
as a thermal component, and
report \lxht~=~$(1.1\pm0.4) \times 10^{39}$~\lunits\
and 
$L_{X,{\rm 0.5-7.0 keV}} = $~\scier{1.5}{39}{+2.8}{-1.0}~\lunits.

Point source catalogs for all sources are presented in Tables
\ref{tab-ae4-7} to \ref{tab-count-92}\footnote{These tables are 
available only in
the online version of the journal.}. 
For completeness,
the tables include sources considered both detected and undetected,
according to our conservative $2\sigma$ criterion above. We
have compiled two sets of tables. The first set 
(Tables \ref{tab-ae4-7} to \ref{tab-ae4-92})
presents
$\Gamma$\ values, fluxes and luminosities in the full, soft and hard
\x\ bands, derived as explained above, 
as well as the flux density at 2 keV.

The second set 
(Tables \ref{tab-count-7} to \ref{tab-count-92})
presents counts
and count rates in the full, soft and hard band.

Apart from the
Gehrels detection criterion,
for completeness in this second set of tables we also include
the two \ace\ detection criteria, namely the \ace\ {\it significance},
which essentially is a traditional signal-to-noise criterion,
as well as the {\it binomial probability, $P_B$, that a source is spurious} 
\citep[\er{equ:binpro} below; see][Appendix B for details]{broos2010}. 
Users of our catalogs are left to choose which 
detection criterion they prefer.

In both sets of
tables, we also include two types of detection/non-detection
flags for each source. As already mentioned, the first type of flag 
(columns 17, 18, 19 in Tables \ref{tab-ae4-7} to \ref{tab-ae4-92};
columns 3, 5, 7 in Tables \ref{tab-count-7} to \ref{tab-count-92}) is related
to the relative numbers of source and background counts, and is equal
to 1 (net counts negative; no detection in band), 5 (net counts
minus Gehrels $2\sigma$ error~$<0$) or 0 otherwise (unambiguous detection
in band). The second type of flag (column 16 in 
Tables \ref{tab-ae4-7} to \ref{tab-ae4-92}
and column 9 in Tables \ref{tab-count-7} to \ref{tab-count-92})
indicates whether the hardness ratio is an upper or lower limit.
From the HR definition (\er{equ:hr}) it follows that, if there is a detection in
the hard but not the soft band, an HR value is a lower limit (flag value equal
to $-1$).
Conversely, if there is a detection in the soft but not in the hard band,
an HR value is an upper limit (flag value equal to 1). If there is
no detection in either band, this flag is equal to $-2$, and if there is
a detection in both bands, the flag is equal to $0$.
 

Finally, in both sets of tables,
for sources which fall within the boundaries of individual
galaxies (defined as explained in \scr{sec:lims}) 
an upper-case letter in column 1 (ID) indicates the galaxy they belong to.
If these are also nuclear sources, this is indicated by
an asterisk next to the galaxy's letter designation.

\begin{deluxetable*}{cccc ccc}
\tablecolumns{7}
\tabletypesize{\scriptsize}
\tablewidth{0pc} 
\tablecaption{X-ray analysis of HCG nuclear sources \label{tab-xray}}
\tablehead{
\colhead{HCG ID}  
&\colhead{\x\ ID}
&\colhead{$\Gamma$}
&\colhead{\lxte}
&\colhead{c($0.5-8.0$~keV)}
&\colhead{HR}
&\colhead{Strong AGN?}
\\
\colhead{(1)} &
\colhead{(2)} &
\colhead{(3)} &
\colhead{(4)} &
\colhead{(5)} &
\colhead{(6)} & 
\colhead{(7)} 
\\
}
\startdata
     7A &        6 &   $2.0$   &   $40.0$   &   	  \aer{  102}{+   11}{  -10}   &    \aer{-0.62}{+0.09}{-0.08} & n \\
      7B &       15 &   $1.9$   &   $39.0$   &   	  \aer{    9}{+    4}{   -3}   &    \aer{-0.58}{+0.00}{-0.42} & n \\
      7C &       44 &   $2.1$   &   $39.1$   &   	  \aer{   12}{+    4}{   -3}   &    \aer{-0.68}{+0.00}{-0.32} & n \\
      7D &    \ldots&   \ldots  &   \ldots   &   	  \ldots                       &    \ldots                    & \ldots\\
     16A &        9 &   $1.2$   &   $40.7$   &   	  \aer{  134}{+   12}{  -11}   &    \aer{-0.37}{+0.09}{-0.08} & n\\
     16B &        4 &   $-0.7$  &   $41.3$   &   	  \aer{  189}{+   14}{  -13}   &    \aer{0.58}{+0.06}{-0.07}  & y\\
     16C &       44 &   $2.0$   &   $39.7$   &   	  \aer{   24}{+    6}{   -4}   &    \aer{-0.69}{+0.21}{-0.15} & n\\
     16D &       57 &   $2.4$   &   $39.8$   &   	  \aer{   37}{+    7}{   -6}   &    \aer{-0.79}{+0.15}{-0.10} & n\\
     22A &    77,76 &   $1.7$($1.1$)   &   $38.7$($38.7$) & \aer{    9}{+    4}{   -3}(\aer{    7}{+    }{   -2})   & \aer{-0.50}{+0.00}{-0.50}(\aer{-0.23}{+0.00}{-0.77})  & n\\
     22B &       22 &   $1.2$   &   $38.7$   &   	  \aer{    7}{+    3}{   -2}   &    \aer{-0.28}{+0.00}{-0.72} & n\\
     22C &    \ldots&   \ldots  &   \ldots   &   	  \ldots     &      \ldots& n\\                                                  
   31ACE &    40,38 &   $1.3$($1.1$)   &   $40.1$($40.3$)   &   	  \aer{   73}{+    9}{   -8}(\aer{  128}{+   1}{  -11})   &    \aer{-0.32}{+0.12}{-0.12}(\aer{-0.21}{+0.09}{-0.09})  & n\\
     31B &    \ldots&   \ldots  &   \ldots   &   	  \ldots     &      \ldots&\ldots \\                                                  
     31F &    \ldots&   \ldots  &   \ldots   &   	  \ldots     &      \ldots& \ldots\\                                                  
     31G &       63 &   $1.6$   &   $40.0$   &   	  \aer{   72}{+    9}{   -8}    &   \aer{-0.43}{+0.12}{-0.11} & n\\
     31Q &       41 &   $1.1$   &   $38.9$   &   	  \aer{    4}{+    3}{   -2}    &   \aer{-0.23}{+0.00}{-0.77} & n\\
     42A &       18 &   $3.3$   &   $40.3$   &   	  \aer{  178}{+   14}{  -13}    &   \aer{-0.91}{+0.04}{-0.03} & n\\
     42B &    \ldots&   \ldots  &   \ldots   &   	  \ldots     &      \ldots& \ldots\\                                                  
     42C &        6 &   $2.4$   &   $39.3$   &   	  \aer{   19}{+    5}{   -4}    &   \aer{-0.76}{+0.00}{-0.24} & n\\
     42D &    \ldots&   \ldots  &   \ldots   &   	  \ldots     &      \ldots& \ldots\\                                                  
     59A &       50 &   $0.8$   &   $40.1$   &   	  \aer{   43}{+    7}{   -6}    &   \aer{-0.09}{+0.17}{-0.16} & n\\
     59B &       26 &   $1.5$   &   $39.3$   &   	  \aer{   10}{+    4}{   -3}    &   \aer{-0.44}{+0.00}{-0.56} & n\\
     59C &    \ldots&   \ldots  &   \ldots   &   	  \ldots     &      \ldots& \ldots \\                                                  
     59D &    \ldots&   \ldots  &   \ldots   &   	  \ldots     &      \ldots& \ldots \\                                                  
     62A &       83 &   $2.5$   &   $39.5$   &   	  \aer{  133}{+   12}{  -11}    &   \aer{-0.79}{+0.07}{-0.05} & n\\
     62B &       67 &   $2.6$   &   $39.4$   &   	  \aer{  116}{+   11}{  -10}    &   \aer{-0.80}{+0.07}{-0.06} & n\\
     62C &      115 &   $1.7$   &   $39.0$   &   	  \aer{   33}{+    6}{   -5}    &   \aer{-0.54}{+0.18}{-0.15} & n\\
     62D &    \ldots&   \ldots  &   \ldots   &   	  \ldots                        &             \ldots & \ldots\\                      
     90A &       88 &   $-1.1$  &   $42.6$   &   	  \aer{23361}{+  153}{ -152}    &   \aer{0.95}{+0.00}{-0.00}  & y\\
     90B &      164 &   $2.3$   &   $39.1$   &   	  \aer{   31}{+    6}{   -5}    &   \aer{-0.64}{+0.18}{-0.14} & n\\
     90C &      108 &   $1.5$   &   $39.1$   &   	  \aer{   23}{+    5}{   -4}    &   \aer{-0.33}{+0.23}{-0.21} & n\\
     90D &    \ldots&   \ldots  &   \ldots   &   	  \ldots                        &      \ldots                 & \ldots\\                    
     92B &       46 &   $1.8$   &   $39.4$   &   	  \aer{   19}{+    5}{   -4}    &   \aer{-0.53}{+0.25}{-0.21} & n\\
     92C &       94 &   $-1.1$  &   $42.3$   &   	  \aer{ 3375}{+   59}{  -58}    &   \aer{0.74}{+0.01}{-0.01}  & y\\
     92D &       36 &   $2.6$   &   $39.8$   &   	  \aer{   75}{+    9}{   -8}    &   \aer{-0.78}{+0.09}{-0.07} & n\\
     92E &       22 &   $2.3$   &   $39.6$   &   	  \aer{   40}{+    7}{   -6}    &   \aer{-0.69}{+0.15}{-0.12} & n\\
     92F &      145 &   $2.9$   &   $39.4$   &   	  \aer{   30}{+    6}{   -5}    &   \aer{-0.83}{+0.00}{-0.17} & n\\        
\enddata
\tablecomments{
Columns are:
(1) HCG galaxy ID;
(2) \x\ source ID in master list;
(3) $\Gamma$ value from hardness ratio;
(4) log \lx\ in full band;
(5) net counts in full band;
(6) hardness ratio;
(7) X-rays indicate strong AGN (\lxte~$\ge 10^{41}$~\lunits).
}
\end{deluxetable*}

\begin{figure*}[ht]
\epsscale{1.4}
\centering
\newcommand{\wid}{9cm}
\subfloat
{
\hspace{-1.5cm}
\includegraphics[width=\wid]{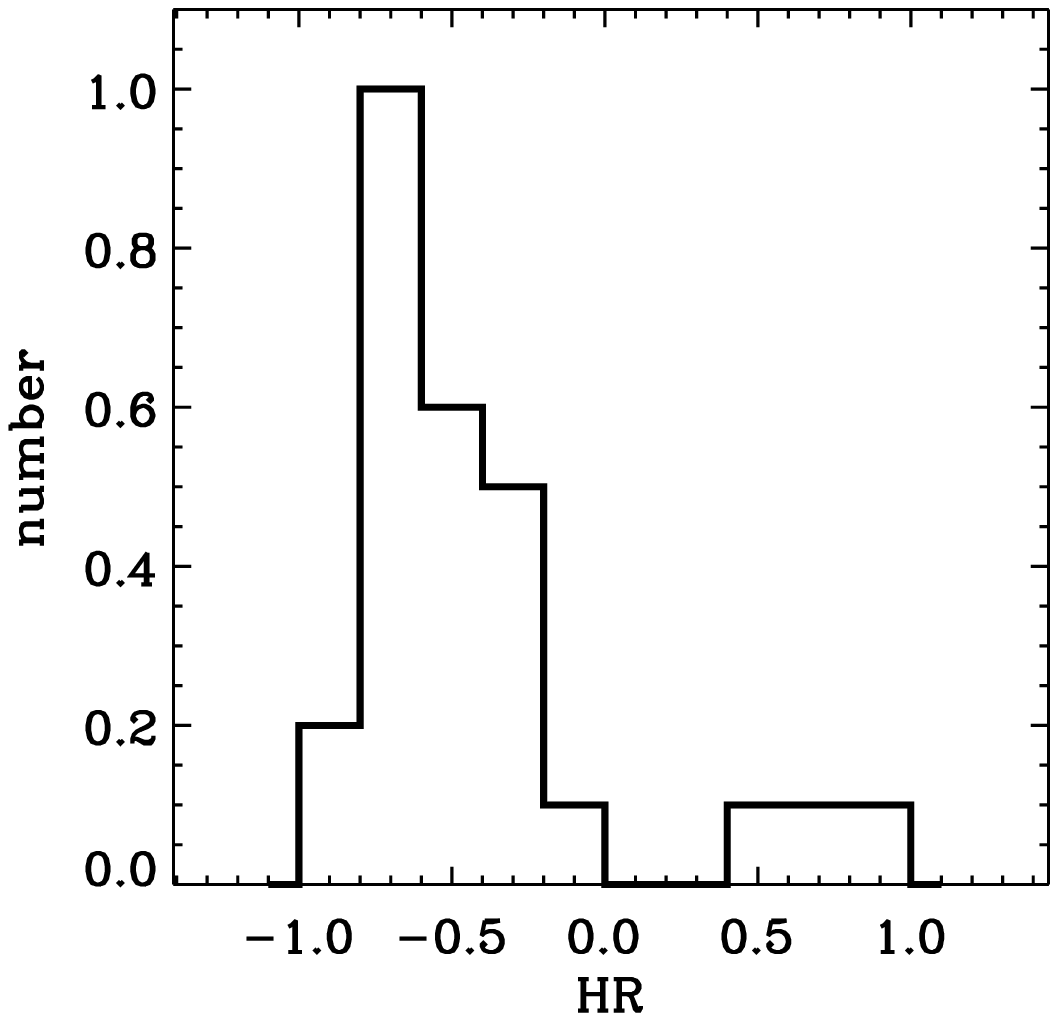}\hspace{-3cm}
\includegraphics[width=\wid]{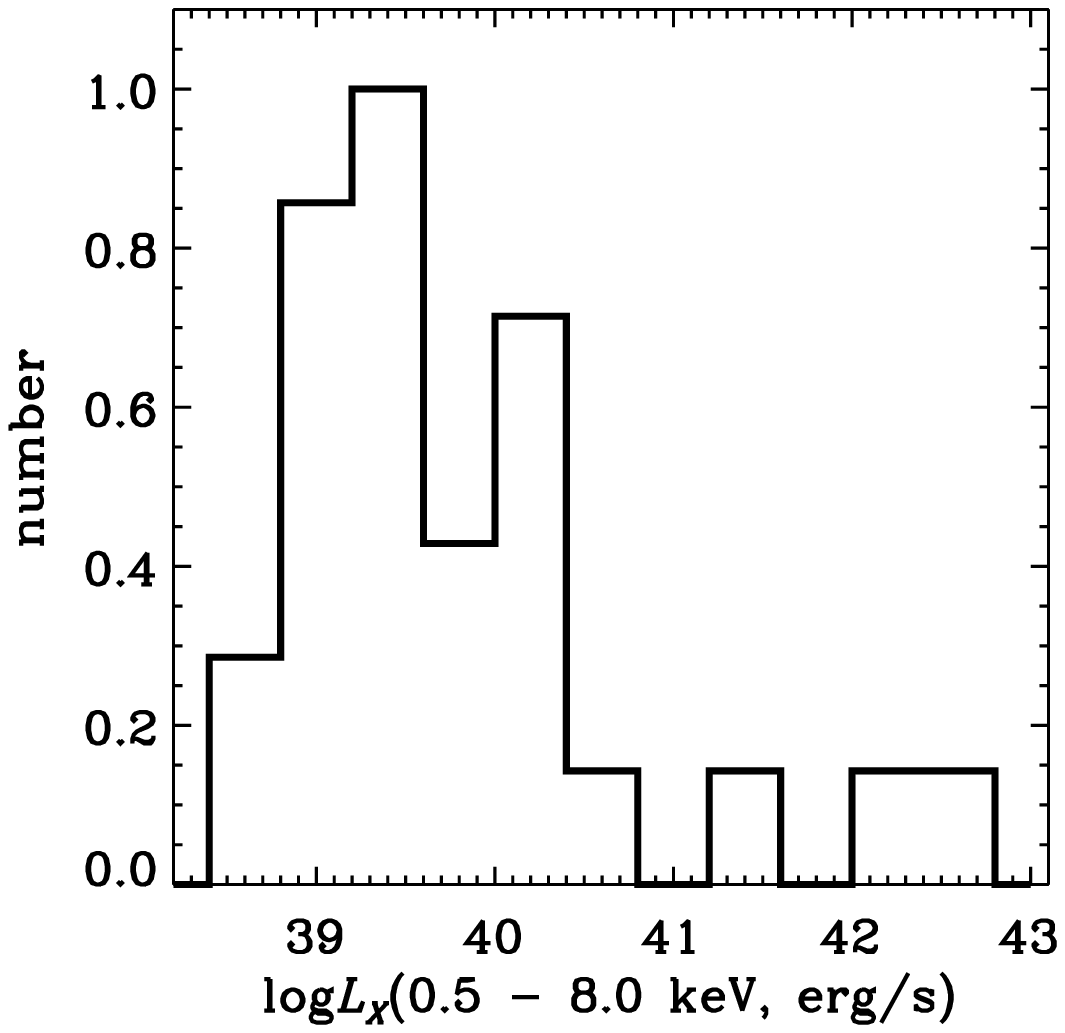}\hspace{-3cm}
\includegraphics[width=\wid]{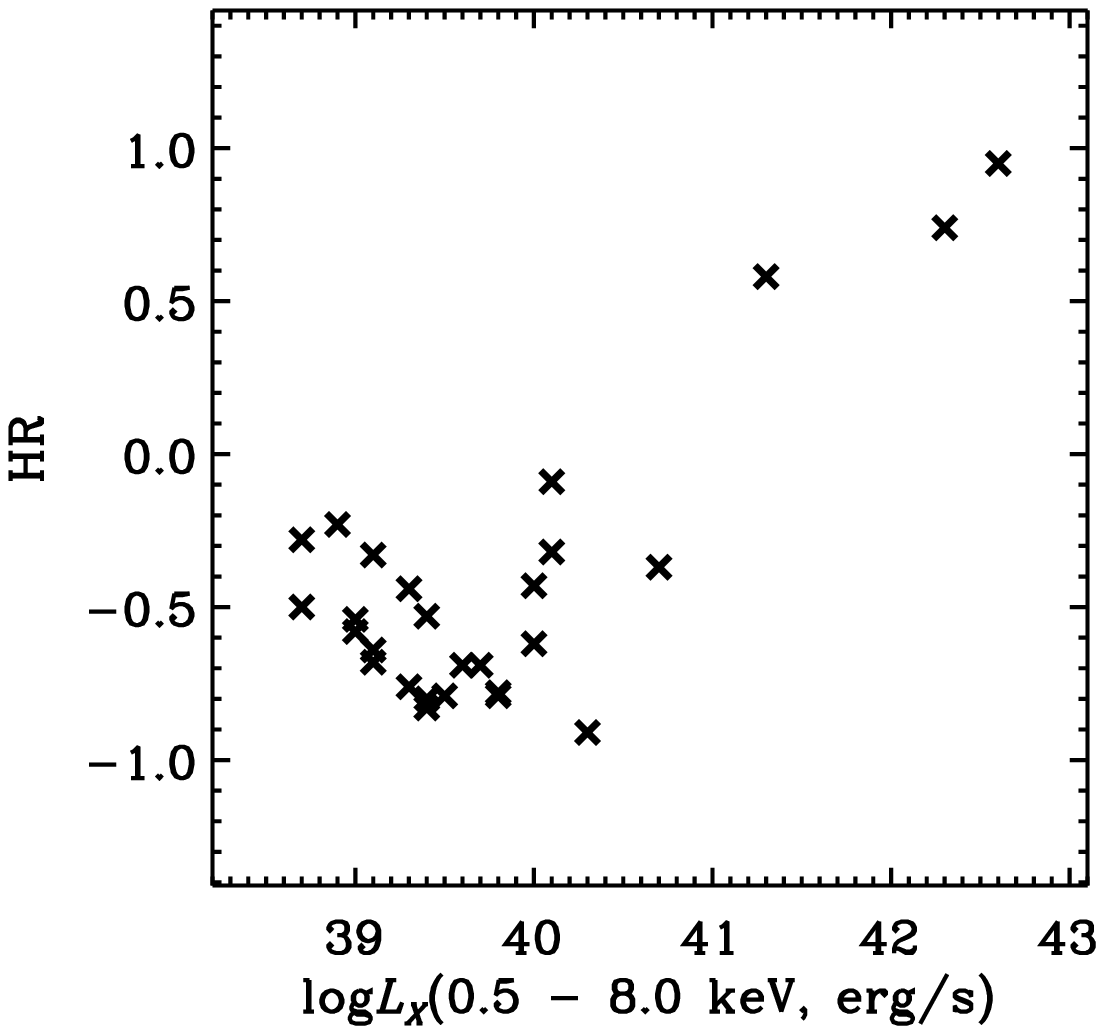}
}
\caption{ 
Hardness ratios and \x\ luminosities for nuclear sources
(\tr{tab-xray}). The left and middle panels show the normalized distributions
of HR and \lxte\ values for \x\ nuclear detections in HCG galaxies.
The right panel plots HR against \lxte\ for the same sources.
With the exception of three sources, HCG nuclei have negative
HR values and low luminosities in the X-rays.
}
\label{fig:lxhr}
\end{figure*}

A summary of key results from the \x\ analysis for all galaxy
nuclei in this sample is presented in \tr{tab-xray}. 
In this table, the ID number in column 2 refers to the master ID running
number in column 1 of Tables \ref{tab-ae4-7} to \ref{tab-count-92}.
For instance, for the nucleus of HCG 7A this number is 6 (first row of
\tr{tab-xray}).
This indicates that this is the sixth source in the master \x\ catalog of group HCG 7.
Since this source belongs to galaxy A and is a nuclear source,
it is listed as 6A$^{\star}$ in Tables \ref{tab-ae4-7} and \ref{tab-count-7}.

\tr{tab-xray} shows that there appear to be nuclear \x\ detections for 
27 out of 37 HCG galaxies. 
According to column 6, the great majority of \x\ sources associated with
galaxy nuclei (24/27 or 89\%) are soft (HR~$<0$).
Comparing with column 4, we see that these are also the sources
with low luminosities, \lxte~$< 10^{41}$~\lunits.
What about the remaining three sources?
The last column indicates whether a galaxy is
a good candidate for being a {\it strong} AGN host, based on 
whether \lxte~$\ge 10^{41}$~\lunits. The three remaining sources
(3/27 or 11\%) are those that fulfill this condition
and are also those that have HR~$>0$. 
These trends are also easy to see in \fr{fig:lxhr}. The left and middle
panels show 
histograms for HR and \lxte, while the right panel plots
HR against \lxte. Again, a small minority of 
positive HR, high \lxte\ sources are good
strong AGN candidates. Note that the last column
of \tr{tab-xray} also appears in \tr{tab-multi}
(its column 8), which combines multiwavelength
nuclear activity diagnostics.

\begin{deluxetable*}{cccc cccc ccc } 
\tablecolumns{11}
\tablewidth{0pc} 
\tablecaption{Flux limit estimates and \x\ 
point sources associated with galaxies.\label{tab-flim}}
\tablehead{ 
\colhead{ID}
& \colhead{\flims}
& \colhead{\lxlims}
& \colhead{\nobss}
& \colhead{\nexcs}
& \colhead{$P_S$}
& \colhead{\flimh}
& \colhead{\lxlimh}
& \colhead{\nobsh}
& \colhead{\nexch}
& \colhead{$P_H$}
\\
\colhead{}
& \colhead{(\funits)}
& \colhead{(\lunits)}
& \colhead{}
& \colhead{}
& \colhead{}
& \colhead{(\funits)}
& \colhead{(\lunits)}
& \colhead{}
& \colhead{}
& \colhead{}
\\
\colhead{(1)}
& \colhead{(2)}
& \colhead{(3)}
& \colhead{(4)}
& \colhead{(5)}
& \colhead{(6)}
& \colhead{(7)}
& \colhead{(8)}
& \colhead{(9)}
& \colhead{(10)}
& \colhead{(11)}
}
\startdata
HCG   7   A &$  5.8\times 10^{16}$&$  2.2\times 10^{38}$ &  4  & 3.5 & 0.9 &$  2.0\times 10^{15} $&$  7.6\times 10^{38}$ &  3  & 2.3 & 0.8 \\
HCG   7   B &$  2.9\times 10^{16}$&$  1.1\times 10^{38}$ &  3  & 2.5 & 0.8 &$  1.5\times 10^{15} $&$  5.7\times 10^{38}$ &  0  & -0.5 & 0.0 \\
HCG   7   C &$  3.9\times 10^{16}$&$  1.5\times 10^{38}$ &  6  & 5.5 & 0.9 &$  2.5\times 10^{15} $&$  9.5\times 10^{38}$ &  2  & 1.6 & 0.8 \\
HCG  16   A &$  1.9\times 10^{15}$&$  6.6\times 10^{38}$ &  2  & 1.9 & 0.9 &$  5.8\times 10^{15} $&$  2.0\times 10^{39}$ &  2  & 1.9 & 0.9 \\
HCG  16   B &$  1.7\times 10^{15}$&$  5.7\times 10^{38}$ &  1  & 0.9 & 0.9 &$  7.2\times 10^{15} $&$  2.5\times 10^{39}$ &  1  & 1.0 & 1.0 \\
HCG  16   C &$  3.9\times 10^{15}$&$  1.3\times 10^{39}$ &  6  & 5.9 & 1.0 &$  5.8\times 10^{15} $&$  2.0\times 10^{39}$ &  2  & 1.8 & 0.9 \\
HCG  16   D &$  1.4\times 10^{15}$&$  4.7\times 10^{38}$ &  1  & 0.9 & 0.9 &$  4.4\times 10^{15} $&$  1.5\times 10^{39}$ &  1  & 0.8 & 0.8 \\
HCG  22   A &$  4.4\times 10^{16}$&$  6.9\times 10^{37}$ & 15  & 13.9 & 0.9&$  1.7\times 10^{15} $&$  2.7\times 10^{38}$ &  4  & 2.7 & 0.7 \\
HCG  22   B &$  3.3\times 10^{16}$&$  5.2\times 10^{37}$ &  1  & 0.8 & 0.8 &$  1.7\times 10^{15} $&$  2.7\times 10^{38}$ &  0  & -0.2 & 0.0 \\
HCG  31 ACE &$  1.1\times 10^{15}$&$  4.3\times 10^{38}$ &  9  & 8.9 & 1.0 &$  2.6\times 10^{15} $&$  1.0\times 10^{39}$ &  9  & 8.8 & 1.0 \\
HCG  31   G &$  5.8\times 10^{16}$&$  2.3\times 10^{38}$ &  5  & 4.9 & 1.0 &$  2.1\times 10^{15} $&$  8.3\times 10^{38}$ &  3  & 2.9 & 1.0 \\
HCG  31   Q &$  2.9\times 10^{16}$&$  1.2\times 10^{38}$ &  1  & 1.0 & 1.0 &$  1.5\times 10^{15} $&$  6.2\times 10^{38}$ &  0  & 0.0 & 0.0 \\
HCG  42   A &$  4.0\times 10^{15}$&$  1.9\times 10^{39}$ &  1  & 0.8 & 0.8 &$  3.4\times 10^{15} $&$  1.6\times 10^{39}$ &  1  & 0.2 & 0.2 \\
HCG  42   C &$  1.2\times 10^{15}$&$  5.6\times 10^{38}$ &  4  & 3.7 & 0.9 &$  3.4\times 10^{15} $&$  1.6\times 10^{39}$ &  1  & 0.7 & 0.7 \\
HCG  59   A &$  5.4\times 10^{16}$&$  2.6\times 10^{38}$ &  1  & 0.9 & 0.9 &$  1.4\times 10^{15} $&$  6.9\times 10^{38}$ &  1  & 0.8 & 0.8 \\
HCG  59   B &$  2.7\times 10^{16}$&$  1.3\times 10^{38}$ &  3  & 2.9 & 1.0 &$  1.4\times 10^{15} $&$  6.9\times 10^{38}$ &  0  & -0.1 & 0.0 \\
HCG  62   A &$  3.5\times 10^{15}$&$  1.7\times 10^{39}$ & 11  & 10.9 & 1.0&$  1.2\times 10^{15} $&$  5.9\times 10^{38}$ &  6  & 5.6 & 0.9 \\
HCG  62   B &$  1.2\times 10^{15}$&$  5.9\times 10^{38}$ &  6  & 5.9 & 1.0 &$  8.7\times 10^{16} $&$  4.3\times 10^{38}$ &  4  & 3.7 & 0.9 \\
HCG  62   C &$  2.5\times 10^{16}$&$  1.2\times 10^{38}$ &  9  & 8.7 & 1.0 &$  6.5\times 10^{16} $&$  3.2\times 10^{38}$ &  8  & 7.5 & 0.9 \\
HCG  90   A &$  1.0\times 10^{15}$&$  1.4\times 10^{38}$ &  1  & 0.7 & 0.7 &$  2.2\times 10^{14} $&$  3.1\times 10^{39}$ &  1  & 1.0 & 1.0 \\
HCG  90   B &$  6.7\times 10^{16}$&$  9.3\times 10^{37}$ &  7  & 6.8 & 1.0 &$  1.7\times 10^{15} $&$  2.3\times 10^{38}$ &  2  & 1.6 & 0.8 \\
HCG  90   C &$  4.5\times 10^{16}$&$  6.2\times 10^{37}$ &  5  & 4.8 & 1.0 &$  1.3\times 10^{15} $&$  1.8\times 10^{38}$ &  5  & 4.8 & 1.0 \\
HCG  92   B &$  1.8\times 10^{16}$&$  1.7\times 10^{38}$ &  2  & 1.9 & 0.9 &$  6.5\times 10^{16} $&$  6.1\times 10^{38}$ &  1  & 0.9 & 0.9 \\
HCG  92   C &$  1.3\times 10^{15}$&$  1.3\times 10^{39}$ &  2  & 2.0 & 1.0 &$  3.1\times 10^{15} $&$  2.9\times 10^{39}$ &  2  & 2.0 & 1.0 \\
HCG  92   D &$  2.1\times 10^{16}$&$  2.0\times 10^{38}$ &  2  & 1.9 & 0.9 &$  8.1\times 10^{16} $&$  7.6\times 10^{38}$ &  1  & 0.9 & 0.9 \\
HCG  92   E &$  1.8\times 10^{16}$&$  1.7\times 10^{38}$ &  2  & 1.8 & 0.9 &$  8.1\times 10^{16} $&$  7.6\times 10^{38}$ &  1  & 0.8 & 0.8 \\
HCG  92   F &$  3.7\times 10^{16}$&$  3.5\times 10^{38}$ &  1  & 0.8 & 0.8 &$  2.9\times 10^{15} $&$  2.7\times 10^{39}$ &  0  & -0.1 & 0.0 \\
\hline
\enddata
\tablecomments{Columns are: 
(1) HCG galaxy ID; 
(2) flux limit estimate in soft band;
(3) luminosity limit estimate in soft band;
(4) number of detected point sources inside galaxy region;
(5) number of point sources in the soft band and 
inside galaxy region that are in 
excess of the number expected from the background $\log N - \log S$;
(6) probability estimate that in the soft band point sources
detected in this galaxy belong to the galaxy and are not background AGN
(column (5) / column (4) );
(7) as column (2) for the hard band;
(8) as column (3) for the hard band;
(9) as column (4) for the hard band;
(10) as column (5) for the hard band;
(11) as column (6) for the hard band.
}
\end{deluxetable*}

\begin{deluxetable}{cccc}
\tablecolumns{4}
\tablecaption{\swift\ UVOT magnitude limits for HCG fields\label{tab-uvmaglim}}
\tablehead{
\colhead{HCG ID}
&\colhead{\wtwo}
&\colhead{\mtwo}
&\colhead{\wone}
\\
\colhead{}
&\colhead{(2030~\AA)}
&\colhead{(2231~\AA)}
&\colhead{(2634~\AA)}
\\
\colhead{(1)}
&\colhead{(2)}
&\colhead{(3)}
&\colhead{(4)}
}
\startdata
7 & 20.5 & 20.4 & 20.6 \\
16 & 21.6 & 20.6 & 20.6 \\
22 & 20.2 & 20.7 & 20.4 \\
31 & 20.4 & 20.3 & 20.5 \\
42 & 20.5 & 20.1 & 20.9 \\
59 & 20.7 & 20.2 & 21.1 \\
62 & 20.6 & 20.3 & 20.9 \\
90 & 22.6 & 22.8 & 21.6 \\
92 & 20.9 & 21.0 & 20.3 \\
\enddata
\tablecomments{
Columns are:
(1) HCG field ID; (2), (3), (4): magnitude limit estimates for the
\wtwo\ (effective wavelength 2030~\AA), \mtwo\ (2231~\AA),
\wone\ (2634~\AA) \uvot\ filters.
}\end{deluxetable}

\subsection{Flux limits and source statistics}\label{sec:lims}
Although we detect a large number of \x\ point sources in our fields,
we have no a priori information which of these sources
are physically associated with HCG galaxies and are not 
just background AGNs. We assess the effect of AGN background contamination as
follows.

The binomial probability, $P_B$, that a source is spurious is given
by the binomial function \citep{broos2010}
\exi
P_B = f_b ( C_s; C_s+C_b, (1+A_b / A_s)^{-1}) \ , \label{equ:binpro}
\exo
where $C_s$ and $C_b$ are the number of counts observed in the
source and background region in a given energy band, and
$A_s$, $A_b$ are the areas of the source and background regions.
In other words, sources with values of $P_B$ {\it less} than a given 
threshold value may be considered as detections. We adopt the threshold
$P_B = 0.004$ established by \citet{xue2011} and use local
background information for each detected nuclear source
to establish a detectability limit
in terms of counts, fluxes and luminosities in the soft and
hard band at the location of each galaxy on the ACIS
CCD. Each of our nuclear
sources has associated background and source extraction regions with
measured background counts. The advantage of using these regions, together
with measured background counts,
is that they have been
constructed and measured by
{\sc acis extract} by taking into account the size of the \chandra\ PSF at
the location of the particular source on the CCD. Thus the detection
limits that we calculate are local, position-dependent and specific
to each galaxy. We fix $C_b$ to the locally measured background
counts and start by setting the source counts $C_s =0$, in which
case $P_B >> 0.004$ (a source with no source counts must be spurious).
We then iteratively increase $C_s$ to estimate the minimum
number of source counts, $C_{s,{\rm lim}}$, 
required to reach our chosen probability threshold
of 0.004.    
We convert the estimated $C_{s,{\rm lim}}$ value to a flux limit
by assuming a power law spectrum
with $\Gamma=1.4$ \citep{hickox2006,steffen2007}
and the Galactic $N_H$ value for each galaxy group in
{\sc pimms} \citep{mukai1993}.
The flux and corresponding luminosity limits for the soft and hard
band are shown in columns 2, 3 (soft band) and 7,8 (hard band)
of \tr{tab-flim}. As expected, the estimated limit fluxes are
lower in the soft band, due to the higher sensitivity of
ACIS in this regime.

We use these flux limits to estimate how many background AGNs
we expect to see inside the galaxy regions of our galaxies.
We use the \lq\lq $\log N - \log S$\rq\rq\ relation of
\citet{cappelluti2007} which relates the number of 
detected point sources per angular area in the soft and hard
band as a function of flux, established over 2 deg$^2$ in the 
COSMOS field.
We thus estimate for each band 
the expected total number of background AGNs
that would be detected over the area of each
of our galaxy regions. These regions
are determined following
\citet{tzanavaris2010}
who use mid-infrared defined galaxy regions \citep{johnson2007} for their
global galaxy photometry.
For each band and galaxy, we then compare the number of expected AGNs to
the number of detected sources, \nobss\ and \nobsh\ (columns 4 and 9
in \tr{tab-flim}), and calculate the number of detected sources which
are in excess of the expected background number, \nexcs\ and
\nexch\ (columns 5 and 10 in \tr{tab-flim}). The ratio of
the number of excess sources over the total number of observed
sources is a rough estimate of the probability that a
source detected inside a galaxy region is not a background AGN.
Although this method cannot tell us whether a {\it specific}
source is likely to be a background source or not, it does
provide
an overall estimate of whether background contamination may be a serious
concern for our galaxies. The probability estimates in 
\tr{tab-flim} (columns 6 and 11) are in general quite high. Note that
in these columns values equal to zero are simply due to non-detections
in one band. Otherwise, rounded values range from 0.7 to 1.0, giving
us confidence that our point source detections are likely to be due
to HCG galaxies.

\section{UV data and data analysis}
\subsection{Data}
For all galaxies in this sample we use 3-band UV data
obtained with UVOT
on NASA's \swift\ Gamma-ray Burst Explorer \citep{2004ApJ...611.1005G}. 
Details on the telescope, filters,
observations, and data reduction can be found in \citet{tzanavaris2010}.
Magnitude limits for each HCG field are estimated using the
zero points specific to each UVOT filter \citep{poole2008} and
shown in \tr{tab-uvmaglim}.
For quick reference, we mention here that the three UV bands
used are \wtwo, \mtwo\ and \wone, with effective wavelengths
2030, 2231 and 2634~\AA, respectively.

As mentioned, in the present work we do include UVOT data for HCG
90. These were excluded in \citet{tzanavaris2010}, as some of the HCG
galaxies were not fully covered by the stacked UVOT exposures. A careful
re-analysis of the UVOT data for this group reveals that they consist
of two exposure \lq\lq stacks\rq\rq. The HCG galaxies are not covered
by three out of eleven individual exposures in stack 00053602001. We
thus combined the eight useful exposures of this stack with the
second stack. The corrected total exposure times for HCG 90 and
the \wtwo, \mtwo\
and \wone\ filters are now 7387s, 6732s and 5525s (c.f. 8601s, 7946s,
6663s in \citet[][Table 3]{tzanavaris2010}).

We also include UVOT data for HCG 92 (Stephan's Quintet).
The UVOT observation ID's for this group are
00035083005, 00035083007, 00035083008 and 00035083009, with total
exposure times for the \wtwo, \mtwo\ and \wone\ filters
of 3054s, 3214s and 1007s, respectively.

\subsection{UV nuclear photometry}
\citet{tzanavaris2010} carried out galaxy-wide photometry 
for their HCG
galaxies in order to calculate galaxy star formation rates.
In this paper we are interested in comparing nuclear fluxes
in the UV and \x\ regime and thus perform nuclear photometry
as described below.

Using the \wone\ ($\sim 2600$~\AA) image, we define circular source
regions of radius 5\arcsec, centered at the central intensity peak of
each galaxy.  The choice of radius is dictated both by the UVOT PSF
\citep[2.37\arcsec, FWHM, for \wone,][]{breeveld2010} and the fact
that the UVOT count rate to flux conversion factors have been
calibrated for such a radius.  We use the \wone\ image because its
effective wavelength is closest to the one traditionally used in
estimating the \x-to-UV spectral index \exi \alpha_{\rm OX} \equiv
0.380 \log ( L_{\nu, {\rm 2 keV}} / L_{\nu, {uvw1}} ) \ , \exo
sometimes referred to as \lqq \x\ loudness\rqq\ \citep{tananbaum1979},
or \x-to-\lq\lq optical\rq\rq\ spectral index\footnote{Note that
  with this definition, in this paper {\it lower} \aox\ values
  are more {\it negative}.}. Note that, since the effective wavelength
of the \wone\ filter is $\sim 100$~\AA\ redward of 2500~\AA, this is
very close, but not identical, to frequent definitions of this index
which use the 2500~\AA\ luminosity instead. However, given that the
spectral slope is essentially flat in the near-to-far UV spectral
region \citep{kennicutt1998}, we expect this discrepancy to have a
negligible effect, especially given the scatter in our \aox\ results
(more than an order of magnitude, \scr{sec:xuv}).

\begin{deluxetable}{cccc ccc}
\tablecolumns{5}
\tablecaption{\swift\ UVOT nuclear photometry \label{tab-uvfl}}
\tablehead{
\colhead{HCG ID}
&\colhead{\fnu(2030~\AA)}
&\colhead{\fnu(2231~\AA)}
&\colhead{\fnu(2634~\AA)}
&\colhead{\nulnu(2634~\AA)}
\\
\colhead{(1)}
&\colhead{(2)}
&\colhead{(3)}
&\colhead{(4)}
&\colhead{(5)}
}
\startdata
7A  & $0.13\pm 0.01$ & $0.16\pm 0.01$ & $0.27\pm 0.01$ & $42.1$ \\
7B  & $0.06\pm 0.00$ & $0.05\pm 0.01$ & $0.17\pm 0.01$ & $41.9$ \\
7C  & $0.20\pm 0.01$ & $0.24\pm 0.01$ & $0.26\pm 0.01$ & $42.1$ \\
7D  & $0.14\pm 0.01$ & $0.16\pm 0.01$ & $0.17\pm 0.01$ & $41.9$ \\
16A  & $0.58\pm 0.03$ & $0.71\pm 0.03$ & $0.97\pm 0.04$ & $42.6$ \\
16B  & $0.04\pm 0.00$ & $0.03\pm 0.01$ & $0.12\pm 0.01$ & $41.7$ \\
16C  & $1.32\pm 0.05$ & $1.52\pm 0.05$ & $1.77\pm 0.07$ & $42.8$ \\
16D  & $0.10\pm 0.01$ & $0.11\pm 0.01$ & $0.20\pm 0.01$ & $41.9$ \\
22A  & $0.12\pm 0.01$ & $0.10\pm 0.01$ & $0.29\pm 0.02$ & $41.7$ \\
22B  & $0.04\pm 0.00$ & $0.04\pm 0.01$ & $0.07\pm 0.01$ & $41.1$ \\
22C  & $0.11\pm 0.01$ & $0.12\pm 0.01$ & $0.10\pm 0.01$ & $41.3$ \\
31ACE  & $2.74\pm 0.13$ & $3.15\pm 0.11$ & $2.59\pm 0.11$ & $43.1$ \\
31B  & $0.35\pm 0.02$ & $0.42\pm 0.02$ & $0.37\pm 0.02$ & $42.2$ \\
31F  & $0.17\pm 0.02$ & $0.19\pm 0.01$ & $0.17\pm 0.01$ & $41.9$ \\
31G  & $1.33\pm 0.07$ & $1.50\pm 0.06$ & $1.28\pm 0.06$ & $42.8$ \\
31Q  & $0.14\pm 0.01$ & $0.15\pm 0.01$ & $0.15\pm 0.01$ & $41.8$ \\
42A  & $0.18\pm 0.01$ & $0.17\pm 0.01$ & $0.36\pm 0.02$ & $42.3$ \\
42B  & $0.05\pm 0.01$ & $0.03\pm 0.01$ & $0.10\pm 0.01$ & $41.7$ \\
42C  & $0.05\pm 0.01$ & $0.04\pm 0.01$ & $0.14\pm 0.01$ & $41.9$ \\
42D  & $0.02\pm 0.00$ & $0.02\pm 0.01$ & $0.05\pm 0.01$ & $41.4$ \\
59A  & $0.06\pm 0.01$ & $0.07\pm 0.01$ & $0.14\pm 0.01$ & $41.9$ \\
59B  & $0.02\pm 0.00$ & $0.02\pm 0.01$ & $0.05\pm 0.01$ & $41.5$ \\
59C  & $0.08\pm 0.01$ & $0.09\pm 0.01$ & $0.09\pm 0.01$ & $41.7$ \\
59D  & $0.19\pm 0.01$ & $0.21\pm 0.01$ & $0.17\pm 0.01$ & $42.0$ \\
62A  & $0.09\pm 0.01$ & $0.08\pm 0.01$ & $0.20\pm 0.01$ & $42.1$ \\
62B  & $0.07\pm 0.01$ & $0.06\pm 0.01$ & $0.17\pm 0.01$ & $42.0$ \\
62C  & $0.03\pm 0.00$ & $0.03\pm 0.01$ & $0.09\pm 0.01$ & $41.7$ \\
62D  & $0.02\pm 0.00$ & $0.02\pm 0.01$ & $0.06\pm 0.01$ & $41.5$ \\
90A  & $0.05\pm 0.00$ & $0.04\pm 0.01$ & $0.11\pm 0.01$ & $41.3$ \\
90B  & $0.15\pm 0.01$ & $0.12\pm 0.01$ & $0.36\pm 0.02$ & $41.8$ \\
90C  & $0.12\pm 0.01$ & $0.10\pm 0.01$ & $0.32\pm 0.02$ & $41.7$ \\
90D  & $0.08\pm 0.01$ & $0.08\pm 0.01$ & $0.14\pm 0.01$ & $41.3$ \\
92B  & $0.04\pm 0.01$ & $0.05\pm 0.01$ & $0.32\pm 0.02$ & $42.5$ \\
92C  & $0.03\pm 0.01$ & $0.04\pm 0.01$ & $0.17\pm 0.01$ & $42.3$ \\
92D  & $0.07\pm 0.01$ & $0.06\pm 0.01$ & $0.38\pm 0.02$ & $42.6$ \\
92E  & $0.04\pm 0.01$ & $0.04\pm 0.01$ & $0.29\pm 0.02$ & $42.5$ \\
92F  & $0.02\pm 0.01$ & $0.02\pm 0.01$ & $0.10\pm 0.01$ & $42.0$ \\
\enddata
\tablecomments{
Columns are:
(1) HCG nucleus ID; (2), (3), (4): flux densities (mJy) for the
\wtwo\ (effective wavelength 2030~\AA), \mtwo\ (2231~\AA),
\wone\ (2634~\AA) \uvot\ filters (corrected for Galactic extinction only); 
(5) log luminosity (\lunits) for
\wone\ filter.  }\end{deluxetable}

At this point we should mention a possible
adverse consequence of the UV resolution and photometric aperture. Since the resolution
is worse and the aperture larger than the corresponding quantities for
the \chandra\ data, and given that we are interested in detecting a possible
signature of AGN activity, there is a risk of an AGN signal being
diluted if it is weak and there is significant star formation. This issue
will be less important for earlier type galaxies, where star formation is
unlikely to play a major role.
We will come back to this effect later in the paper, but at this point we are
cautioning that it is difficult to quantify properly.

To estimate background emission we construct
background regions interactively to ensure that no emission
either from galaxies or foreground stars was included.
In a minority of cases, concentric annuli at radii 50\arcsec\ and 60\arcsec\ from the
source centers are appropriate. However, compact group galaxies
are often very close to each other, so that in most cases we have to
construct a background region manually to avoid contamination.
We then obtain net count rates
in all three UV bands. Finally, using the UVOT-specific flux conversion factors
\citep{poole2008}, we obtain flux and luminosity densities for
all galaxies (see \tr{tab-uvfl}).
The tabulated values have been corrected for Galactic
extinction, using the maps of \citet{schlegel1998} and the extinction
curve of \citet{cardelli1989}.

For each galaxy, UV fluxes and luminosities have further been corrected for
intrinsic extinction by using the 
UV and 24\micron\ SFR components
in \citet{tzanavaris2010} and assuming that
SFR$_{24\mu}$ = SFR$_{\rm UV, unobscured}$. 
As HCG 90 and 92 were not
analyzed in that work, 
for these groups we correct for extinction 
by adopting the highest $A_{UV}$ value
for HCG 92 in \citet{xu2005}, i.e. $A_{UV} = 2$.


\section{Multiwavelength nuclear analysis}\label{sec:multiw}
We investigate the nature of nuclear activity in our HCG
galaxies by combining diagnostics using the \x, UV, optical
and radio regions. 
An overview of the main multiwavelength results is presented
in \tr{tab-multi},
which we discuss in greater detail later.

\subsection{Optical}
\subsubsection{Emission Line Ratio Classifications}\label{sec_class}
M10 have carried out spectroscopic observations
and obtained emission line ratios for 200 HCG galaxies. They have also
obtained emission line ratios from archival spectra and the literature
for a further 70 HCG galaxies, bringing the total to 270. Their
primary classification criterion is the location of
a galaxy in the \ohb\ vs. \niha\ diagram of K06
\citep[hereafter K06-a; see K06 Figure 1a and 4a and also][]{baldwin1981,veilleux1987}. 
As explained
by K06, galaxies located below the \citet[][hereafter Ka03]{kauffmann2003}
line (lower curve) are purely star forming, while galaxies above the
\citet[][hereafter Ke01]{kewley2001} line are purely AGNs.
Galaxies that fall between the two lines are considered
composite or transition objects \citep{ho1997}, in which circumnuclear
star formation effectively dilutes the high-ionization
emission line ratio signal.
Note that this diagram cannot distinguish between
LINERs and AGNs in any way.
M10 assume that LINERs are just an AGN subcategory and
use this diagram to separate SF from AGN systems.
Although K06 also consider LINERs to be AGNs, their
additional, \ohb\ vs. \siha\ and \ohb\ vs. \oha, diagrams 
can be used
to establish
a well defined dividing line (their Figures 4b and 4c, hereafter
K06-b and K06-c, respectively) 
separating galaxies
dominated by LINER vs. those dominated by Seyfert (i.e. AGN) activity. 
The distinction between LINERs and AGNs is not a mere
matter of semantics for two reasons. First, although a majority
of LINERs harbor weak AGNs, not all LINERs have AGNs.
Second, the energetics of LINERs cannot be understood in terms
of AGN activity, as any weak AGNs in LINERs cannot fully
account for the observed
emission lines. Thus LINERs, even those that host weak AGNs, are not
a scaled-down AGN; in particular, they are not just a LLAGN.
They should be considered as an activity class in
their own right beside star-forming and AGN systems.
 
\begin{deluxetable}{ccccc}
\tablecolumns{5}
\tablewidth{0pc}
\tablecaption{Nuclear optical spectroscopic classification.\label{tab-class}}
\tablehead{
\colhead{ID}
& \colhead{K06(a)}
& \colhead{K06(b)}
& \colhead{K06(c)}
& \colhead{Type}
\\
\colhead{(1)}
& \colhead{(2)}
& \colhead{(3)}
& \colhead{(4)}
& \colhead{(5)}
\\
}
\startdata
        7A         &        TO         &        SF         &        SF         &        TO/SF     \\
        7B         &    \ldots         &    \ldots         &    \ldots         &    \ldots        \\
        7C         &        SF         &        SF         &    \ldots         &        SF        \\
        7D         &        SF         &        SF         &    \ldots         &        SF        \\
       16A         &     nonSF         &    LNR/AGN        &    LNR/AGN        &    LNR/AGN        \\
       16B         &     nonSF         &       LNR         &       LNR         &       LNR        \\
       16C         &        TO         &        SF         &        SF         &        TO/SF      \\
       16D         &        SF         &        SF         &        SF         &        SF        \\
       22A         &     nonSF         &        AGN        &        AGN        &       AGN        \\
       22B         &    \ldots         &    \ldots         &    \ldots         &    \ldots        \\
       22C         &        SF         &    \ldots         &    \ldots         &        SF        \\
     31ACE         &        SF         &        SF         &        SF         &        SF        \\
       31B         &        SF         &        SF         &        SF         &        SF        \\
       31F         &    \ldots         &    \ldots         &    \ldots         &    \ldots        \\
       31G         &        SF         &        SF         &    \ldots         &        SF        \\
       31Q         &        SF         &    \ldots         &    \ldots         &        SF        \\
       42A         &     nonSF         &    \ldots         &    \ldots         &    nonSF\tm{1}        \\
       42B         &    \ldots         &    \ldots         &    \ldots         &    \ldots        \\
       42C         &    \ldots         &    \ldots         &    \ldots         &    \ldots        \\
       42D         &    \ldots         &    \ldots         &    \ldots         &    \ldots        \\
       59A         &        TO         &       LNR         &       LNR         &    TO/LNR        \\
       59B         &        TO         &    \ldots         &    \ldots         &    TO\tm{1}         \\
       59C         &        SF         &    \ldots         &    \ldots         &    SF\tm{1}         \\
       59D         &        SF         &       SF?         &       SF?         &        SF?        \\
       62A         &        TO         &        SF         &    LNR/SF         &    TO/LNR?        \\
       62B         &    \ldots         &    \ldots         &    \ldots         &    \ldots        \\
       62C         &    \ldots         &    \ldots         &    \ldots         &    \ldots        \\
       62D         &    \ldots         &    \ldots         &    \ldots         &    \ldots        \\
       90A         &     nonSF         &       AGN         &       AGN         &        AGN        \\
       90B         &    \ldots         &    \ldots         &    \ldots         &    \ldots        \\
       90C         &    \ldots         &    \ldots         &    \ldots         &    \ldots        \\
       90D         &        TO         &        SF         &      LNR?         &   TO/LNR?        \\
       92B         &    \ldots         &    \ldots         &    \ldots         &    \ldots        \\
       92C         &     nonSF         &        AGN        &   \ldots         &      AGN        \\
       92D         &    \ldots         &    \ldots         &    \ldots         &    \ldots        \\
       92E         &    \ldots         &    \ldots         &    \ldots         &    \ldots        \\
       92F         &    \ldots         &    \ldots         &    \ldots         &    \ldots        \\
\enddata
\tx{1}{Based on \niha\ only.}
\tablecomments{
Columns give nuclear classification results based on
K06 emission line ratio diagrams as follows:
(1) \ohb\ vs. \niha\ (K06 - a); 
(2) \ohb\ vs. \siha\ (K06 - b);
(3) \ohb\ vs. \oha\ (K06 - c).
Nuclear activity classifications are TO (transition object),
SF (star forming), LNR (LINER), AGN (supermassive black hole accretion),
nonSF (either LNR or AGN).
}
\end{deluxetable}

\begin{figure*}\vspace{-4.5cm}
\epsscale{1.0}
\plotone{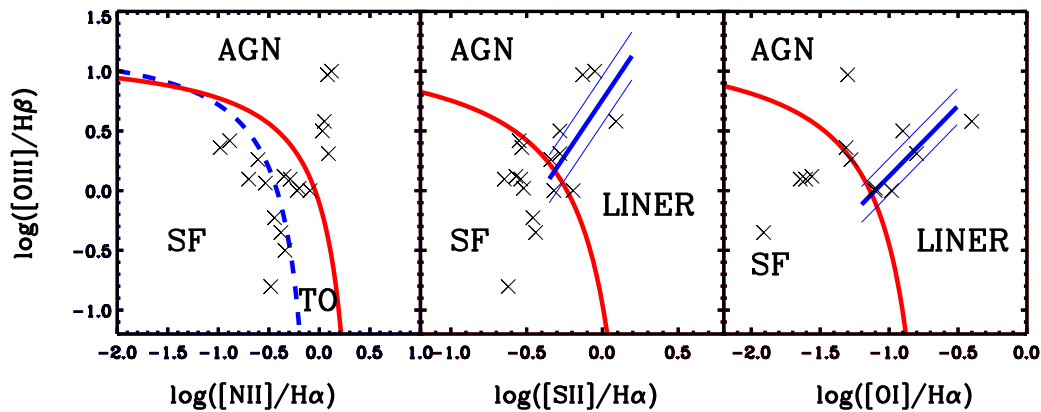}
\caption{Diagnostic optical emission line ratio diagrams for HCG nuclei
in this paper. The diagrams are based on the classification scheme of
\citet{kewley2006}. See the text for details.}
\label{fig:class}
\end{figure*}
We thus use
all three diagnostic diagrams in K06 
and the emission line ratios of M10 as our primary
criterion for obtaining new
nuclear classifications for galaxies in our sample. Our
classification scheme
explicitly includes LINERs.
Our new classifications are presented in 
\tr{tab-class} and \fr{fig:class}.
Columns 2, 3 and 4 of this table give the classification
based on each of the emission line ratio diagrams K06-a, b and c.
As we do not consider LINERs to be just an AGN subclass, we 
can only use K06-a to classify galaxies
as star forming (SF), non star-forming (nonSF, either
Seyfert of LINER) or transition objects (TO).
We use K06-b and K06-c to classify
galaxies as SF, AGN or LINER. 
Our final classification is given in column 5 and is
based on the results of the previous
three columns. If in any of the diagrams a galaxy
falls on the dividing line, this is a borderline
case indicated by a question mark in \tr{tab-class}.

Finally, for three galaxies in our sample there is
only \niha\ information. In this case we
adopt the classification criterion of M10,
who classify galaxies with $\log ($~\niha~$)\le -0.4$ as SF, 
those with $\log ($~\niha~$)>-0.1$ as AGN, and those
in between as TO \citep{stasinska2006}.
Of course, as explained, this
precludes the possibility of identifying LINERs.
Note that we include column 5 from \tr{tab-class}
in \tr{tab-multi} (its column 7), which presents
together multiwavelength nuclear activity diagnostics.

According to \tr{tab-class}, there exists emission-line
ratio information for 22 galaxies in our sample. Out of these 10
are classified optSF (45.5\%), 2 are optTO/SF (9\%), 1 is optTO (4.5\%),
3 are optAGN (13.6\%),
1 is optLINER/AGN (4.5\%), 1 is opt-nonSF (4.5\%), 3 are optTO/LINER (13.6\%) 
and
1 is optLINER (4.5\%). Thus, with the caveat of
small-number statistics, we see that
the most clear result of this classification is that
star forming systems represent the most numerous
class, followed by AGNs. Otherwise, there appears to be only
one clear LINER, as well as a substantial number of mixed
classifications.

\begin{deluxetable*}{c c c c c c c c}
\tablecolumns{8}
\tablecaption{\hst\ data for galaxies in this paper.\label{tab-hst}}
\tablehead{
\colhead{Group}
&\colhead{Galaxy}
&\colhead{\hst}
&\colhead{Filter}
&\colhead{Exp.~time}
&\colhead{Date}
&\colhead{Program}
&\colhead{PI}
\\
\colhead{}
&\colhead{}
&\colhead{Instrument}
&\colhead{}
&\colhead{(s)}
&\colhead{}
&\colhead{ID}
&\colhead{}
\\
\colhead{(1)}
&\colhead{(2)}
&\colhead{(3)}
&\colhead{(4)}
&\colhead{(5)}
&\colhead{(6)}
&\colhead{(7)}
&\colhead{(8)}
}
\startdata
HCG 07 & A, B, D    & ACS/WFC & F606W & 1230 & Sept 2006 & 10787 & Charlton\\
       & C          & ACS/WFC & F606W & 1230 & Sept 2006 & \\
HCG 16 & A          & WFPC2   & F606W & 1900 & July 2007 & \\
	   & C          & WFPC2   & F606W & 1900 & Aug 2007  & \\
	   & D          & WFPC2   & F606W & 1900 & Aug 2007  & \\
	   & A, B       & WFPC2   & F606W & 1900 & Sept 2007 & \\	   
HCG 22 & B          & WFPC2   & F606W & 1900 & Aug 2007  & \\
       & C          & WFPC2   & F606W & 1900 & Sept 2007 & \\
       & A          & WFPC2   & F606W & 1900 & Sept 2007 & \\
HCG 31 & A--C, E--H & ACS/WFC & F606W & 1230 & Aug 2006  & \\
HCG 42 & B          & WFPC2   & F606W & 4200 & Dec 2007  & \\
       & D          & WFPC2   & F606W & 4200 & Dec 2007  & \\
       & A, C       & ACS/WFC & F606W & 1230 & Dec 2007  & \\
HCG 59 & A, C, D    & ACS/WFC & F606W & 1230 & Nov 2006  & \\
       & B, I       & ACS/WFC & F606W & 1230 & Nov 2006  & \\
HCG 62 & \multicolumn{7}{c}{No HST data}\\
HCG 90 & \multicolumn{7}{c}{No HST data}\\
HCG 92 & B, D       & WFC3    & F606W & 1395 & Aug 2009  & 11502 & Noll\\
       & C, B       & WFC3    & F606W & 1395 & Aug 2009  & \\
       & E          & WFC3    & F606W & 1395 & July 2009 & \\
\enddata
\tablecomments{
Columns give (1) compact group ID; 
(2) galaxies observed;
(3) instrument used;
(4) filter;
(5) exposure time;
(6) observation date;
(7) program ID;
(8) program PI. 
}
\end{deluxetable*}

\subsubsection{Optical nuclear excess}\label{sec:hst}
For a subset of our galaxy nuclei in seven HCGs, \hst\ data are available.
An observation log is given in \tr{tab-hst}.
The high angular resolution ($\leq 0.1\arcsec$) of \hst\ may provide a
complementary means of identifying nuclear point sources. We identify
these sources by examining the median divided image of each galaxy (we
use a $13 \times 13$ pixel smoothing window and divide the original image by
the smoothed one). Additionally, we
use {\tt GALFIT} \citep{peng2010} to fit surface brightness profiles
and identify nuclear point sources. We compare the 
{\tt GALFIT}-derived centers with the central sources detected in median
divided images. The findings are summarized in columns 10 and 11 of
\tr{tab-multi}. A \lqq y\rqq\ 
in column 10 indicates that a nuclear point source in
the median-divided image is detected, while a \lqq y\rqq\ 
in column 11 indicates that the source coincides with the {\tt GALFIT} center,
within the positional uncertainties (3 pixels). 
It turns out that, due to the mostly disturbed nature of these
galaxies, \galfit\ is unable to either converge or provide 
good fits. We discuss this issue further later (\scr{sec:over}).

\begin{figure}
\epsscale{1.0}
\plotone{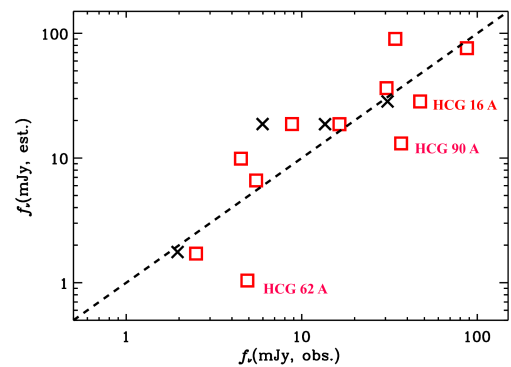}
\caption{
1.4 GHz flux density, estimated from the radio - SFR correlation
of \citet{bell2003a}, against observed 1.4 GHz flux density.
Red squares correspond to NVSS and black crosses
to FIRST data (see Table 5). The dashed line
is the locus of equal estimated and observed flux densities.
The three labeled HCG galaxies that show
an excess in observed flux density are all known AGN.
}
\label{fig:f14-f14}
\end{figure}

\begin{deluxetable*}{cccc cccc cccc cccc} 
\tablecolumns{8}
\tablewidth{0pc} 
\tablecaption{Radio detections of HCG nuclei in this sample.\label{tab-radio}}
\tablehead{ 
\colhead{}
& \multicolumn{5}{c}{FIRST\tablenotemark{a}}
& \multicolumn{3}{c}{NVSS\tablenotemark{b}}
& \multicolumn{3}{c}{SUMMS\tablenotemark{c}}
& \multicolumn{4}{c}{WISH\tablenotemark{d}}
\\
\colhead{} 
& \multicolumn{5}{c}{\rule{4.5cm}{.02cm}}
& \multicolumn{3}{c}{\rule{3cm}{.02cm}}
& \multicolumn{3}{c}{\rule{3cm}{.02cm}}
& \multicolumn{4}{c}{\rule{4cm}{.02cm}}
\\
\colhead{}
& \colhead{$P$}
& \colhead{\ftwp}
& \colhead{$\pm$}
& \colhead{\ftwi}
& \colhead{$\Delta$pos}
& \colhead{\ftwi}
& \colhead{$\pm$}
& \colhead{$\Delta$pos}
& \colhead{\ftsi}
& \colhead{$\pm$}
& \colhead{$\Delta$pos}
& \colhead{\ftftp}
& \colhead{\ftfti}
& \colhead{$\pm$}
& \colhead{$\Delta$pos}
\\
\colhead{HCG ID}
& 
& \colhead{(mJy)}
& \colhead{(mJy)}
& \colhead{(mJy)}
& (\arcsec)
& \colhead{(mJy)}
& \colhead{(mJy)}
& (\arcsec)
& \colhead{(mJy)}
& \colhead{(mJy)}
& (\arcsec)
& \colhead{(mJy)}
& \colhead{(mJy)}
& \colhead{(mJy)}
& (\arcsec)
\\
\colhead{(1)}
& \colhead{(2)}
& \colhead{(3)}
& \colhead{(4)}
& \colhead{(5)}
& \colhead{(6)}
& \colhead{(7)}
& \colhead{(8)}
& \colhead{(9)}
& \colhead{(10)}
& \colhead{(11)}
& \colhead{(12)}
& \colhead{(13)}
& \colhead{(14)}
& \colhead{(15)}
& \colhead{(16)}
}
\startdata
	    7A &   0.014 &       7.43 &   0.131 & 13.57 & 0.84               &	        16.4 &              0.7 & 1.86                &\nodata&\nodata&\nodata&\nodata&\nodata&\nodata&\nodata\\
	    7C & \nodata &   \nodata  & \nodata &\nodata & \nodata            &	         4.5 &              0.5 & 8.04                &\nodata&\nodata&\nodata&\nodata&\nodata&\nodata&\nodata\\
           16A &   0.014 &      12.80 &   0.159 & 30.81 & 2.52               &	        47.3 &              2.1 & 1.98                &\nodata&\nodata&\nodata& 93    & 100   & 4.6   & 4.5 \\
	   16B &   0.014 &       1.98 &   0.158 &  1.96 & 0.6                &	   \nodata   & \nodata     & \nodata                  &\nodata&\nodata&\nodata& 51    &  53   & 4.8   & 0.54 \\
           16C & \nodata &   \nodata  & \nodata &\nodata & \nodata            &	        87.3 &              3.2 & 2.04                &\nodata&\nodata&\nodata&\nodata&\nodata&\nodata&\nodata\\
           16D & \nodata &   \nodata  & \nodata &\nodata & \nodata            &	        34.1 &              1.1 & 0.78                &\nodata&\nodata&\nodata&\nodata&\nodata&\nodata&\nodata\\
          31ACE& \nodata &   \nodata  & \nodata &\nodata & \nodata            &	        30.4 &              1.6 & 6.9                 &\nodata&\nodata&\nodata&\nodata&\nodata&\nodata&\nodata\\
           31G & \nodata &   \nodata  & \nodata &\nodata & \nodata            &	         5.5 &              0.5 & 4.14                &\nodata&\nodata&\nodata&\nodata&\nodata&\nodata&\nodata\\
           42A & \nodata &   \nodata  & \nodata &\nodata & \nodata            &	         2.5 &              0.5 & 9                   &\nodata&\nodata&\nodata&\nodata&\nodata&\nodata&\nodata\\
	   59A &   0.014 &       6.18 &   0.187 &  5.98 & 1.44               & 	         8.8 &              1.0 & 8.4                 &\nodata&\nodata&\nodata&\nodata&\nodata&\nodata&\nodata\\
           62A & \nodata &   \nodata  & \nodata &\nodata & \nodata            &	         4.9 &              0.5 & 6.78                &\nodata&\nodata&\nodata&\nodata&\nodata&\nodata&\nodata\\
           90A & \nodata &   \nodata  & \nodata &\nodata & \nodata            &	        36.8 &              1.5 & 3.54                &  45.6 & 1.7 & 6.6     &\nodata&\nodata&\nodata&\nodata\\
\enddata
\tablenotetext{a}{Faint Images of the Radio Sky at Twenty-cm survey \citep{becker1995}.
}
\tablenotetext{b}{NRAO-VLA Sky Survey \citep{condon1998}.
}
\tablenotetext{c}{Sydney University Molonglo Sky Survey \citep{bock1999}.
}
\tablenotetext{d}{Westerbork in the Southern Hemisphere survey \citep{debreuck2002},
}
\tablecomments{Columns are: (1) HCG galaxy ID; 
(2) FIRST probability that the source is spurious;
(3) FIRST peak flux at 20 cm;
(4) FIRST error on peak flux;
(5) FIRST integrated 20 cm flux;
(6) FIRST positional uncertainty in arcseconds;
(7) NVSS integrated 20 cm flux;
(8) NVSS flux error;
(9) NVSS positional uncertainty in arcseconds;
(10) SUMMS integrated 36 cm flux;
(11) SUMMS flux error;
(12) SUMMS positional uncertainty in arcseconds;
(13) WISH 352 MHz peak flux;
(14) WISH 352 MHz integrated flux;
(15) WISH flux error;
(16) WISH positional uncertainty in arcseconds.
}
\end{deluxetable*}

\begin{deluxetable*}{cccc cccc ccc}
\centering
\tablecolumns{11}
\tabletypesize{\scriptsize} 
\tablewidth{0pc}
\tablecaption{Multiwavelength analysis of HCG nuclear sources \label{tab-multi}}
\tablehead{
\colhead{HCG ID}  
&\colhead{Morphology}
&\colhead{$\Delta\theta$}
&\colhead{\x\ - UV}
&\colhead{\aox}
&\colhead{\aox}
&\colhead{Optical}
&\colhead{X-ray}
&\colhead{Radio}
&\multicolumn{2}{c}{Optical Nuclear}
\\
\colhead{}  
&\colhead{}
&\colhead{}
&\colhead{counterparts}
&\colhead{(corrected)}
&\colhead{(uncorrected)}
&\colhead{Type}
&\colhead{Strong AGN}
&\colhead{Excess}
&\multicolumn{2}{c}{Source}
\\
\colhead{(1)} &
\colhead{(2)} &
\colhead{(3)} &
\colhead{(4)} &
\colhead{(5)} &
\colhead{(6)} &
\colhead{(7)} &
\colhead{(8)} & 
\colhead{(9)} &
\colhead{(10)} &
\colhead{(11)}
\\
}
\startdata
      7A & Sb  &1.2       &   y  &   -2.23 & -1.96 &   TO/SF &  n     &        n &  y & y                    \\
      7B & SB0 &1.1       &   y  &   -2.31 & -2.26 &  \ldots &  n     &   \ldots &  y & y                    \\
      7C & SBc &0.5       &   y  &   -2.39 & -2.30 &      SF &  n     &        n &  y & n                    \\
      7D & SBc &\ldots    &   n  &   -2.51 & -2.46 &      SF &  n     &   \ldots &  y & y                    \\
     16A & SBab&3.0       &   y  &   -2.16 & -1.92 &LINER/AGN&  n     &        y &  y & \ldots               \\	  
     16B & Sab &0.8       &   y  &   -1.68 & -1.57 &   LINER &  y     &        n &  y & \ldots               \\	  
     16C & Im  &3.0       &   y  &   -2.75 & -2.35 &   TO/SF &  n     &        n &  y & \ldots           \\	  
     16D & Im  &1.1       &   y  &   -2.61 & -1.96 &      SF &  n     &        n &  y & \ldots           \\	  
     22A & E2  &0.8, 0.7  &   ym &   -2.35 & -2.33 &     AGN &  n     &   \ldots &  y & \ldots               \\	  
     22B & Sa  &1.7       &   y  &   -2.13 & -2.11 &  \ldots &  n     &   \ldots &  y & \ldots               \\	  
     22C & SBcd&\ldots    &   n  &   -2.42 & -2.36 &      SF &  n     &   \ldots &  n & \ldots               \\	  
   31ACE & Sdm &1.7, 0.7  &   ym &   -2.50 & -2.34 &      SF &  n     &        n &  n  & \ldots              \\	  
     31B & Sm  &\ldots    &   n  &   -2.74 & -2.69 &      SF &  n     &   \ldots &  n  & \ldots              \\	  
     31F & Im  &\ldots    &   n  &   -2.57 & -2.56 &  \ldots &  n     &   \ldots &  n  & \ldots              \\	  
     31G & Im  &0.8       &   y  &   -2.30 & -2.23 &      SF &  n     &        n &  y  & \ldots              \\	  
     31Q & Im  &2.3       &   n? &   -2.35 & -2.32 &      SF &  n     &   \ldots &  n  & \ldots              \\	  
     42A & E3  &1.8       &   y  &   -2.05 & -2.01 &   nonSF &  n     &        n &  y  & y                   \\
     42B & SB0 &\ldots    &   n  &   -2.32 & -2.28 &  \ldots &  n     &   \ldots &  y & \ldots               \\	  
     42C & E2  &1.0       &   y  &   -2.16 & -2.14 &  \ldots &  n     &   \ldots &  y & \ldots               \\	  
     42D & E2  &\ldots    &   n  &   -2.12 & -2.09 &  \ldots &  n     &   \ldots &  y & \ldots               \\	  
     59A & Sa  &1.8       &   y  &   -2.47 & -1.90 &TO/LINER &  n     &        n &  y  &  n                  \\
     59B & E0  &2.0       &   y? &   -2.01 & -1.98 &      TO &  n     &   \ldots &  y  &  y                  \\
     59C & Sc  &\ldots    &   n  &   -2.55 & -2.49 &      SF &  n     &   \ldots &  y  &  y                  \\
     59D & Im  &\ldots    &   n  &   -2.63 & -2.58 &      SF? &  n     &   \ldots &  n  &  n                  \\
     62A & E3  &1.1       &   y  &   -2.16 & -2.14 &TO/LINER? &  n     &        y &  \ldots & \ldots          \\	  
     62B & S0  &0.1       &   y  &   -2.16 & -2.15 &  \ldots &  n     &   \ldots &  \ldots & \ldots          \\	  
     62C & S0  &0.4       &   y  &   -2.21 & -2.20 &  \ldots &  n     &   \ldots &  \ldots & \ldots          \\	  
     62D & E2  &\ldots    &   n  &   -2.17 & -2.08 &  \ldots &  n     &   \ldots &  \ldots & \ldots          \\	  
     90A & Sa  &3.3       &   y  &   -1.31 & -1.01 &     AGN &  y     &        y &  \ldots & \ldots          \\	  
     90B & E0  &0.4       &   y  &   -2.49 & -2.18 &  \ldots &  n     &   \ldots &  \ldots & \ldots          \\	  
     90C & E0  &0.9       &   y  &   -2.49 & -2.18 &  \ldots &  n     &   \ldots &  \ldots & \ldots          \\	  
     90D & Im  &\ldots    &   n  &   -2.91 & -2.60 &TO/LINER? &  n     &   \ldots &  \ldots & \ldots          \\	  
     92B & Sbc &0.6       &   y  &   -2.68 & -2.37 &  \ldots &  n     &   \ldots &  y  & \ldots              \\	  
     92C & Sbc &0.9       &   y  &   -1.81 & -1.51 &     AGN &  y     &   \ldots &  y  & \ldots              \\	  
     92D & E2  &0.7       &   y  &   -2.54 & -2.23 &  \ldots &  n     &   \ldots &  y  & \ldots              \\	  
     92E & E1  &1.5       &   y  &   -2.57 & -2.27 &  \ldots &  n     &   \ldots &  y  & \ldots              \\	  
     92F & SAB0&2.1       &   y? &   -2.49 & -2.19 &  \ldots &  n     &   \ldots &  \ldots & \ldots          \\	  
\enddata 
\tablecomments{
Columns are:
(1) HCG galaxy ID;
(2) Morphological type \citep{hickson1989};
(3) offset between peak of nuclear UV emission in the UVOT \wone\
filter and peak of \x\ emission in the \x s;
(4) \x\ -- UV counterparts flag based on PSF overlap:
y = detected \x\ -- UV overlap; n = no overlapping PSFs - no counterparts; 
? = ambiguous; m = multiple \x\ PSFs overlap with UV PSF;
(5) \aox\ value corrected for intrinsic extinction;
(6) \aox\ value uncorrected for intrinsic extinction;
(7) nuclear type according to optical spectroscopy classification
(\tr{tab-class}, column 5):
SF = star-forming, TO = transition object, \ldots\ = no classification;
(8) \lxte~$\ge 10^{41}$~\lunits, suggestive of strong AGN: y = yes, n = no;
(9) radio excess flag: y = observed 1.4 GHz flux density in excess
of that expected based on the \citet{bell2003a} radio - SFR correlation; 
(10) \hst\ nuclear detection flag for median-divided images: y = 
central point source detected in \hst\ median-divided image; n = no detection;
(11) comparison flag for \hst\ nuclear detections:
y = nuclear point source in \hst\ median-divided image 
coincides with \galfit\ center
within 3 pixels; n = no coincidence within 3 pixels.
}
\end{deluxetable*}

\subsection{Radio}\label{sec:radio}
Radio detections of nuclei constitute possible
complementary evidence of
AGN activity (in the case of radio-loud AGN).
Conversely, in star forming galaxies there is a well-known
correlation between the 1.4 GHz (21 cm) luminosity density, \lof, and
SFR
\citep[e.g.][]{bell2003a}. Thus, if the SFR is known,
one can use this correlation to obtain an estimate for the
\lof, as well as the 1.4 GHz flux density, \fofe.
Any significant excess between the observed flux density, \fofo,
and \fofe\ may be an indication of AGN activity.

We searched the archives of all publically available
major radio surveys, and found detections for 12 of our galaxies in
four of the surveys.  Details on the radio detections are given in
\tr{tab-radio}.  
As none of the catalogs 
has full sky coverage, 
a non-detection does not necessarily imply
the lack of radio emission. The catalogs also vary
in sensitivity, resolution and
positional accuracy. 
We use both the NVSS and FIRST detections, together
with UV+IR SFR values of \citet{tzanavaris2010} to estimate
the \fofe\ by means of the \citet{bell2003a} correlation.
This correlation has two different forms, depending on whether
a galaxy has $M_V > -21 $ or not.
We estimated $M_V < -21$ for all of our galaxies by using the $B$ and $R$-band
values for HCG galaxies in \citet{hickson1989} and the color transformations of
\citet{fukugita1995}. Using the \citet{bell2003a} correlation, we thus
calculated \fofe. We plot \fofe\ against \fofo\ in \fr{fig:f14-f14} for
the NVSS and FIRST detected galaxies. For 9 out of 12 galaxies with
radio detections the estimated
and observed flux density appear to be consistent with each other.
The absence of a radio excess for these 9 galaxies 
is indicated by \lq\lq n\rq\rq\ in column 9 of
\tr{tab-multi}. 
We note though that for galaxies HCG 16~A, 90~A and 62~A,
there is an indication of an excess in \fofo\
(\lq\lq y\rq\rq\ in column 9 of \tr{tab-multi}).
For the first two this
is entirely consistent with
their AGN classification in the optical. 
In addition, HCG 90~A has high \x\ luminosity, further
consistent with AGN activity.
Our optical classification for
the third is LINER, and the radio excess may suggest that it also hosts
an AGN (recall that, as mentioned, a majority of LINERs do harbor AGNs).
In spite of this consistency, note
that for these galaxies \fofe\ is still within a factor of $\sim 2$
from the corresponding \fofo\ value.
This factor also corresponds to the uncertainty of the radio - SFR correlation, so this should
not be considered as a robust result. Unfortunately, we are unable to 
say anything for galaxies for which we do not have any other 
classification, as none of these are detected in the radio.

\begin{figure}
\epsscale{1.0}
\plotone{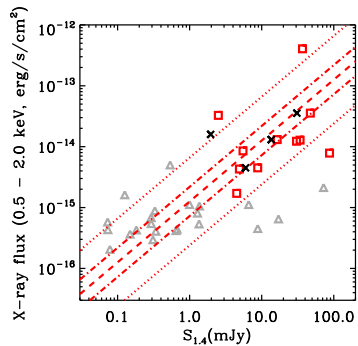}
\caption{Radio vs. X-ray soft band fluxes for HCG nuclei in this paper.
HCG radio fluxes are shown as crosses (FIRST) or red squares (NVSS).
Grey triangles are VLA-COSMOS galaxies classified as star-forming by
\citet[][\lq\lq c1\rq\rq\ class from their tables 1 and 2.]{ranalli2012}.
The red dashed line is the radio-\x\ correlation of \citet[][their equation (8) converted to flux using the average distance of our nine HCG nuclei.]{ranalli2003}}.
\label{fig:ranalco}
\end{figure}

We also compare our combined \x-radio results for HCG nuclei
with the work of 
\citet{ranalli2003,ranalli2012} for entire galaxies.
Following \citet{ranalli2012}, 
in \fr{fig:ranalco} we plot 
soft band
\x\ luminosities against 1.4 GHz fluxes 
from NVSS and FIRST (red squares
and crosses, respectively) for HCG nuclei. We also
show the observational \x-radio correlation for local star-forming galaxies
established by \citet{ranalli2003} converted to flux using the mean
distance of the HCG nuclei. For comparison, we further plot VLA COSMOS sources
detected in C-COSMOS and classified
  as star-forming by \citet[][class \lq\lq c1\rq\rq, their
tables 1 and 2]{ranalli2012}. In this figure, HCG nuclei appear to blend
smoothly with the lower flux, more distant COSMOS star-forming sources, although with
considerable scatter. The topic of \x\ emission in HCGs as a function
of star formation rate will be examined in greater detail in a forthcoming 
publication.

\subsection{X-ray -- UV}\label{sec:xuv}
Unlike in the
X-rays, each galaxy's nucleus is well defined in the UV regime as
an emission peak in the central region of the galaxy.
To identify nuclear \x\ point sources for each galaxy we thus
visually inspect all \x\ point sources for
spatial coincidence with the central intensity peak of UV sources at 2600~\AA.
Specifically, we examine the degree of overlap between the
\ace-determined \chandra\ PSF and a UVOT \lq\lq PSF\rq\rq, defined as
a circular region 3\arcsec\ in diameter centered at the
\wone\ intensity peak.  Unlike the case for photometry, the choice of
UV circular region size here is solely determined by the \wone\ PSF
FWHM of 2.37\arcsec.
Note that the
absolute astrometry for both UVOT and \chandra\ is very good, as
measured relative to the International Celestrial Reference System
\citep[ICRS,][]{ma1998}. We estimate that in the worst case scenario
this could lead to a maximum spurious offset between two coincident sources of
$\sim 0.7$\arcsec\ (combining in quadrature 
$\sim 0.4$\arcsec\ from UVOT, \citet{breeveld2010},
and $\sim 0.6$\arcsec\ from \chandra,
http://cxc.harvard.edu/cal/ASPECT/celmon/).

We also calculate the offset, $\Delta\theta$,
in arcseconds between the \x\ and the UV galaxy nucleus
as determined by the \ace\ center and UV peak emission, respectively.
This is shown in column 3 of \tr{tab-multi}, 
while column 4 reports if, based on 
an overlap between the \x\ and UV PSFs, we consider that there
exist X-ray -- UV nuclear counterparts for a given galaxy.
Due to the different resolution between
the UV and the \x s, our primary criterion 
for the existence of counterparts
is the PSF overlap, rather
than $\Delta\theta$. We do note that all offsets are less than $\sim 3$\arcsec. 
If the PSF overlap criterion indicates that
there exist \x\ -- UV counterparts,
this is indicated by \lqq y\rqq\ in column 4. 
There are 22 such cases, which are best candidates
for \x\ -- UV counterparts. This reduces by 5 the original number of 27 nuclear X-ray sources estimated without strict use of the PSF overlap criterion.
Conversely, an \lqq n\rqq\ indicates no counterparts
(10 cases).
In two cases
(22A, 31ACE) it is not clear which \x\ source is the best counterpart,
as two \x\ PSFs overlap with the UV PSF.  This is indicated
by an \lq\lq m\rq\rq\ (for \lq\lq multiple\rq\rq) in column 3.  A question
mark
indicates that the \x\ -- UV counterpart is uncertain, as
described 
in \scr{sec:over} (three cases).

\begin{figure*}
\epsscale{1.2}
\newcommand{\wid}{9cm}
\includegraphics[width=\wid]{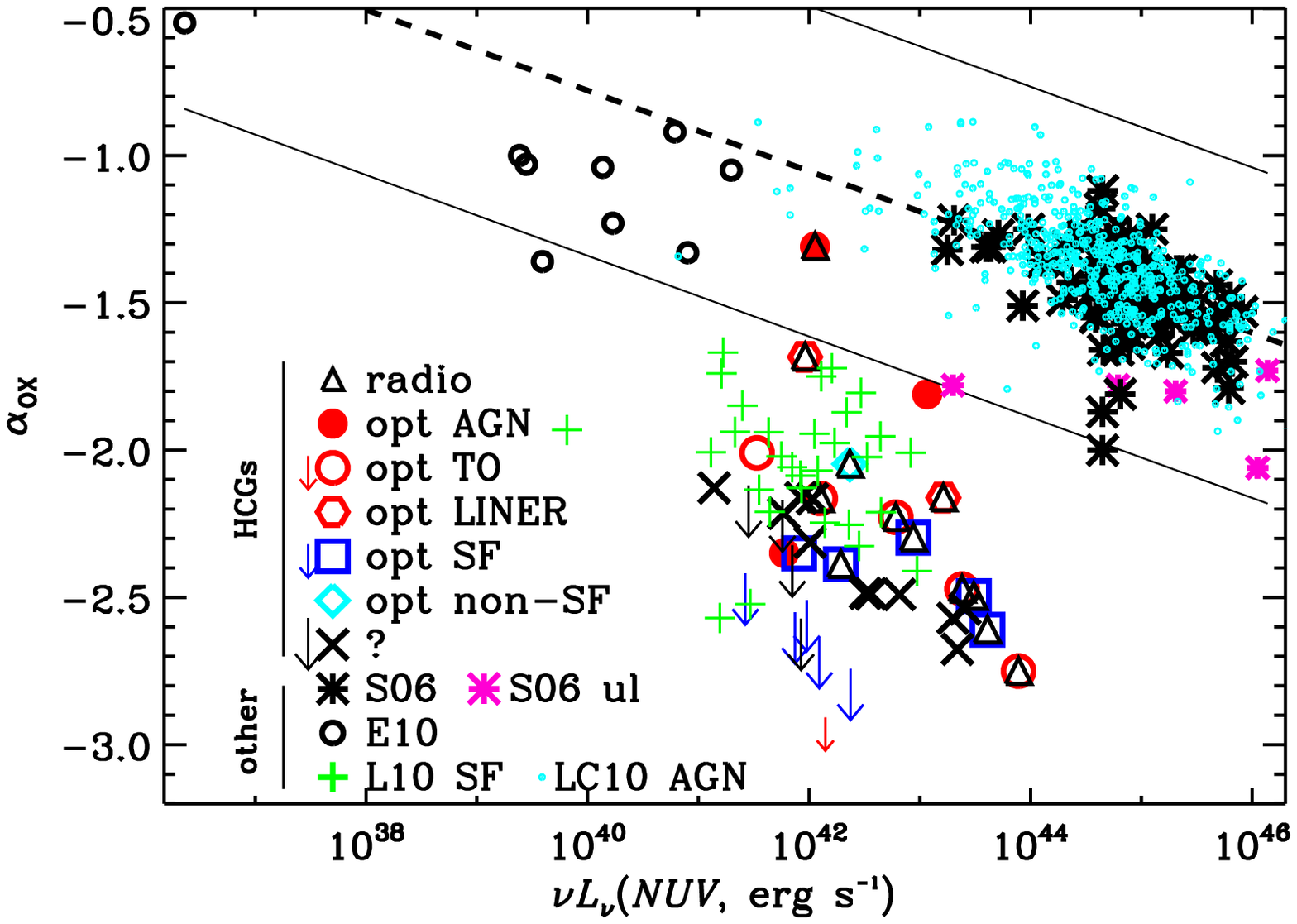}\hspace{-0.5cm}
\includegraphics[width=\wid]{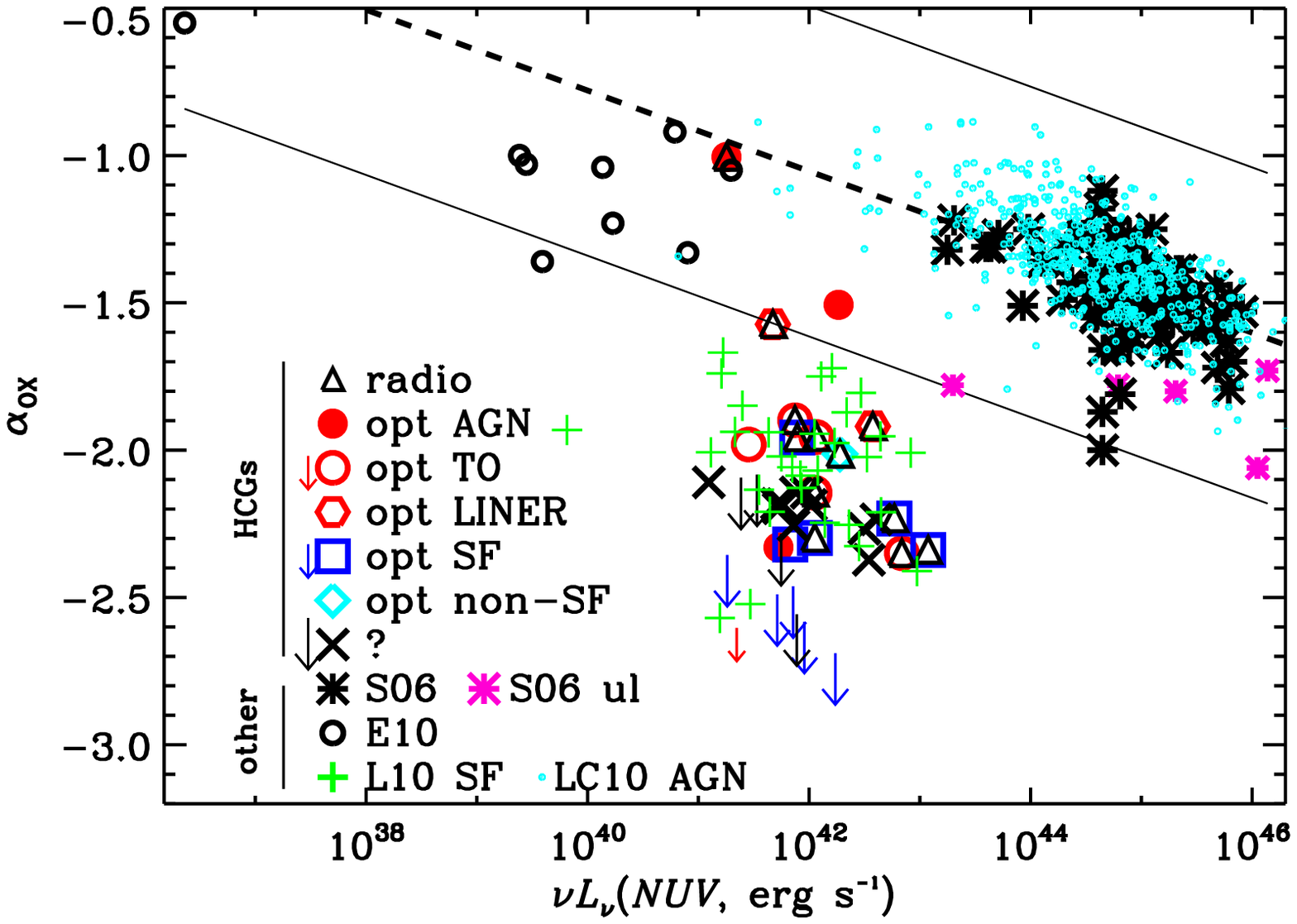}
\caption{UV-to-\x\ spectral index, \aox, vs. UV luminosity at $\sim
  2600$\AA.  The left panel uses values corrected for intrinsic
  extinction, while the right panel uses uncorrected values.  Symbols
  indicate optical nuclear classification (\scr{sec_class}) as
  indicated in the legend: Triangles, large open and filled circles,
  squares, diamonds, pentagons and crosses are for HCG nuclei.
  Downward pointing arrows are upper limits for HCG nuclei based on
  upper limit \x\ flux estimates for \x\ non-detections and are
  color-coded to indicate optical nuclear classifications (black if
  there are none).  Note that in the interest of clarity, mixed
  classifications have been simplified, as explained in the text.
  Non-HCG data points are from \citet[][S06, ul = upper
    limits]{steffen2006}, \citet[][E10]{eracleous2010}
  \citet[][L10]{lehmer2010}, 
\citet[][LC10]{lusso2010}.  The dashed line is the correlation for
  strong AGNs from S06. The two solid lines indicate the $\pm 3\sigma$
  region for AGN (S06) and LINERs (E10).  The green crosses are
  star-forming galaxies from L10, with SFRs in the same range as our
  HCG galaxies.  These plots clearly suggest that most HCG nuclei do
  not follow the correlation for strong AGN but are more similar to
  star-forming systems.  }
\label{fig:aox_nulnu}
\end{figure*}

To quantify the relative contributions coming from the \x s and the UV
we calculate the \aox\ index defined above 
and tabulate our results in columns 5 and 6 of \tr{tab-multi}, corrected
and uncorrected for intrinsic extinction, respectively.
The mean \aox\ values and 1$\sigma$ standard deviations are 
\aox~$=-2.33 \pm 0.31$ (corrected) and 
\aox~$=-2.17 \pm 0.32$ (uncorrected). Thus, overall, the extinction
correction does not significantly affect \aox\ values.
We plot \aox\
vs.
2600~\AA\ luminosity, \nulnu$_{2600}$, in \fr{fig:aox_nulnu}. The well-known
correlation for 333 
moderate- to high-luminosity AGNs
by \citet[][S06]{steffen2006}
is shown by the dashed line and the
$\pm 3\sigma$ scatter of the data points on the correlation by the
solid lines.  The 
S06 AGNs are shown as black (magenta
for upper limits) stars
(their tables 1 and 2). 
We also show the AGNs of \citet[][herafter LC10]{lusso2010} as cyan dots.
The black open circles are the LINERs
of \citet{eracleous2010}. Our HCG nuclei are shown in color, according to
their optical spectroscopic classification (\scr{sec_class}),
or as black crosses if there is no such classification. HCG upper limits
are shown by downward pointing arrows. In the interest of clarity, in this
and subsequent plots
mixed classifications (\tr{tab-class}, column 5, and \tr{tab-multi}, column 7) 
have been simplified as follows:
AGN? $\rightarrow$ AGN; SF? $\rightarrow$ SF; TO/AGN, TO/SF, TO/LNR 
$\rightarrow$ TO; LNR/AGN $\rightarrow$ LNR.
It can be seen that HCG nuclei do not occupy the same locus as strong AGNs and LINERs.
This suggests that HCG nuclei are not likely to harbor strong AGNs
or LINERs.
To further explore
whether this is also consistent with star formation being dominant
in HCG nuclei, we compare with star forming galaxies from the compilation of
\citet[][hereafter L10]{lehmer2010}. 
These authors use multiwavelength criteria to select
star forming galaxies among nearby ($<60$~Mpc) 
Luminous Infrared Galaxies (LIRGs).
We select a sub-sample of 30 galaxies from this sample 
whose SFRs are in the range
0.011 to 17~\msuny, matching the SFR range of our HCG galaxies
\citep{tzanavaris2010}. Note that we do not need nuclear photometry of these
galaxies in order to test our results. What we wish to investigate is
whether the \x\ and UV contributions in a galactic environment dominated
by star formation are similar to our findings for the majority of HCG nuclei.
We thus calculate \aox\ for these systems and
indicate them in \fr{fig:aox_nulnu} as green crosses. The mean \aox\   
for these sources is $-1.90 \pm 0.33$ and the figure shows that there
is significant overlap with our nuclei. In particular,
the most relevant comparison is with HCG nuclei data {\it not}
corrected for intrinsic extinction (right panel in \fr{fig:aox_nulnu}), 
as L10 did not carry out any such
correction. At the same time, compared
to strong AGNs, these
galaxies seem to occupy a completely different region.

\begin{figure*}\vspace{-5cm}\hspace{-1cm}
\epsscale{1}
\hspace{-1cm}
\includegraphics[width=8in]{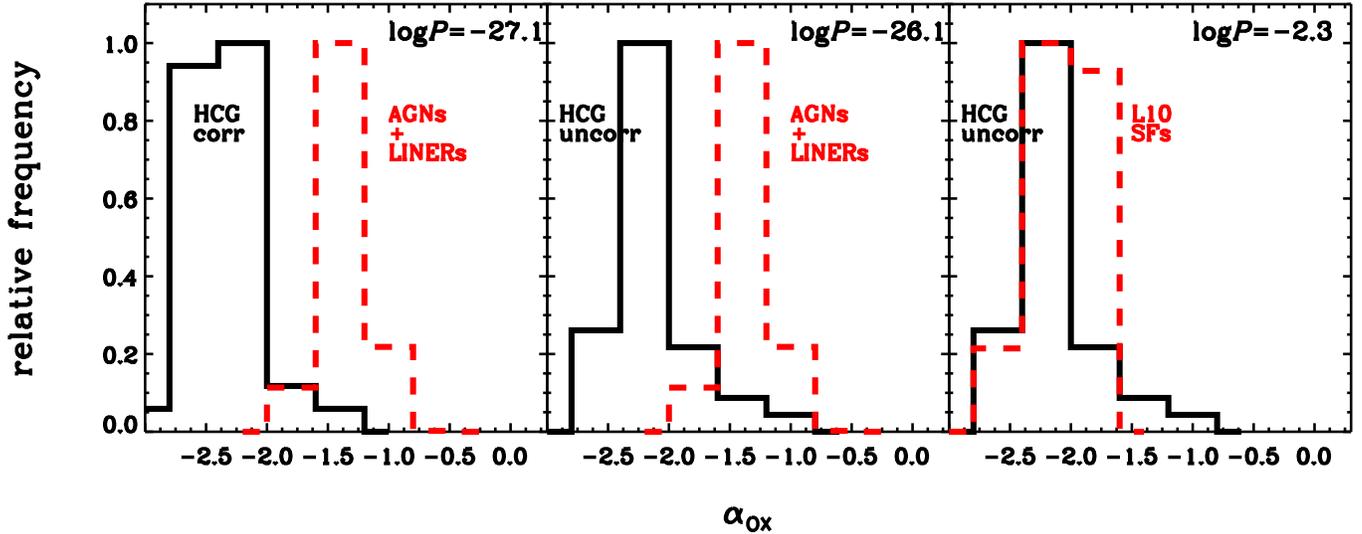}
\caption{
Normalized distributions of \x-to-UV spectral index, \aox, and KS probability
for different 
galaxy sample pairs. In all panels, the distribution for HCG nuclei is indicated by the solid black line, corrected for intrinsic extinction in the left
panel, and uncorrected in the other two. In the left and middle panels, the red dashed
line is for all strong AGNs and LINERs from other samples (S06, LC10, 
Eracleous 2010a,b). In the right panel, the red dashed line
is for the star-forming galaxies of L10. The logarithm of KS probability
that two distributions come from the same parent distribution is
shown at the top right of each panel.
}
\label{fig:aox_hist}
\end{figure*}

We quantify the comparisons between \aox\ values for
different datasets by carrying out a standard two-sided
KS test, the results of which are
shown in \fr{fig:aox_hist}. The probability that the
\aox\ values for HCG nuclei come from the same distribution
as the values for strong AGNs and LINERs is extremely small
($8\times 10^{-28}$ if corrected
and $8\times 10^{-27}$ if uncorrected for intrinsic extinction).
In contrast, the probability that the \aox\ values for
HCG nuclei uncorrected for intrinsic extinction come from the
same distribution as those for the L10 star-forming
galaxies is substantially larger at $0.005$. This number is
not extremely large, as, after all, the L10 sample represents
an extragalactic environment (LIRGs) 
distinct from compact groups. However this probability may be
high enough to suggest that HCG nuclei
are at least more similar to star-forming galaxies than they
are to strong AGNs. This is also evident from the overlap between
the two distributions in the histogram of the rightmost panel
in \fr{fig:aox_hist}.
We thus conclude that,
 generally speaking,
HCG nuclei do not follow the strong AGN correlation {\it and}
are consistent with star formation being dominant.

\begin{figure*}[ht]
\epsscale{1.0}
\newcommand{\wid}{9cm}
\centering
\subfloat
{
\includegraphics[width=\wid]{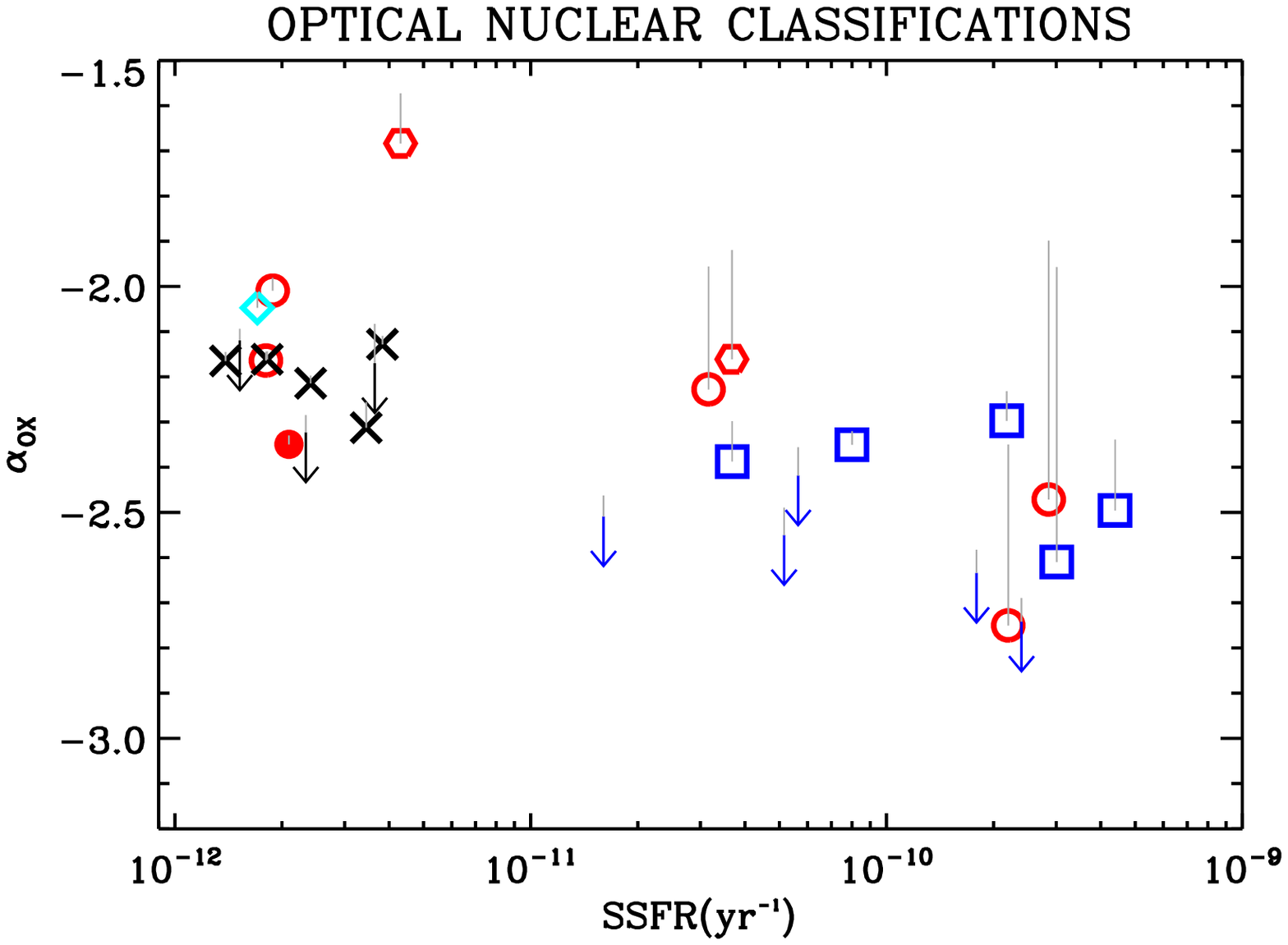}\hspace{-0.5cm}
\includegraphics[width=\wid]{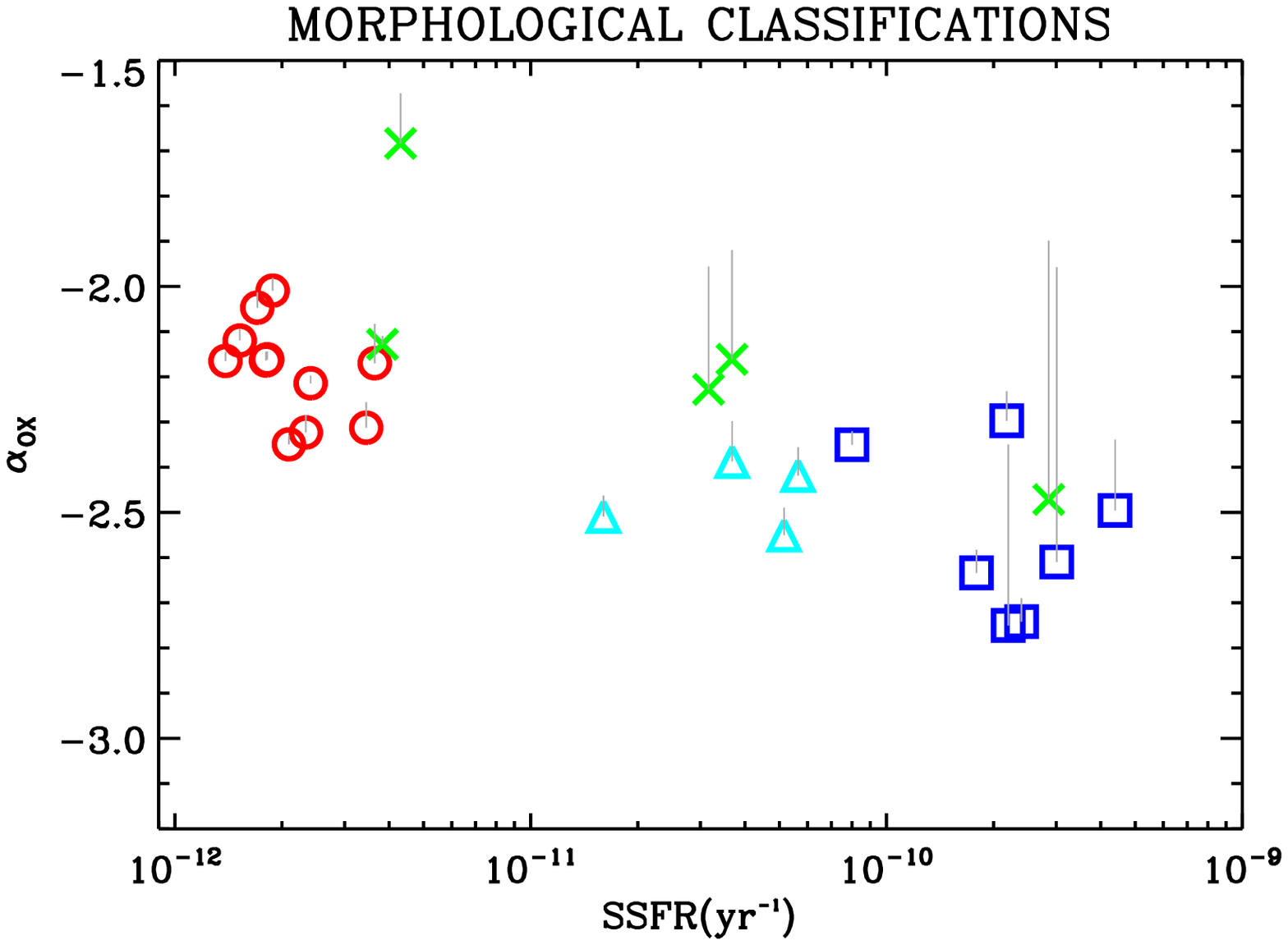}
}
{\hspace{1cm}
\includegraphics[width=\wid]{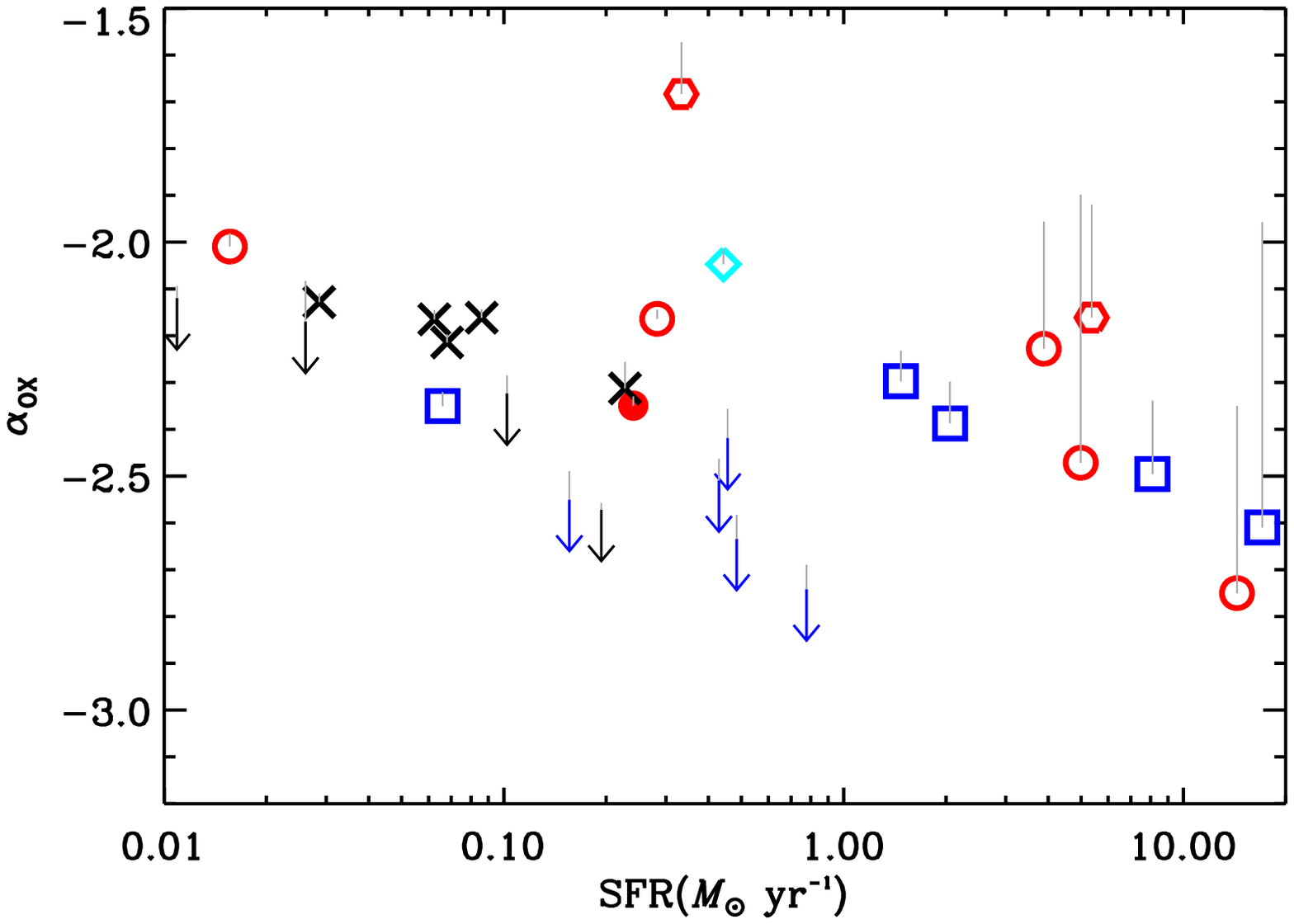}\hspace{-0.5cm}
\includegraphics[width=\wid]{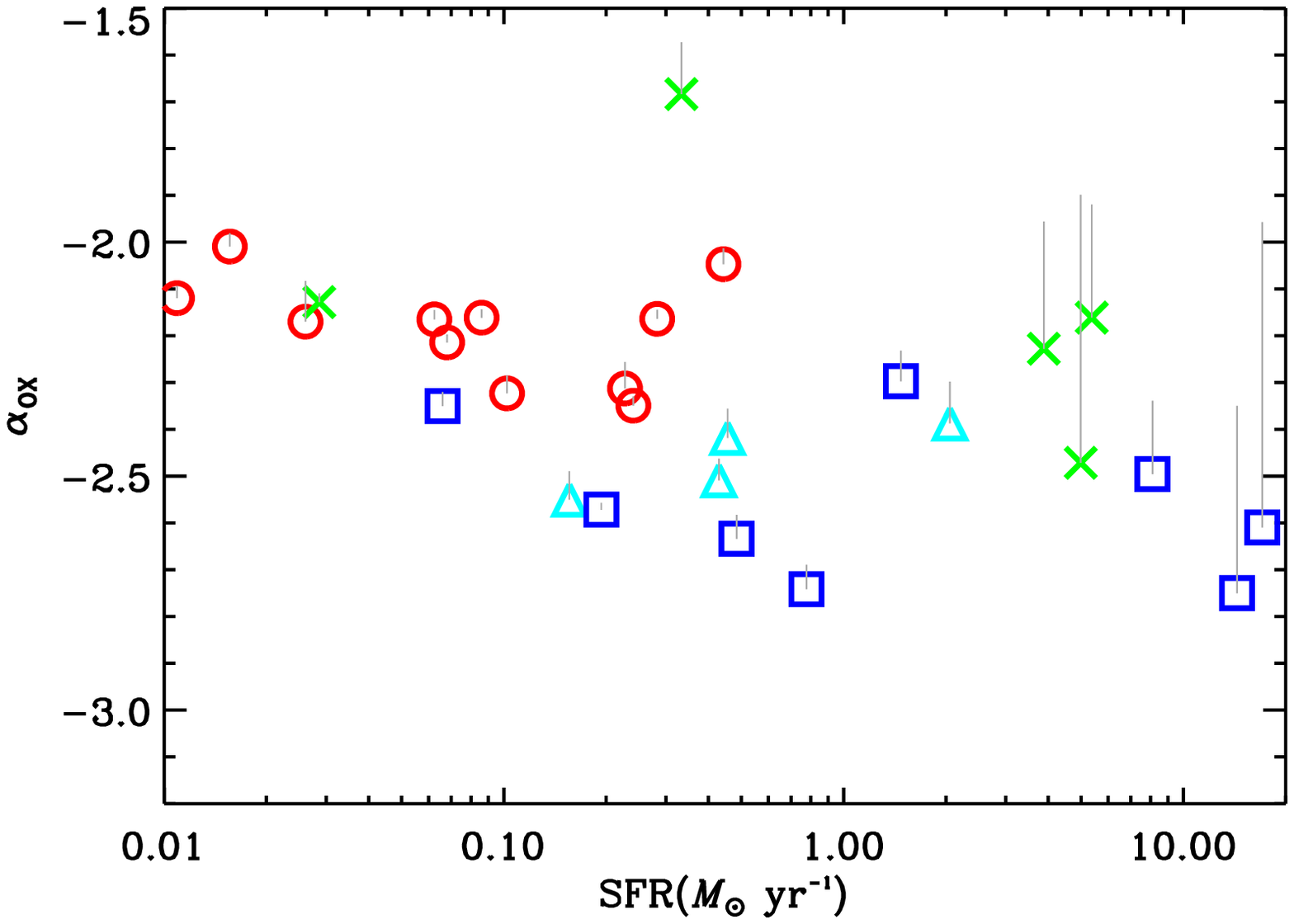}
}
{
\includegraphics[width=\wid]{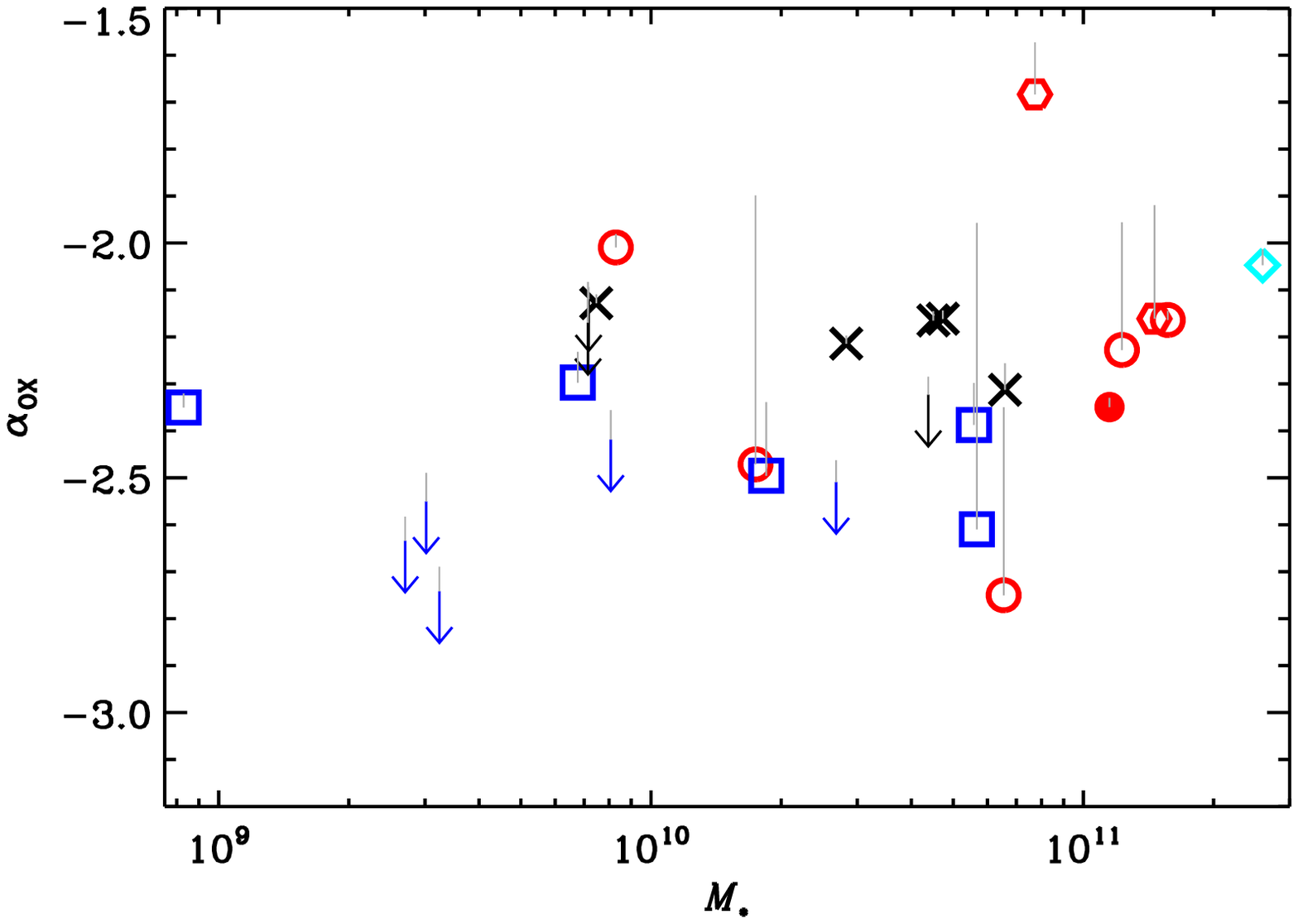}\hspace{-0.5cm}
\includegraphics[width=\wid]{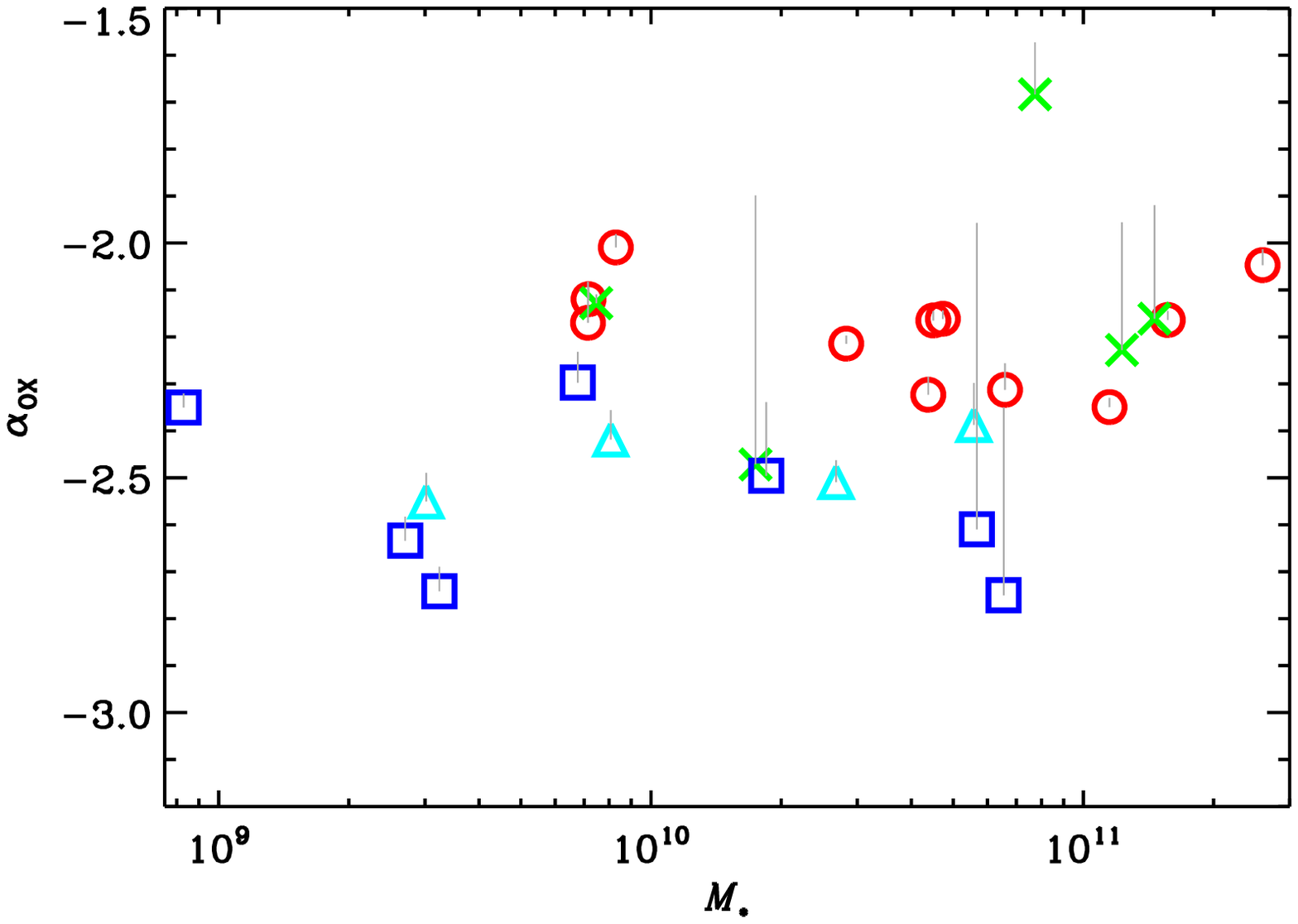}
}
\caption{UV-to-\x\ spectral index, \aox, vs SSFR (top panels),
SFR (middle panels) and \mstar\ (lower panels).
The data points in each panel pair are identical and
upper limits are only indicated in the left panel.
All symbols use \aox\ values corrected for intrinsic
extinction as explained in the text. The vertical grey
lines show the effect of extinction correction (\aox\
values are more positive if not corrected).
{\it Left,} symbols indicate
optical nuclear classifications (same as in \fr{fig:aox_nulnu}).
{\it Right,} symbols indicate morphological T-types, 
where T increases from ellipticals to spirals to
irregulars:
Red circles are T~$\le 0$,
green crosses are $1 \le {\rm T} \le 3$,
cyan triangles are $4 \le {\rm T} \le 7$,
blue squares are ${\rm T}\ge 8$.
The SSFR panel pairs suggest that 
nuclei with lower \aox\ are
highly star-forming and are hosted by later morphological types.
This trend is less pronounced in the SFR panels.
The \mstar\ panels show that AGNs are hosted 
by the more massive systems and the earlier morphological types.}
\label{fig:aox3}
\end{figure*}

Further, in \fr{fig:aox_nulnu} 
there is a trend
with increasing \nulnu$_{2600}$ and decreasing \aox\ for nuclei to be both
further below the strong AGN locus defined by the correlation and
to be optically classified as either SF or TO.
To further investigate this possible
connection between \aox\ and star formation,
we plot \aox\ versus 
SSFR
in the top row of \fr{fig:aox3} using two different
color and symbol schemes. In the left panel
symbols are coded to indicate the optically-based nuclear activity
classification (where available). 
In this panel optSFs preferentially occupy the high-SSFR/low-\aox\
region of parameter space, where \aox~$\lesssim -2.3$. In contrast,
optLINERs, opt-nonSFs, optAGNs and some optTOs inhabit the
low-SSFR/high-\aox\ region of parameter space. The fact that
two opt-TOs are found in the optSF region is consistent with their
definition, as one would expect transition objects to appear in both regions.
If this apparent distribution pattern in \aox -- SSFR space is
a real effect, then we
predict that the nuclei for which there is no optical classification
(marked by black crosses or downward pointing arrows in this panel) 
are most likely AGNs, LINERs or transition objects.

On the other hand,
symbols in the right panel indicate the RC3-based morphological
classifications of the host galaxies. A value of \aox~$\sim -2.3$ here
roughly separates ellipticals 
(with low SSFRs) from galaxies with progressively more spiral morphologies (and higher SSFRs). This is consistent with our prediction for 
unknown nuclear activity systems in the left panel, as these
nuclei are hosted by morphologically
earlier-type systems.

Since SSFR is SFR normalized by stellar mass, \mstar, it is
interesting to also investigate possible trends of \aox\ with these
quantities separately.  
In the middle and bottom rows of \fr{fig:aox3}
we show pairs of plots
for \aox\ versus SFR and
\aox\ versus \mstar.
Although there is a trend for lower SFR (and higher \mstar) systems
to have larger values of \aox, this is not as pronounced as
the trend with SSFR.

\subsection{Multi-wavelength Classifications (\tr{tab-multi})}\label{sec:over}
\tr{tab-multi} brings together results on the type of
activity of HCG nuclei from different wavelength ranges.
The information presented includes \x\ -- UV nuclear counterparts
(columns 3 and 4, \scr{sec:xuv}), UV-to-X-ray spectral indices
(columns 5 and 6, \scr{sec:xuv}), optical emission line ratio
classifications (column 7, \scr{sec_class}), level of
\x\ nuclear activity (column 8, \scr{sec:xsrc}),
radio nuclear excess (column 9, \scr{sec:radio}) and
\hst -- based nuclear detections (columns 10 and 11, \scr{sec:hst}).

Considering this table as a whole, we can draw a number of useful conclusions:

1. 
For about 60\%\ (22/37) of HCG galaxies we have
detected a single, nuclear \x\ point source, as indicated
by the overlap between the \x\ and UV PSFs in the central region of a galaxy
(\lq\lq y\rq\rq\ in column 4).
Two of these galaxies are classified optAGN, one opt-nonSF, one
optLINER/AGN, one optLINER, two optTO/LINER, two optTO/SF and
three optSF. Nine are have no emission line ratios and are
unclassified.
Given \chandra's resolution, the detection of these nuclear
\x\ sources is consistent with {\it some} level of AGN
activity in these systems, although an XRB origin cannot
be excluded (and could, in fact, dominate).

2.  
For 27\%\ (10/37) of galaxies there is no detected
nuclear \x\ emission at all (\lq\lq n\rq\rq\ in column 4).
Five of these nuclei are classified optSF, four are
unclassified and one is classified optTO/LINER. This is HCG 90 D, which
has a very irregular, clumpy appearance, likely related to
its close interaction with 90 B. Although we formally
identify the center of galaxies
with a UV intensity peak,
there is likely no well-defined nucleus for such a
morphology. Depending on the slit position, it is unclear
whether the optical spectroscopic classification corresponds to the same
source.

3.
For three galaxies ($\sim 8$\%), the identification 
of counterparts is questionable (\lq\lq ?\rq\rq\ in column 4). 
Specifically, for HCG
31~Q, the \chandra\ PSF overlaps slightly with the UV PSF, but the
source positions are also $\sim 2.3$\arcsec\ apart, so the lack of
coincidence is unlikely to be due to astrometric errors 
(\lq\lq n?\rq\rq\ in column 4).  For HCG 59~B
and 92~F, the \x\ sources are $\sim 2$\arcsec\ from the UV intensity
peak with
little PSF overlap. These can only tentatively be considered
as \x\ counterparts
to the UV emission (\lq\lq y?\rq\rq\ in column 4).

4.  As mentioned, in two cases ($\sim 5$\%) 
there are two possible \x\ counterparts
for a single UV nuclear source.  In particular, for HCG 31~ACE, the
\chandra\ PSFs for \x\ sources 38 and 40 both overlap with the
UV PSF. Since source 40 is only
0.7\arcsec\ away from the UV peak (c.f. 1.7\arcsec\ for 38),
we chose source 40 as the most likely counterpart. In the case of
22~A both \x\ PSFs are almost symmetrical about the UV peak, so not
even a tentative choice is possible.

5.
\x\ luminosity is often used as a direct diagnostic
for the presence of an AGN, with $\sim$\ten{1}{41-42}~\lunits\ taken
as a fiducial threshold for
unambiguously strong AGN activity \citep[e.g.][]{reynolds2003,bauer2004}.
Using \lxte~$=10^{41}$~\lunits\ as our threshold, we see that 
there are only three sources which are candidates for strong AGNs
(column 8 in \tr{tab-multi}).
These are HCGs 16~B, 90~A and 92~C. The first is
classified as optLINER and the other two optAGN. Thus together with
62~A, HCG 16~B is a second case where a LINER and an AGN may coexist.

7. 
According to column 7, there is one optLINER/AGN, one optAGN and
one opt-nonSF
which, according to column 8, are not strong \x\ AGNs. 
Combining these results, we believe that these are
good LLAGN candidates.

8.  
Columns 10 and 11 suggest that our \hst\ data do not
help us discriminate clearly between AGN and SF systems.  
For 22 sources the \hst\ median-divided image
suggests a nuclear detection
(\lqq y\rqq\ in column 10). However, there is no 1-to-1
correspondence between these 22 sources and the 22 sources for which
X-ray -- UV counterparts exist. On the one hand, it is encouraging that
for four of these, namely the nuclei of HCG 16~A, 22~A, 42~A and 92~C,
classified as optLINER/AGN, optAGN, opt-nonSF, and optAGN,
respectively (column 7), an optical nuclear source seems to be present
in the median-divided image (column 10). 
However, combining this result with \galfit\ is problematic for 
two main reasons. First, in general \galfit\ leads to fits that
are poor and
in most cases do not converge. This is because most of the galaxies
in our sample are disturbed and thus full Sersic profile fitting is usually
unable to fit the diffuse light. Second, only one of the AGN candidate
galaxies mentioned, HCG 42~A, has a 
\galfit\ center that lies within the 3-pixel positional uncertainty from
the median-divided center. For the other three galaxies, we are 
unable to obtain a satisfactory \galfit\ fit.  
On the other hand, note that five galaxies whose optical classification is
not AGN, nevertheless have their \galfit\ and median-divided centers 
within the 3-pixel positional uncertainty.
Given the lack of good \galfit\ results, it is hard to assess
the significance of this result.



\section{AGN fraction}
For the purpose of comparison with galaxy clusters, we calculate the
AGN fraction in our compact group sample by following
\citet{martini2006}. We estimate 3$\sigma$ uncertainties using the
method of \citet{gehrels1986}.  We use $R$-band magnitudes from
\citet{hickson1989} to establish that in our sample there are 26
galaxies with $M_R < -20$. For 4 galaxies for which
\citet{hickson1989} do not provide $R$-band data we obtain magnitude
information from the NASA/IPAC Extragalactic Database.  We assume that the
AGN hosts among these 26 galaxies are those that are optically identified
as such (Column 7 in \tr{tab-multi}). This includes mixed
(e.g. LINER/AGN) and tentative (AGN?)  classifications. The inclusion of
these latter cases does not actually matter, as we also use the
further criterion that \lxte~$ > 10^{41}$ \lunits. This only leaves
two galaxies, which are the two strongest AGN cases, namely HCG~90~A
and HCG~92~C.  The AGN fraction defined in this way for two out of 26
galaxies is then $f_A (M_R
< -20, L_X > 10^{41}) = $~\aer{0.08}{+0.35}{-0.01}.  This is close,
but higher than the $\sim 5$\%\ fraction of \citet{martini2006} in
galaxy clusters.

\section{Discussion}
The main result of this paper is the significant detection of nuclear
\x\ point sources in 60\%\ of the 37 HCG galaxies in this
sample. We detect these sources by taking advantage of \chandra's
excellent angular resolution, while also taking into account the size of
the PSF at each source's detected position on the CCD and the overlap
with the UV PSF.

\begin{deluxetable}{ccc}
\tablecolumns{3}
\tablewidth{0pc} 
\tablecaption{Physical size estimate (kpc) of areas used for UV and \x\ nuclear
photometry.\label{tab-psf}}
\tablehead{ 
\colhead{group ID}
& \colhead{\swift-UVOT (10\arcsec)}
& \colhead{\chandra\ (1\arcsec)}
\\
\colhead{(1)}
& \colhead{(2)}
& \colhead{(3)}
}
\startdata
HCG   7 & 2.7 & 0.3  \\
HCG  16 & 2.6 & 0.3  \\
HCG  22 & 1.8 & 0.2  \\
HCG  31 & 2.8 & 0.3  \\
HCG  42 & 3.0 & 0.3  \\
HCG  59 & 3.1 & 0.3  \\
HCG  62 & 3.1 & 0.3  \\
HCG  90 & 1.6 & 0.2  \\
HCG  92 & 4.3 & 0.4  \\
\enddata
\tablecomments{Columns are: (1) HCG group ID; 
(2) Extent in kpc of 10\arcsec\ at the galaxy group
distance. 10\arcsec\ is the diameter of the circular regions
used for nuclear UVOT photometry.
(3) Extent in kpc of 1\arcsec, the on-axis \chandra\ PSF, at
the galaxy group distance.
}
\end{deluxetable}

%
%
Unfortunately, the mere detection of an X-ray point source
that appears to coincide with the UV galactic nucleus 
is no strong evidence that the 
origin
of the nuclear emission is an AGN, rather than
circumnuclear star formation,
unless the source also has a high \x\
luminosity. The threshold we use in this paper is
\lxte~$\ge 10^{41}$~\lunits\ and most of our nuclear sources
are fainter.
We provide a quantitative illustration of this caveat in \tr{tab-psf},
where we give estimates of the physical sizes (in kpc) 
of the areas used for \chandra\ and UVOT photometry
at the distances of our
galaxies.  Taking the case of \chandra, 1\arcsec\ at the distances of
these galaxies corresponds to physical sizes ranging from $\sim 0.16$
(for HCG 90, our nearest group) to $\sim 0.43$ kpc (for HCG 92).  The
known detections of nuclear star clusters 
with sizes of this order
and ages as young as 10~Myr
demonstrates that regions of this physical size can be the sites of compact and
intense circumnuclear star formation \citep[e.g.][]{rossa2006} that,
in turn, can harbor unresolved XRB populations.  Since most of our nuclear
sources have low
\lx\ estimates, such a possibility
should be seriously considered.
The optical emission-line ratio classifications
for 22 galaxies in the sample are in support of the lack of
strong AGN activity and the prevalence
of star formation in HCG nuclei. We caution that these are
small numbers, but, taken at face value, almost half of 
our systems (45.5\%) are classified as pure optSF, while more are classified
as optTO or have mixed classifications. Only a minority (13.6\%) 
are optAGN, and two of these also have high \x\ luminosities.

Two of the three nuclei 
that fulfill the \x\ luminosity criterion for harboring strong AGNs
have 
\lxte~$> 10^{42}$ \lunits\ (90 A and 92 C, both also optAGN).
The result for HCG~90 A is also independently confirmed
by \citet{lamassa2011} who, using \xmm\ data and detailed
spectral fitting, report \lxht~$=10^{42.96}$~\lunits\
(compare with \lxte~$=10^{42.6}$~\lunits\ in our \tr{tab-xray}).
These are the {\it only} nuclear sources in our sample that
fulfill optical spectroscopic, \x\ luminosity and, for 90~A,
radio excess criteria for being unambiguous and strong AGNs.

As \x\ luminosity decreases, so does its discriminative power 
as an AGN diagnostic. Two nuclei have  $ 10^{40.5} <$~\lx~$< 10^{42}$
(16A and 16B, optLINER/AGN and optLINER, respectively), but
the optLINER/AGN is the \x\ {\it fainter} one. 
The rest of the systems are progressively fainter in the
X-rays, precluding any conclusions on the nature of 
the nuclear activity from \x\ information alone.
The conservative conclusion based on \x\ luminosity is to
consider
these as candidates either for nuclear star clusters and XRB hosts or
LLAGN hosts (or, possibly, some combination of the two).
Although \x\ faint (\lxht~$\sim 10^{38}$~\lunits) nuclei 
have been identified as LLAGNs with \chandra\ data, e.g. 
in the Palomar sample of
\citet{ho2001}, the comparison is not entirely fair
as that sample contained early type galaxies where star formation
is unlikely to strongly dilute the AGN emission.

We have also combined \x\ and UV nuclear photometry to calculate 
\aox\ values and compare with those for strong AGNs and star forming
galaxies. The KS test shows that HCG nuclei are completely
distinct from strong AGNs and fairly similar, though not
identical, to star forming galaxies. 
The correlation in \aox\ -- \nulnu$_{2600}$ space (\fr{fig:aox_nulnu})
for strong AGNs is well established over $\sim 4-6$ orders
of magnitude \citep{strateva2005,steffen2006,just2007}.
\citet{lusso2010} also confirm this correlation, slightly extending it to
fainter systems.
The location of HCG nuclei in \fr{fig:aox_nulnu}
is consistent with their 
\lx\ values: The highest \lx\ nuclei, i.e. those that are more likely
to harbor strong AGNs, do, in fact, follow the 
correlation. For lower nuclear \lx\ values,
nuclei are located further below the
correlation. Notably, this region is also inhabited by the normal,
star-forming galaxies of L10. Since this latter sample was compiled
independently from our HCG sample, as well as the other comparison samples
shown in \fr{fig:aox_nulnu}, this result is in support of our conclusion that
strong AGN activity is absent and
star formation may be dominant in most HCG nuclei.

Taking into account the optical AGN/TO/SF classification information,
in \fr{fig:aox_nulnu} there appears to be a broad transition region
at \aox~$\sim -2.3$. At lower \aox\ values most of the nuclear emission
is likely due exclusively to
star-formation. 
This is consistent with the top row of \fr{fig:aox3}.  The left
panel shows that it is precisely the high SSFR systems that have
\aox~$\lesssim -2.3$, and the right panel shows that these have 
the most spiral-like morphologies. In addition, the right panel shows a
very smooth progression with morphological type towards values of
\aox~$\lesssim -2.3$.  Although we do not claim that
\aox~$\gtrsim -2.3$
{\it guarantees} AGN activity, the fact that these systems are
preferentially early type (as well as the most massive, see bottom
row of \fr{fig:aox3})
is consistent with earlier results that AGNs
in compact groups are located preferentially in early type galaxies
\citep{coziol1998a,coziol1998b}.  Note also that the trend with
decreasing \aox, although still present,
is not as clear in
the lower two panel rows
of \fr{fig:aox3}. This simply 
reflects the fact that SSFR is 
by construction
a better discriminator between
highly star forming late types and more quiescent (and more massive)
early types.

One possible caveat in this work is the mismatch
in resolution and, hence, photometric aperture between the \x\ and UV data. 
The use of a 10\arcsec\ diameter for UV photometry,
corresponding to physical sizes between
1.6 and 4.3~kpc,
means that some of the UV emission
may not be co-spatial with emission
from the
\x\ source inside the smaller \x\ aperture. This would lead
to an overestimate of the UV luminosity, and, hence,
an underestimate of \aox. In such a case, 
data points corresponding to the \lq\lq true\rq\rq\
\aox\ and UV luminosity values in \fr{fig:aox_nulnu} would be
shifted
to the left (towards lower \nulnu$_{2600}$ values)
and up (towards higher \aox) relatively to the data points
shown.
Note though that, with the exception of
possible nuclear star clusters, the UV
aperture mostly traces galactic bulges,
where old stellar populations dominate. 
Contamination from non-nuclear UV emission due
to massive young stars might then be not so important.
Also, contamination from {\it old} stars
(the so-called UV upturn) is not important
at the near-UV 
wavelengths studied because it only becomes dominant
blueward of 2000~\AA\ \citep{oconnell1999}.
These observations are in line with the
results of the tests by \citet{grupe2010}
who performed AGN UV photometry with
\swift\ UVOT. They performed photometry
both with the standard 5\arcsec-radius UVOT 
aperture and a narrower 3\arcsec-radius that included
aperture corrections. They found that host galaxy
contamination is most serious in
the optical UVOT bands, while in the UV
bands the magnitude difference was on the
order of 0.05 mag. We should caution, however, that since
\citet{grupe2010} had bright AGNs, they were more likely
to have the AGNs dominating both in the UV and the \x, which
is more likely to drive their conclusions.
Similarly, the \aox\ values of our known strong
AGN systems do fall close to the strong AGN
correlation.
Finally, we test the effect of photometric aperture by performing
\x\ photometry using the larger UV apertures. This leads, on
average to $L_{\nu, {\rm 2 keV}}$ values that are higher by
0.48 \lnuunits\ or \aox\ values that increase by 0.18. Such
changes do not significantly affect the results of this work.

We are thus led to the conclusion that 
the observed trends of \aox\ with
\nulnu$_{2600}$ are telling us two things:
First, HCG nuclei {\it as a class} generally do not harbor
strong AGNs. Instead, the UV and \x\ emission we see
is in most cases 
likely
dominated by nuclear star formation.
Second, considering {\it early and late-type morphologies},
the most likely interpretation is that
low SSFR, early-type, systems are the ones
harboring weak AGNs, while high SSFR, late-type, systems
are dominated by star formation.

The radio and \hst-based diagnostics do not provide
much additional insight regarding the nuclear activity
in this galaxy sample. For three galaxies, the possible detection of excess
radio emission, above
that expected from star formation alone, is
consistent with the \x\ luminosity (for HCG 90~A), as
well as the optAGN classification (for HCG 16~A and 90~A).
However, the excess is not greater than the uncertainty
in the radio-SFR correlation used. We also have no radio
data for any of the galaxies that lack optical classifications.
Thus the radio results are not particularly useful.

Using \hst-data for a subset of our galaxies, we detect what 
seem to be nuclear
sources after subtracting median-smoothed images for 22 galaxies.
However, these sources do not correspond to the optical
emission line ratio classifications in any systematic way.
Due to the disturbed nature of most galaxies, we are also unable to obtain reliable and consistent surface
brightness fits with \galfit\ in most cases. The fact that some
sources have their median divided center within 3-pixels of
the \galfit\ center is thus of limited significance. 

Our AGN fraction result for compact groups suffers from small
number statistics and needs to be tested with larger samples.
However, taken at face value and given that we have applied
the same criteria used by others to obtain the AGN fraction in 
galaxy clusters, our result is somewhat higher than
the one for clusters. This might suggest that the lower
velocity dispersions and shorter crossing times
in compact groups, which make this
environment physically distinct from that of clusters,
also have an effect on AGN activity, just as they seem to
do for star formation.

We have no emission line ratios, and thus no
optical spectroscopic classifications for 15 of our
galaxies, indicated by black crosses and arrows in the Figures.
These systems include HCG~7B, 22B, 31F, 42B, 42C, 42D,
62B, 62C, 62D, 90B, 90C, 92B, 92D and 92E.
Based on the trends discussed above and the morphologies of these
systems, we tentatively predict the nature of their nuclear activity.
There are two late type systems, HCG~31F (morphological type Im) 
and HCG~92B (type Sbc).
Our prediction is that these will be dominated by star formation.
The rest of the systems have elliptical/S0 morphologies, and may harbor a
weak LLAGN.

\section{Conclusions}
This paper presents the first compilation of \x\ detected
point sources in the fields of 9 HCGs, for which we provide
an extensive compilation of source characteristics.

We have used multi-wavelength diagnostics
(\x, UV, optical, and radio) to assess the levels
of AGN, SF and LINER activity in the compact group
environment. Our main results are the following:
\begin{enumerate}

\item In 60\%\ of 37 galaxies we detect single, nuclear \x\ sources 
that have nuclear UV counterparts.
We detect no nuclear \x\ emission for 27\%\ of our galaxies.
The rest of the systems have more uncertain \x\ nuclear detections.

\item Out of the 22 galaxies for which emission line ratios are available in the literature, we classify a clear plurality (45.5\%) asn optSF.
Our criteria allow us, for the first time, to also classify
five systems as LINERs, although
four of these are mixed (LINER/AGN, TO/LINER). 
Thus any LINER activity is associated with a minority (22.5\%) of systems.
Only three nuclei (13.6\%) are classified optAGN.

\item Only three systems (HCG 16~B, 90~A, 92~C) are candidates for 
hosting an \x\ strong AGN
(\lxte~$\ge  10^{41}$~\lunits).

\item When several criteria are taken into account
(optical spectroscopic classification, excess radio emission,
\x\ luminosity, location in \aox--\nulnu$_{2600}$ parameter space)
 only two HCG nuclei (90~A, 92~C) fulfill several criteria
and are classified as strong, unambiguous AGNs.

\item In \aox--\nulnu$_{2600}$ space, HCG nuclei occupy a region which
  is distinct from that occupied by strong AGNs and largely overlaps
  with that occupied by other, nearby star-forming galaxies not known
  to harbor AGNs (\fr{fig:aox_nulnu}).  The only exceptions are the
  two strong AGNs which do fall in the AGN region. We thus tentatively
  make the prediction that HCG nuclei without optical nuclear-type
  classifications are dominated by star formation (if they have late-type
morphologies) or may harbor {\it low} luminosity AGNs (especially if they
have early-type morphologies).

\item \aox\ anticorrelates with galaxy-wide SSFR and spiral morphology
  so that the star formation contribution is strongest in highest
  SSFR and later type morphology galaxies\fr{fig:aox3}. The detected
  anticorrelation (correlation) with SFR (\mstar) is weaker.

\item Using the same criterion used in galaxy clusters
\citep{martini2006},
the AGN fraction of HCG galaxies more luminous both than
$M_R = -20$ and \lxte~$=10^{41}$~\lunits\ is \aer{0.08}{+0.35}{-0.01}, which is
close but higher to that in clusters.

\end{enumerate}

Our general conclusion is that overall the CG environment has a
mitigating effect on the level of AGN activity but not 
AGN numbers. With future expanded \x\ and UV
samples as well as deeper observations we will be able to better
assess the nature and statistics of HCG nuclei. The comparison with
galaxy clusters suggests that environment plays a key role
for the overall level of AGN activity. In this respect, 
it is imperative to carry out detailed comparisons
with samples from other group, cluster and field environments.

\acknowledgments We gratefully acknowledge the support of the ACIS
Instrument Team contract SV4-74018 (PI: G.~P.~ Garmire).  P.T.
acknowledges support through a NASA Postdoctoral Program Fellowship at
NASA Goddard Space Flight Center, administered by Oak Ridge Associated
Universities through a contract with NASA.  A.H. and P.T. were also
supported by NASA ADAP 09-ADP09-0071 (P.I. Hornschemeier).  We thank
Kip Kuntz for discussions and assistance with
matching of sources in three \x\ bands. We thank
Bret Lehmer for making his catalog of star forming galaxies available
to us.  K.F., T.D.D. and S.C.G thank the Natural Science and Engineering
Research Council of Canada and the Ontario Early Researcher Award
Program for support.  W.N.B. acknowledges support from NASA ADP grant
NNX10AC99G.  J.C.C. and C.G. acknowledge funding that was provided
through \chandra\ Award No. GO8-9124B issued by the \chandra\ X-Ray
Observatory Center, which is operated by the Smithsonian Astrophysical
Observatory under NASA contract NAS8-03060, grant number
HST-GO-10787.15-A from the Space Telescope Science Institute which is
operated by AURA, Inc., under NASA contract NAS 5-26555, and by the
National Science Foundation under award 090894.  The Institute for
Gravitation and the Cosmos is supported by the Eberly College of
Science and the Office of the Senior Vice President for Research at
the Pennsylvania State University.  K.F. acknowledges support from the
Queen Elizabeth II Graduate Scholarships in Science and Technology.
This research has made use of data obtained from the High Energy
Astrophysics Science Archive Research Center (HEASARC), provided by
NASA's Goddard Space Flight Center.  This research has made use of the
NASA/IPAC Extragalactic Database (NED) which is operated by the Jet
Propulsion Laboratory, California Institute of Technology, under
contract with the National Aeronautics and Space Administration.

{\it Facilities:} \facility{\chandra}, \facility{\swift}


\begin{thebibliography}{106}
\expandafter\ifx\csname natexlab\endcsname\relax\def\natexlab#1{#1}\fi

\bibitem[{{Arnaud}(1996)}]{arnaud1996}
{Arnaud}, K.~A. 1996, in Astronomical Society of the Pacific Conference Series,
  Vol. 101, Astronomical Data Analysis Software and Systems V, ed.
  {G.~H.~Jacoby \& J.~Barnes}, 17

\bibitem[{{Baldwin} {et~al.}(1981){Baldwin}, {Phillips}, \&
  {Terlevich}}]{baldwin1981}
{Baldwin}, A., {Phillips}, M.~M., \& {Terlevich}, R. 1981, \pasp, 93, 817

\bibitem[{{Barth} {et~al.}(1998){Barth}, {Ho}, {Filippenko}, \&
  {Sargent}}]{barth1998}
{Barth}, A.~J., {Ho}, L.~C., {Filippenko}, A.~V., \& {Sargent}, W.~L.~W. 1998,
  \apj, 496, 133

\bibitem[{{Bauer} {et~al.}(2004){Bauer}, {Alexander}, {Brandt}, {Schneider},
  {Treister}, {Hornschemeier}, \& {Garmire}}]{bauer2004}
{Bauer}, F.~E., {Alexander}, D.~M., {Brandt}, W.~N., {Schneider}, D.~P.,
  {Treister}, E., {Hornschemeier}, A.~E., \& {Garmire}, G.~P. 2004, \aj, 128,
  2048

\bibitem[{{Becker} {et~al.}(1995){Becker}, {White}, \& {Helfand}}]{becker1995}
{Becker}, R.~H., {White}, R.~L., \& {Helfand}, D.~J. 1995, \apj, 450, 559

\bibitem[{{Bell}(2003)}]{bell2003a}
{Bell}, E.~F. 2003, \apj, 586, 794

\bibitem[{{Bock} {et~al.}(1999){Bock}, {Large}, \& {Sadler}}]{bock1999}
{Bock}, D.~C.-J., {Large}, M.~I., \& {Sadler}, E.~M. 1999, \aj, 117, 1578

\bibitem[{{Breeveld} {et~al.}(2010){Breeveld}, {Curran}, {Hoversten}, {Koch},
  {Landsman}, {Marshall}, {Page}, {Poole}, {Roming}, {Smith}, {Still},
  {Yershov}, {Blustin}, {Brown}, {Gronwall}, {Holland}, {Kuin}, {McGowan},
  {Rosen}, {Boyd}, {Broos}, {Carter}, {Chester}, {Hancock}, {Huckle}, {Immler},
  {Ivanushkina}, {Kennedy}, {Mason}, {Morgan}, {Oates}, {de Pasquale},
  {Schady}, {Siegel}, \& {vanden Berk}}]{breeveld2010}
{Breeveld}, A.~A., {Curran}, P.~A., {Hoversten}, E.~A., {Koch}, S., {Landsman},
  W., {Marshall}, F.~E., {Page}, M.~J., {Poole}, T.~S., {Roming}, P., {Smith},
  P.~J., {Still}, M., {Yershov}, V., {Blustin}, A.~J., {Brown}, P.~J.,
  {Gronwall}, C., {Holland}, S.~T., {Kuin}, N.~P.~M., {McGowan}, K., {Rosen},
  S., {Boyd}, P., {Broos}, P., {Carter}, M., {Chester}, M.~M., {Hancock}, B.,
  {Huckle}, H., {Immler}, S., {Ivanushkina}, M., {Kennedy}, T., {Mason}, K.~O.,
  {Morgan}, A.~N., {Oates}, S., {de Pasquale}, M., {Schady}, P., {Siegel}, M.,
  \& {vanden Berk}, D. 2010, \mnras, 406, 1687

\bibitem[{{Broos} {et~al.}(2010){Broos}, {Townsley}, {Feigelson}, {Getman},
  {Bauer}, \& {Garmire}}]{broos2010}
{Broos}, P.~S., {Townsley}, L.~K., {Feigelson}, E.~D., {Getman}, K.~V.,
  {Bauer}, F.~E., \& {Garmire}, G.~P. 2010, \apj, 714, 1582

\bibitem[{{Cappelluti} {et~al.}(2007){Cappelluti}, {Hasinger}, {Brusa},
  {Comastri}, {Zamorani}, {B{\"o}hringer}, {Brunner}, {Civano}, {Finoguenov},
  {Fiore}, {Gilli}, {Griffiths}, {Mainieri}, {Matute}, {Miyaji}, \&
  {Silverman}}]{cappelluti2007}
{Cappelluti}, N., {Hasinger}, G., {Brusa}, M., {Comastri}, A., {Zamorani}, G.,
  {B{\"o}hringer}, H., {Brunner}, H., {Civano}, F., {Finoguenov}, A., {Fiore},
  F., {Gilli}, R., {Griffiths}, R.~E., {Mainieri}, V., {Matute}, I., {Miyaji},
  T., \& {Silverman}, J. 2007, \apjs, 172, 341

\bibitem[{{Cardelli} {et~al.}(1989){Cardelli}, {Clayton}, \&
  {Mathis}}]{cardelli1989}
{Cardelli}, J.~A., {Clayton}, G.~C., \& {Mathis}, J.~S. 1989, \apj, 345, 245

\bibitem[{{Cisternas} {et~al.}(2011){Cisternas}, {Jahnke}, {Inskip},
  {Kartaltepe}, {Koekemoer}, {Lisker}, {Robaina}, {Scodeggio}, {Sheth},
  {Trump}, {Andrae}, {Miyaji}, {Lusso}, {Brusa}, {Capak}, {Cappelluti},
  {Civano}, {Ilbert}, {Impey}, {Leauthaud}, {Lilly}, {Salvato}, {Scoville}, \&
  {Taniguchi}}]{cisternas2011}
{Cisternas}, M., {Jahnke}, K., {Inskip}, K.~J., {Kartaltepe}, J., {Koekemoer},
  A.~M., {Lisker}, T., {Robaina}, A.~R., {Scodeggio}, M., {Sheth}, K., {Trump},
  J.~R., {Andrae}, R., {Miyaji}, T., {Lusso}, E., {Brusa}, M., {Capak}, P.,
  {Cappelluti}, N., {Civano}, F., {Ilbert}, O., {Impey}, C.~D., {Leauthaud},
  A., {Lilly}, S.~J., {Salvato}, M., {Scoville}, N.~Z., \& {Taniguchi}, Y.
  2011, \apj, 726, 57

\bibitem[{{Condon} {et~al.}(1998){Condon}, {Cotton}, {Greisen}, {Yin},
  {Perley}, {Taylor}, \& {Broderick}}]{condon1998}
{Condon}, J.~J., {Cotton}, W.~D., {Greisen}, E.~W., {Yin}, Q.~F., {Perley},
  R.~A., {Taylor}, G.~B., \& {Broderick}, J.~J. 1998, \aj, 115, 1693

\bibitem[{{Coziol} {et~al.}(2004){Coziol}, {Brinks}, \&
  {Bravo-Alfaro}}]{coziol2004}
{Coziol}, R., {Brinks}, E., \& {Bravo-Alfaro}, H. 2004, \aj, 128, 68

\bibitem[{{Coziol} {et~al.}(1998{\natexlab{a}}){Coziol}, {de Carvalho},
  {Capelato}, \& {Ribeiro}}]{coziol1998b}
{Coziol}, R., {de Carvalho}, R.~R., {Capelato}, H.~V., \& {Ribeiro}, A.~L.~B.
  1998{\natexlab{a}}, \apj, 506, 545

\bibitem[{{Coziol} {et~al.}(1998{\natexlab{b}}){Coziol}, {Ribeiro}, {de
  Carvalho}, \& {Capelato}}]{coziol1998a}
{Coziol}, R., {Ribeiro}, A.~L.~B., {de Carvalho}, R.~R., \& {Capelato}, H.~V.
  1998{\natexlab{b}}, \apj, 493, 563

\bibitem[{{De Breuck} {et~al.}(2002){De Breuck}, {Tang}, {de Bruyn},
  {R{\"o}ttgering}, \& {van Breugel}}]{debreuck2002}
{De Breuck}, C., {Tang}, Y., {de Bruyn}, A.~G., {R{\"o}ttgering}, H., \& {van
  Breugel}, W. 2002, \aap, 394, 59

\bibitem[{{Deng} {et~al.}(2013){Deng}, {Yu}, \& {Jiang}}]{deng2013}
{Deng}, X.-F., {Yu}, G., \& {Jiang}, P. 2013, PASA, 30, 18

\bibitem[{{Desjardins} {et~al.}(2013){Desjardins}, {Gallagher}, {Tzanavaris},
  {Mulchaey}, {Brandt}, {Charlton}, {Garmire}, {Gronwall}, {Hornschemeier},
  {Johnson}, {Konstantopoulos}, \& {Zabludoff}}]{desjardins2013}
{Desjardins}, T.~D., {Gallagher}, S.~C., {Tzanavaris}, P., {Mulchaey}, J.~S.,
  {Brandt}, W.~N., {Charlton}, J.~C., {Garmire}, G.~P., {Gronwall}, C.,
  {Hornschemeier}, A.~E., {Johnson}, K.~E., {Konstantopoulos}, I.~S., \&
  {Zabludoff}, A.~I. 2013, \apj, 763, 121

\bibitem[{{Dopita} {et~al.}(1996){Dopita}, {Koratkar}, {Evans}, {Allen},
  {Bicknell}, {Sutherland}, {Hawley}, \& {Sadler}}]{dopita1996}
{Dopita}, M.~A., {Koratkar}, A.~P., {Evans}, I.~N., {Allen}, M., {Bicknell},
  G.~V., {Sutherland}, R.~S., {Hawley}, J.~F., \& {Sadler}, E. 1996, in
  Astronomical Society of the Pacific Conference Series, Vol. 103, The Physics
  of Liners in View of Recent Observations, ed. M.~{Eracleous}, A.~{Koratkar},
  C.~{Leitherer}, \& L.~{Ho}, 44

\bibitem[{{Dressler} {et~al.}(1985){Dressler}, {Thompson}, \&
  {Shectman}}]{dressler1985}
{Dressler}, A., {Thompson}, I.~B., \& {Shectman}, S.~A. 1985, \apj, 288, 481

\bibitem[{{Ehlert} {et~al.}(2013){Ehlert}, {Allen}, {Brandt}, {Xue}, {Luo},
  {von der Linden}, {Mantz}, \& {Morris}}]{ehlert2013}
{Ehlert}, S., {Allen}, S.~W., {Brandt}, W.~N., {Xue}, Y.~Q., {Luo}, B., {von
  der Linden}, A., {Mantz}, A., \& {Morris}, R.~G. 2013, \mnras, 428, 3509

\bibitem[{{Eracleous} {et~al.}(2010{\natexlab{a}}){Eracleous}, {Hwang}, \&
  {Flohic}}]{eracleous2010a}
{Eracleous}, M., {Hwang}, J.~A., \& {Flohic}, H.~M.~L.~G. 2010{\natexlab{a}},
  \apj, 711, 796

\bibitem[{{Eracleous} {et~al.}(2010{\natexlab{b}}){Eracleous}, {Hwang}, \&
  {Flohic}}]{eracleous2010}
---. 2010{\natexlab{b}}, \apjs, 187, 135

\bibitem[{{Fedotov} {et~al.}(2011){Fedotov}, {Gallagher}, {Konstantopoulos},
  {Chandar}, {Bastian}, {Charlton}, {Whitmore}, \& {Trancho}}]{fedotov2011}
{Fedotov}, K., {Gallagher}, S.~C., {Konstantopoulos}, I.~S., {Chandar}, R.,
  {Bastian}, N., {Charlton}, J.~C., {Whitmore}, B., \& {Trancho}, G. 2011, \aj,
  142, 42

\bibitem[{{Ferland} \& {Netzer}(1983)}]{ferland1983}
{Ferland}, G.~J., \& {Netzer}, H. 1983, \apj, 264, 105

\bibitem[{{Filho} {et~al.}(2004){Filho}, {Fraternali}, {Markoff}, {Nagar},
  {Barthel}, {Ho}, \& {Yuan}}]{filho2004}
{Filho}, M.~E., {Fraternali}, F., {Markoff}, S., {Nagar}, N.~M., {Barthel},
  P.~D., {Ho}, L.~C., \& {Yuan}, F. 2004, \aap, 418, 429

\bibitem[{{Filippenko} \& {Terlevich}(1992)}]{filippenko1992}
{Filippenko}, A.~V., \& {Terlevich}, R. 1992, \apjl, 397, L79

\bibitem[{{Flohic} {et~al.}(2006){Flohic}, {Eracleous}, {Chartas}, {Shields},
  \& {Moran}}]{flohic2006}
{Flohic}, H.~M.~L.~G., {Eracleous}, M., {Chartas}, G., {Shields}, J.~C., \&
  {Moran}, E.~C. 2006, \apj, 647, 140

\bibitem[{{Freeman} {et~al.}(2002){Freeman}, {Kashyap}, {Rosner}, \&
  {Lamb}}]{freeman2002}
{Freeman}, P.~E., {Kashyap}, V., {Rosner}, R., \& {Lamb}, D.~Q. 2002, \apjs,
  138, 185

\bibitem[{{Fukugita} {et~al.}(1995){Fukugita}, {Shimasaku}, \&
  {Ichikawa}}]{fukugita1995}
{Fukugita}, M., {Shimasaku}, K., \& {Ichikawa}, T. 1995, \pasp, 107, 945

\bibitem[{{Gallagher} {et~al.}(2010){Gallagher}, {Durrell}, {Elmegreen},
  {Chandar}, {English}, {Charlton}, {Gronwall}, {Young}, {Tzanavaris},
  {Johnson}, {Mendes de Oliveira}, {Whitmore}, {Hornschemeier}, {Maybhate}, \&
  {Zabludoff}}]{gallagher2010}
{Gallagher}, S.~C., {Durrell}, P.~R., {Elmegreen}, D.~M., {Chandar}, R.,
  {English}, J., {Charlton}, J.~C., {Gronwall}, C., {Young}, J., {Tzanavaris},
  P., {Johnson}, K.~E., {Mendes de Oliveira}, C., {Whitmore}, B.,
  {Hornschemeier}, A.~E., {Maybhate}, A., \& {Zabludoff}, A. 2010, \aj, 139,
  545

\bibitem[{{Gallagher} {et~al.}(2008){Gallagher}, {Johnson}, {Hornschemeier},
  {Charlton}, \& {Hibbard}}]{gallagher2008}
{Gallagher}, S.~C., {Johnson}, K.~E., {Hornschemeier}, A.~E., {Charlton},
  J.~C., \& {Hibbard}, J.~E. 2008, \apj, 673, 730

\bibitem[{{Gallagher} {et~al.}(2005){Gallagher}, {Richards}, {Hall}, {Brandt},
  {Schneider}, \& {Vanden Berk}}]{gallagher2005}
{Gallagher}, S.~C., {Richards}, G.~T., {Hall}, P.~B., {Brandt}, W.~N.,
  {Schneider}, D.~P., \& {Vanden Berk}, D.~E. 2005, \aj, 129, 567

\bibitem[{{Gehrels}(1986)}]{gehrels1986}
{Gehrels}, N. 1986, \apj, 303, 336

\bibitem[{{Gehrels} {et~al.}(2004){Gehrels}, {Chincarini}, {Giommi}, {Mason},
  {Nousek}, {Wells}, {White}, {Barthelmy}, {Burrows}, {Cominsky}, {Hurley},
  {Marshall}, {M{\'e}sz{\'a}ros}, {Roming}, {Angelini}, {Barbier}, {Belloni},
  {Campana}, {Caraveo}, {Chester}, {Citterio}, {Cline}, {Cropper}, {Cummings},
  {Dean}, {Feigelson}, {Fenimore}, {Frail}, {Fruchter}, {Garmire}, {Gendreau},
  {Ghisellini}, {Greiner}, {Hill}, {Hunsberger}, {Krimm}, {Kulkarni}, {Kumar},
  {Lebrun}, {Lloyd-Ronning}, {Markwardt}, {Mattson}, {Mushotzky}, {Norris},
  {Osborne}, {Paczynski}, {Palmer}, {Park}, {Parsons}, {Paul}, {Rees},
  {Reynolds}, {Rhoads}, {Sasseen}, {Schaefer}, {Short}, {Smale}, {Smith},
  {Stella}, {Tagliaferri}, {Takahashi}, {Tashiro}, {Townsley}, {Tueller},
  {Turner}, {Vietri}, {Voges}, {Ward}, {Willingale}, {Zerbi}, \&
  {Zhang}}]{2004ApJ...611.1005G}
{Gehrels}, N., {Chincarini}, G., {Giommi}, P., {Mason}, K.~O., {Nousek}, J.~A.,
  {Wells}, A.~A., {White}, N.~E., {Barthelmy}, S.~D., {Burrows}, D.~N.,
  {Cominsky}, L.~R., {Hurley}, K.~C., {Marshall}, F.~E., {M{\'e}sz{\'a}ros},
  P., {Roming}, P.~W.~A., {Angelini}, L., {Barbier}, L.~M., {Belloni}, T.,
  {Campana}, S., {Caraveo}, P.~A., {Chester}, M.~M., {Citterio}, O., {Cline},
  T.~L., {Cropper}, M.~S., {Cummings}, J.~R., {Dean}, A.~J., {Feigelson},
  E.~D., {Fenimore}, E.~E., {Frail}, D.~A., {Fruchter}, A.~S., {Garmire},
  G.~P., {Gendreau}, K., {Ghisellini}, G., {Greiner}, J., {Hill}, J.~E.,
  {Hunsberger}, S.~D., {Krimm}, H.~A., {Kulkarni}, S.~R., {Kumar}, P.,
  {Lebrun}, F., {Lloyd-Ronning}, N.~M., {Markwardt}, C.~B., {Mattson}, B.~J.,
  {Mushotzky}, R.~F., {Norris}, J.~P., {Osborne}, J., {Paczynski}, B.,
  {Palmer}, D.~M., {Park}, H.-S., {Parsons}, A.~M., {Paul}, J., {Rees}, M.~J.,
  {Reynolds}, C.~S., {Rhoads}, J.~E., {Sasseen}, T.~P., {Schaefer}, B.~E.,
  {Short}, A.~T., {Smale}, A.~P., {Smith}, I.~A., {Stella}, L., {Tagliaferri},
  G., {Takahashi}, T., {Tashiro}, M., {Townsley}, L.~K., {Tueller}, J.,
  {Turner}, M.~J.~L., {Vietri}, M., {Voges}, W., {Ward}, M.~J., {Willingale},
  R., {Zerbi}, F.~M., \& {Zhang}, W.~W. 2004, \apj, 611, 1005

\bibitem[{{Georgakakis} {et~al.}(2009){Georgakakis}, {Coil}, {Laird},
  {Griffith}, {Nandra}, {Lotz}, {Pierce}, {Cooper}, {Newman}, \&
  {Koekemoer}}]{georgakakis2009}
{Georgakakis}, A., {Coil}, A.~L., {Laird}, E.~S., {Griffith}, R.~L., {Nandra},
  K., {Lotz}, J.~M., {Pierce}, C.~M., {Cooper}, M.~C., {Newman}, J.~A., \&
  {Koekemoer}, A.~M. 2009, \mnras, 397, 623

\bibitem[{{Gonz{\'a}lez-Mart{\'{\i}}n}
  {et~al.}(2009){Gonz{\'a}lez-Mart{\'{\i}}n}, {Masegosa}, {M{\'a}rquez},
  {Guainazzi}, \& {Jim{\'e}nez-Bail{\'o}n}}]{gonzalez-martin2009}
{Gonz{\'a}lez-Mart{\'{\i}}n}, O., {Masegosa}, J., {M{\'a}rquez}, I.,
  {Guainazzi}, M., \& {Jim{\'e}nez-Bail{\'o}n}, E. 2009, \aap, 506, 1107

\bibitem[{{Grogin} {et~al.}(2005){Grogin}, {Conselice}, {Chatzichristou},
  {Alexander}, {Bauer}, {Hornschemeier}, {Jogee}, {Koekemoer}, {Laidler},
  {Livio}, {Lucas}, {Paolillo}, {Ravindranath}, {Schreier}, {Simmons}, \&
  {Urry}}]{grogin2005}
{Grogin}, N.~A., {Conselice}, C.~J., {Chatzichristou}, E., {Alexander}, D.~M.,
  {Bauer}, F.~E., {Hornschemeier}, A.~E., {Jogee}, S., {Koekemoer}, A.~M.,
  {Laidler}, V.~G., {Livio}, M., {Lucas}, R.~A., {Paolillo}, M.,
  {Ravindranath}, S., {Schreier}, E.~J., {Simmons}, B.~D., \& {Urry}, C.~M.
  2005, \apjl, 627, L97

\bibitem[{{Grupe} {et~al.}(2010){Grupe}, {Komossa}, {Leighly}, \&
  {Page}}]{grupe2010}
{Grupe}, D., {Komossa}, S., {Leighly}, K.~M., \& {Page}, K.~L. 2010, \apjs,
  187, 64

\bibitem[{{Halpern} \& {Steiner}(1983)}]{halpern1983}
{Halpern}, J.~P., \& {Steiner}, J.~E. 1983, \apjl, 269, L37

\bibitem[{{Heckman}(1980)}]{heckman1980}
{Heckman}, T.~M. 1980, \aap, 87, 152

\bibitem[{{Hickox} \& {Markevitch}(2006)}]{hickox2006}
{Hickox}, R.~C., \& {Markevitch}, M. 2006, \apj, 645, 95

\bibitem[{{Hickson}(1982)}]{hickson1982}
{Hickson}, P. 1982, \apj, 259, 930

\bibitem[{{Hickson} {et~al.}(1989){Hickson}, {Kindl}, \& {Auman}}]{hickson1989}
{Hickson}, P., {Kindl}, E., \& {Auman}, J.~R. 1989, \apjs, 70, 687

\bibitem[{{Hickson} {et~al.}(1992){Hickson}, {Mendes de Oliveira}, {Huchra}, \&
  {Palumbo}}]{hickson1992}
{Hickson}, P., {Mendes de Oliveira}, C., {Huchra}, J.~P., \& {Palumbo}, G.~G.
  1992, \apj, 399, 353

\bibitem[{{Ho} {et~al.}(2001){Ho}, {Feigelson}, {Townsley}, {Sambruna},
  {Garmire}, {Brandt}, {Filippenko}, {Griffiths}, {Ptak}, \&
  {Sargent}}]{ho2001}
{Ho}, L.~C., {Feigelson}, E.~D., {Townsley}, L.~K., {Sambruna}, R.~M.,
  {Garmire}, G.~P., {Brandt}, W.~N., {Filippenko}, A.~V., {Griffiths}, R.~E.,
  {Ptak}, A.~F., \& {Sargent}, W.~L.~W. 2001, \apjl, 549, L51

\bibitem[{{Ho} {et~al.}(1997){Ho}, {Filippenko}, \& {Sargent}}]{ho1997}
{Ho}, L.~C., {Filippenko}, A.~V., \& {Sargent}, W.~L.~W. 1997, \apj, 487, 568

\bibitem[{{Hopkins} \& {Quataert}(2010)}]{hopkins2010}
{Hopkins}, P.~F., \& {Quataert}, E. 2010, \mnras, 407, 1529

\bibitem[{{Jeltema} {et~al.}(2008){Jeltema}, {Binder}, \&
  {Mulchaey}}]{jeltema2008}
{Jeltema}, T.~E., {Binder}, B., \& {Mulchaey}, J.~S. 2008, \apj, 679, 1162

\bibitem[{{Johnson} {et~al.}(2007){Johnson}, {Hibbard}, {Gallagher},
  {Charlton}, {Hornschemeier}, {Jarrett}, \& {Reines}}]{johnson2007}
{Johnson}, K.~E., {Hibbard}, J.~E., {Gallagher}, S.~C., {Charlton}, J.~C.,
  {Hornschemeier}, A.~E., {Jarrett}, T.~H., \& {Reines}, A.~E. 2007, \aj, 134,
  1522

\bibitem[{{Just} {et~al.}(2007){Just}, {Brandt}, {Shemmer}, {Steffen},
  {Schneider}, {Chartas}, \& {Garmire}}]{just2007}
{Just}, D.~W., {Brandt}, W.~N., {Shemmer}, O., {Steffen}, A.~T., {Schneider},
  D.~P., {Chartas}, G., \& {Garmire}, G.~P. 2007, \apj, 665, 1004

\bibitem[{{Kartaltepe} {et~al.}(2010){Kartaltepe}, {Sanders}, {Le Floc'h},
  {Frayer}, {Aussel}, {Arnouts}, {Ilbert}, {Salvato}, {Scoville}, {Surace},
  {Yan}, {Capak}, {Caputi}, {Carollo}, {Cassata}, {Civano}, {Hasinger},
  {Koekemoer}, {Le F{\`e}vre}, {Lilly}, {Liu}, {McCracken}, {Schinnerer},
  {Smol{\v c}i{\'c}}, {Taniguchi}, {Thompson}, {Trump}, {Baldassare}, \&
  {Fiorenza}}]{kartaltepe2010}
{Kartaltepe}, J.~S., {Sanders}, D.~B., {Le Floc'h}, E., {Frayer}, D.~T.,
  {Aussel}, H., {Arnouts}, S., {Ilbert}, O., {Salvato}, M., {Scoville}, N.~Z.,
  {Surace}, J., {Yan}, L., {Capak}, P., {Caputi}, K., {Carollo}, C.~M.,
  {Cassata}, P., {Civano}, F., {Hasinger}, G., {Koekemoer}, A.~M., {Le
  F{\`e}vre}, O., {Lilly}, S., {Liu}, C.~T., {McCracken}, H.~J., {Schinnerer},
  E., {Smol{\v c}i{\'c}}, V., {Taniguchi}, Y., {Thompson}, D.~J., {Trump}, J.,
  {Baldassare}, V.~F., \& {Fiorenza}, S.~L. 2010, \apj, 721, 98

\bibitem[{{Kashyap} {et~al.}(2010){Kashyap}, {van Dyk}, {Connors}, {Freeman},
  {Siemiginowska}, {Xu}, \& {Zezas}}]{kashyap2010}
{Kashyap}, V.~L., {van Dyk}, D.~A., {Connors}, A., {Freeman}, P.~E.,
  {Siemiginowska}, A., {Xu}, J., \& {Zezas}, A. 2010, \apj, 719, 900

\bibitem[{{Kauffmann} {et~al.}(2003){Kauffmann}, {Heckman}, {Tremonti},
  {Brinchmann}, {Charlot}, {White}, {Ridgway}, {Brinkmann}, {Fukugita}, {Hall},
  {Ivezi{\'c}}, {Richards}, \& {Schneider}}]{kauffmann2003}
{Kauffmann}, G., {Heckman}, T.~M., {Tremonti}, C., {Brinchmann}, J., {Charlot},
  S., {White}, S.~D.~M., {Ridgway}, S.~E., {Brinkmann}, J., {Fukugita}, M.,
  {Hall}, P.~B., {Ivezi{\'c}}, {\v Z}., {Richards}, G.~T., \& {Schneider},
  D.~P. 2003, \mnras, 346, 1055

\bibitem[{{Kennicutt}(1998)}]{kennicutt1998}
{Kennicutt}, R.~C. 1998, \araa, 36, 189

\bibitem[{{Kewley} {et~al.}(2001){Kewley}, {Dopita}, {Sutherland}, {Heisler},
  \& {Trevena}}]{kewley2001}
{Kewley}, L.~J., {Dopita}, M.~A., {Sutherland}, R.~S., {Heisler}, C.~A., \&
  {Trevena}, J. 2001, \apj, 556, 121

\bibitem[{{Kewley} {et~al.}(2006){Kewley}, {Groves}, {Kauffmann}, \&
  {Heckman}}]{kewley2006}
{Kewley}, L.~J., {Groves}, B., {Kauffmann}, G., \& {Heckman}, T. 2006, \mnras,
  372, 961

\bibitem[{{Kim} {et~al.}(2004){Kim}, {Cameron}, {Drake}, {Evans}, {Freeman},
  {Gaetz}, {Ghosh}, {Green}, {Harnden}, {Karovska}, {Kashyap}, {Maksym},
  {Ratzlaff}, {Schlegel}, {Silverman}, {Tananbaum}, {Vikhlinin}, {Wilkes}, \&
  {Grimes}}]{kim2004}
{Kim}, D.-W., {Cameron}, R.~A., {Drake}, J.~J., {Evans}, N.~R., {Freeman}, P.,
  {Gaetz}, T.~J., {Ghosh}, H., {Green}, P.~J., {Harnden}, Jr., F.~R.,
  {Karovska}, M., {Kashyap}, V., {Maksym}, P.~W., {Ratzlaff}, P.~W.,
  {Schlegel}, E.~M., {Silverman}, J.~D., {Tananbaum}, H.~D., {Vikhlinin},
  A.~A., {Wilkes}, B.~J., \& {Grimes}, J.~P. 2004, \apjs, 150, 19

\bibitem[{{Konstantopoulos} {et~al.}(2010){Konstantopoulos}, {Gallagher},
  {Fedotov}, {Durrell}, {Heiderman}, {Elmegreen}, {Charlton}, {Hibbard},
  {Tzanavaris}, {Chandar}, {Johnson}, {Maybhate}, {Zabludoff}, {Gronwall},
  {Szathmary}, {Hornschemeier}, {English}, {Whitmore}, {Mendes de Oliveira}, \&
  {Mulchaey}}]{konstantopoulos2010}
{Konstantopoulos}, I.~S., {Gallagher}, S.~C., {Fedotov}, K., {Durrell}, P.~R.,
  {Heiderman}, A., {Elmegreen}, D.~M., {Charlton}, J.~C., {Hibbard}, J.~E.,
  {Tzanavaris}, P., {Chandar}, R., {Johnson}, K.~E., {Maybhate}, A.,
  {Zabludoff}, A.~E., {Gronwall}, C., {Szathmary}, D., {Hornschemeier}, A.~E.,
  {English}, J., {Whitmore}, B., {Mendes de Oliveira}, C., \& {Mulchaey}, J.~S.
  2010, \apj, 723, 197

\bibitem[{{Kraft} {et~al.}(1991){Kraft}, {Burrows}, \& {Nousek}}]{kraft1991}
{Kraft}, R.~P., {Burrows}, D.~N., \& {Nousek}, J.~A. 1991, \apj, 374, 344

\bibitem[{{LaMassa} {et~al.}(2011){LaMassa}, {Heckman}, {Ptak}, {Martins},
  {Wild}, {Sonnentrucker}, \& {Hornschemeier}}]{lamassa2011}
{LaMassa}, S.~M., {Heckman}, T.~M., {Ptak}, A., {Martins}, L., {Wild}, V.,
  {Sonnentrucker}, P., \& {Hornschemeier}, A. 2011, \apj, 729, 52

\bibitem[{{Lee} {et~al.}(2004){Lee}, {Allam}, {Tucker}, {Annis}, {Johnston},
  {Scranton}, {Acebo}, {Bahcall}, {Bartelmann}, {B{\"o}hringer}, {Ellman},
  {Grebel}, {Infante}, {Loveday}, {McKay}, {Prada}, {Schneider}, {Stoughton},
  {Szalay}, {Vogeley}, {Voges}, \& {Yanny}}]{lee2004}
{Lee}, B.~C., {Allam}, S.~S., {Tucker}, D.~L., {Annis}, J., {Johnston}, D.~E.,
  {Scranton}, R., {Acebo}, Y., {Bahcall}, N.~A., {Bartelmann}, M.,
  {B{\"o}hringer}, H., {Ellman}, N., {Grebel}, E.~K., {Infante}, L., {Loveday},
  J., {McKay}, T.~A., {Prada}, F., {Schneider}, D.~P., {Stoughton}, C.,
  {Szalay}, A.~S., {Vogeley}, M.~S., {Voges}, W., \& {Yanny}, B. 2004, \aj,
  127, 1811

\bibitem[{{Lehmer} {et~al.}(2010){Lehmer}, {Alexander}, {Bauer}, {Brandt},
  {Goulding}, {Jenkins}, {Ptak}, \& {Roberts}}]{lehmer2010}
{Lehmer}, B.~D., {Alexander}, D.~M., {Bauer}, F.~E., {Brandt}, W.~N.,
  {Goulding}, A.~D., {Jenkins}, L.~P., {Ptak}, A., \& {Roberts}, T.~P. 2010,
  \apj, 724, 559

\bibitem[{{Lusso} {et~al.}(2010){Lusso}, {Comastri}, {Vignali}, {Zamorani},
  {Brusa}, {Gilli}, {Iwasawa}, {Salvato}, {Civano}, {Elvis}, {Merloni},
  {Bongiorno}, {Trump}, {Koekemoer}, {Schinnerer}, {Le Floc'h}, {Cappelluti},
  {Jahnke}, {Sargent}, {Silverman}, {Mainieri}, {Fiore}, {Bolzonella}, {Le
  F{\`e}vre}, {Garilli}, {Iovino}, {Kneib}, {Lamareille}, {Lilly}, {Mignoli},
  {Scodeggio}, \& {Vergani}}]{lusso2010}
{Lusso}, E., {Comastri}, A., {Vignali}, C., {Zamorani}, G., {Brusa}, M.,
  {Gilli}, R., {Iwasawa}, K., {Salvato}, M., {Civano}, F., {Elvis}, M.,
  {Merloni}, A., {Bongiorno}, A., {Trump}, J.~R., {Koekemoer}, A.~M.,
  {Schinnerer}, E., {Le Floc'h}, E., {Cappelluti}, N., {Jahnke}, K., {Sargent},
  M., {Silverman}, J., {Mainieri}, V., {Fiore}, F., {Bolzonella}, M., {Le
  F{\`e}vre}, O., {Garilli}, B., {Iovino}, A., {Kneib}, J.~P., {Lamareille},
  F., {Lilly}, S., {Mignoli}, M., {Scodeggio}, M., \& {Vergani}, D. 2010, \aap,
  512, A34

\bibitem[{{Ma} \& {Feissel}(1998)}]{ma1998}
{Ma}, C., \& {Feissel}, M. 1998, VizieR Online Data Catalog, 1251, 0

\bibitem[{{Mamon}(1994)}]{mamon1994}
{Mamon}, G.~A. 1994, in Clusters of Galaxies, ed. F.~{Durret}, A.~{Mazure}, \&
  J.~{Tran Thanh Van}, 291

\bibitem[{{Maoz} {et~al.}(2005){Maoz}, {Nagar}, {Falcke}, \&
  {Wilson}}]{maoz2005}
{Maoz}, D., {Nagar}, N.~M., {Falcke}, H., \& {Wilson}, A.~S. 2005, \apj, 625,
  699

\bibitem[{{Mart{\'{\i}}nez} {et~al.}(2010){Mart{\'{\i}}nez}, {Del Olmo},
  {Coziol}, \& {Perea}}]{martinez2010}
{Mart{\'{\i}}nez}, M.~A., {Del Olmo}, A., {Coziol}, R., \& {Perea}, J. 2010,
  \aj, 139, 1199

\bibitem[{{Martini} {et~al.}(2006){Martini}, {Kelson}, {Kim}, {Mulchaey}, \&
  {Athey}}]{martini2006}
{Martini}, P., {Kelson}, D.~D., {Kim}, E., {Mulchaey}, J.~S., \& {Athey}, A.~A.
  2006, \apj, 644, 116

\bibitem[{{Mihos} \& {Hernquist}(1996)}]{mihos1996}
{Mihos}, J.~C., \& {Hernquist}, L. 1996, \apj, 464, 641

\bibitem[{{Mukai}(1993)}]{mukai1993}
{Mukai}, K. 1993, Legacy, vol.~3, p.21-31, 3, 21

\bibitem[{{Mulchaey} {et~al.}(2003){Mulchaey}, {Davis}, {Mushotzky}, \&
  {Burstein}}]{mulchaey2003}
{Mulchaey}, J.~S., {Davis}, D.~S., {Mushotzky}, R.~F., \& {Burstein}, D. 2003,
  \apjs, 145, 39

\bibitem[{{Nagar} {et~al.}(2005){Nagar}, {Falcke}, \& {Wilson}}]{nagar2005}
{Nagar}, N.~M., {Falcke}, H., \& {Wilson}, A.~S. 2005, \aap, 435, 521

\bibitem[{{O'Connell}(1999)}]{oconnell1999}
{O'Connell}, R.~W. 1999, \araa, 37, 603

\bibitem[{{O'Sullivan} {et~al.}(2009){O'Sullivan}, {Giacintucci}, {Vrtilek},
  {Raychaudhury}, \& {David}}]{osullivan2009}
{O'Sullivan}, E., {Giacintucci}, S., {Vrtilek}, J.~M., {Raychaudhury}, S., \&
  {David}, L.~P. 2009, \apj, 701, 1560

\bibitem[{{Peng} {et~al.}(2010){Peng}, {Ho}, {Impey}, \& {Rix}}]{peng2010}
{Peng}, C.~Y., {Ho}, L.~C., {Impey}, C.~D., \& {Rix}, H.-W. 2010, \aj, 139,
  2097

\bibitem[{{Ponman} {et~al.}(1996){Ponman}, {Bourner}, {Ebeling}, \&
  {B{\"o}hringer}}]{ponman1996}
{Ponman}, T.~J., {Bourner}, P.~D.~J., {Ebeling}, H., \& {B{\"o}hringer}, H.
  1996, \mnras, 283, 690

\bibitem[{{Poole} {et~al.}(2008){Poole}, {Breeveld}, {Page}, {Landsman},
  {Holland}, {Roming}, {Kuin}, {Brown}, {Gronwall}, {Hunsberger}, {Koch},
  {Mason}, {Schady}, {vanden Berk}, {Blustin}, {Boyd}, {Broos}, {Carter},
  {Chester}, {Cucchiara}, {Hancock}, {Huckle}, {Immler}, {Ivanushkina},
  {Kennedy}, {Marshall}, {Morgan}, {Pandey}, {de Pasquale}, {Smith}, \&
  {Still}}]{poole2008}
{Poole}, T.~S., {Breeveld}, A.~A., {Page}, M.~J., {Landsman}, W., {Holland},
  S.~T., {Roming}, P., {Kuin}, N.~P.~M., {Brown}, P.~J., {Gronwall}, C.,
  {Hunsberger}, S., {Koch}, S., {Mason}, K.~O., {Schady}, P., {vanden Berk},
  D., {Blustin}, A.~J., {Boyd}, P., {Broos}, P., {Carter}, M., {Chester},
  M.~M., {Cucchiara}, A., {Hancock}, B., {Huckle}, H., {Immler}, S.,
  {Ivanushkina}, M., {Kennedy}, T., {Marshall}, F., {Morgan}, A., {Pandey},
  S.~B., {de Pasquale}, M., {Smith}, P.~J., \& {Still}, M. 2008, \mnras, 383,
  627

\bibitem[{{Rafferty} {et~al.}(2013){Rafferty}, {B{\^i}rzan}, {Nulsen},
  {McNamara}, {Brandt}, {Wise}, \& {R{\"o}ttgering}}]{rafferty2013}
{Rafferty}, D.~A., {B{\^i}rzan}, L., {Nulsen}, P.~E.~J., {McNamara}, B.~R.,
  {Brandt}, W.~N., {Wise}, M.~W., \& {R{\"o}ttgering}, H.~J.~A. 2013, \mnras,
  428, 58

\bibitem[{{Ranalli} {et~al.}(2003){Ranalli}, {Comastri}, \&
  {Setti}}]{ranalli2003}
{Ranalli}, P., {Comastri}, A., \& {Setti}, G. 2003, \aap, 399, 39

\bibitem[{{Ranalli} {et~al.}(2012){Ranalli}, {Comastri}, {Zamorani},
  {Cappelluti}, {Civano}, {Georgantopoulos}, {Gilli}, {Schinnerer}, {Smol{\v
  c}i{\'c}}, \& {Vignali}}]{ranalli2012}
{Ranalli}, P., {Comastri}, A., {Zamorani}, G., {Cappelluti}, N., {Civano}, F.,
  {Georgantopoulos}, I., {Gilli}, R., {Schinnerer}, E., {Smol{\v c}i{\'c}}, V.,
  \& {Vignali}, C. 2012, \aap, 542, A16

\bibitem[{{Rasmussen} {et~al.}(2008){Rasmussen}, {Ponman}, {Verdes-Montenegro},
  {Yun}, \& {Borthakur}}]{rasmussen2008}
{Rasmussen}, J., {Ponman}, T.~J., {Verdes-Montenegro}, L., {Yun}, M.~S., \&
  {Borthakur}, S. 2008, \mnras, 388, 1245

\bibitem[{{Reynolds} \& {Nowak}(2003)}]{reynolds2003}
{Reynolds}, C.~S., \& {Nowak}, M.~A. 2003, \physrep, 377, 389

\bibitem[{{Ribeiro} {et~al.}(1998){Ribeiro}, {de Carvalho}, {Capelato}, \&
  {Zepf}}]{ribeiro1998}
{Ribeiro}, A.~L.~B., {de Carvalho}, R.~R., {Capelato}, H.~V., \& {Zepf}, S.~E.
  1998, \apj, 497, 72

\bibitem[{{Roche} {et~al.}(1991){Roche}, {Aitken}, {Smith}, \&
  {Ward}}]{roche1991}
{Roche}, P.~F., {Aitken}, D.~K., {Smith}, C.~H., \& {Ward}, M.~J. 1991, \mnras,
  248, 606

\bibitem[{{Roming} {et~al.}(2005){Roming}, {Kennedy}, {Mason}, {Nousek}, {Ahr},
  {Bingham}, {Broos}, {Carter}, {Hancock}, {Huckle}, {Hunsberger}, {Kawakami},
  {Killough}, {Koch}, {McLelland}, {Smith}, {Smith}, {Soto}, {Boyd},
  {Breeveld}, {Holland}, {Ivanushkina}, {Pryzby}, {Still}, \&
  {Stock}}]{2005SSRv..120...95R}
{Roming}, P.~W.~A., {Kennedy}, T.~E., {Mason}, K.~O., {Nousek}, J.~A., {Ahr},
  L., {Bingham}, R.~E., {Broos}, P.~S., {Carter}, M.~J., {Hancock}, B.~K.,
  {Huckle}, H.~E., {Hunsberger}, S.~D., {Kawakami}, H., {Killough}, R., {Koch},
  T.~S., {McLelland}, M.~K., {Smith}, K., {Smith}, P.~J., {Soto}, J.~C.,
  {Boyd}, P.~T., {Breeveld}, A.~A., {Holland}, S.~T., {Ivanushkina}, M.,
  {Pryzby}, M.~S., {Still}, M.~D., \& {Stock}, J. 2005, Space Science Reviews,
  120, 95

\bibitem[{{Rossa} {et~al.}(2006){Rossa}, {van der Marel}, {B{\"o}ker},
  {Gerssen}, {Ho}, {Rix}, {Shields}, \& {Walcher}}]{rossa2006}
{Rossa}, J., {van der Marel}, R.~P., {B{\"o}ker}, T., {Gerssen}, J., {Ho},
  L.~C., {Rix}, H.-W., {Shields}, J.~C., \& {Walcher}, C.-J. 2006, \aj, 132,
  1074

\bibitem[{{Schlegel} {et~al.}(1998){Schlegel}, {Finkbeiner}, \&
  {Davis}}]{schlegel1998}
{Schlegel}, D.~J., {Finkbeiner}, D.~P., \& {Davis}, M. 1998, \apj, 500, 525

\bibitem[{{Shen} {et~al.}(2007){Shen}, {Mulchaey}, {Raychaudhury}, {Rasmussen},
  \& {Ponman}}]{shen2007}
{Shen}, Y., {Mulchaey}, J.~S., {Raychaudhury}, S., {Rasmussen}, J., \&
  {Ponman}, T.~J. 2007, \apjl, 654, L115

\bibitem[{{Shields}(1992)}]{shields1992}
{Shields}, J.~C. 1992, \apjl, 399, L27

\bibitem[{{Silverman} {et~al.}(2011){Silverman}, {Kampczyk}, {Jahnke},
  {Andrae}, {Lilly}, {Elvis}, {Civano}, {Mainieri}, {Vignali}, {Zamorani},
  {Nair}, {Le F{\`e}vre}, {de Ravel}, {Bardelli}, {Bongiorno}, {Bolzonella},
  {Cappi}, {Caputi}, {Carollo}, {Contini}, {Coppa}, {Cucciati}, {de la Torre},
  {Franzetti}, {Garilli}, {Halliday}, {Hasinger}, {Iovino}, {Knobel},
  {Koekemoer}, {Kova{\v c}}, {Lamareille}, {Le Borgne}, {Le Brun}, {Maier},
  {Mignoli}, {Pello}, {P{\'e}rez-Montero}, {Ricciardelli}, {Peng}, {Scodeggio},
  {Tanaka}, {Tasca}, {Tresse}, {Vergani}, {Zucca}, {Brusa}, {Cappelluti},
  {Comastri}, {Finoguenov}, {Fu}, {Gilli}, {Hao}, {Ho}, \&
  {Salvato}}]{silverman2011}
{Silverman}, J.~D., {Kampczyk}, P., {Jahnke}, K., {Andrae}, R., {Lilly}, S.~J.,
  {Elvis}, M., {Civano}, F., {Mainieri}, V., {Vignali}, C., {Zamorani}, G.,
  {Nair}, P., {Le F{\`e}vre}, O., {de Ravel}, L., {Bardelli}, S., {Bongiorno},
  A., {Bolzonella}, M., {Cappi}, A., {Caputi}, K., {Carollo}, C.~M., {Contini},
  T., {Coppa}, G., {Cucciati}, O., {de la Torre}, S., {Franzetti}, P.,
  {Garilli}, B., {Halliday}, C., {Hasinger}, G., {Iovino}, A., {Knobel}, C.,
  {Koekemoer}, A.~M., {Kova{\v c}}, K., {Lamareille}, F., {Le Borgne}, J.-F.,
  {Le Brun}, V., {Maier}, C., {Mignoli}, M., {Pello}, R., {P{\'e}rez-Montero},
  E., {Ricciardelli}, E., {Peng}, Y., {Scodeggio}, M., {Tanaka}, M., {Tasca},
  L., {Tresse}, L., {Vergani}, D., {Zucca}, E., {Brusa}, M., {Cappelluti}, N.,
  {Comastri}, A., {Finoguenov}, A., {Fu}, H., {Gilli}, R., {Hao}, H., {Ho},
  L.~C., \& {Salvato}, M. 2011, \apj, 743, 2

\bibitem[{{Smith} {et~al.}(2012){Smith}, {Swartz}, {Miller}, {Burleson},
  {Nowak}, \& {Struck}}]{smith2012}
{Smith}, B.~J., {Swartz}, D.~A., {Miller}, O., {Burleson}, J.~A., {Nowak},
  M.~A., \& {Struck}, C. 2012, \aj, 143, 144

\bibitem[{{Stasi{\'n}ska} {et~al.}(2006){Stasi{\'n}ska}, {Cid Fernandes},
  {Mateus}, {Sodr{\'e}}, \& {Asari}}]{stasinska2006}
{Stasi{\'n}ska}, G., {Cid Fernandes}, R., {Mateus}, A., {Sodr{\'e}}, L., \&
  {Asari}, N.~V. 2006, \mnras, 371, 972

\bibitem[{{Steffen} {et~al.}(2007){Steffen}, {Brandt}, {Alexander},
  {Gallagher}, \& {Lehmer}}]{steffen2007}
{Steffen}, A.~T., {Brandt}, W.~N., {Alexander}, D.~M., {Gallagher}, S.~C., \&
  {Lehmer}, B.~D. 2007, \apjl, 667, L25

\bibitem[{{Steffen} {et~al.}(2006){Steffen}, {Strateva}, {Brandt}, {Alexander},
  {Koekemoer}, {Lehmer}, {Schneider}, \& {Vignali}}]{steffen2006}
{Steffen}, A.~T., {Strateva}, I., {Brandt}, W.~N., {Alexander}, D.~M.,
  {Koekemoer}, A.~M., {Lehmer}, B.~D., {Schneider}, D.~P., \& {Vignali}, C.
  2006, \aj, 131, 2826

\bibitem[{{Strateva} {et~al.}(2005){Strateva}, {Brandt}, {Schneider}, {Vanden
  Berk}, \& {Vignali}}]{strateva2005}
{Strateva}, I.~V., {Brandt}, W.~N., {Schneider}, D.~P., {Vanden Berk}, D.~G.,
  \& {Vignali}, C. 2005, \aj, 130, 387

\bibitem[{{Tananbaum} {et~al.}(1979){Tananbaum}, {Avni}, {Branduardi}, {Elvis},
  {Fabbiano}, {Feigelson}, {Giacconi}, {Henry}, {Pye}, {Soltan}, \&
  {Zamorani}}]{tananbaum1979}
{Tananbaum}, H., {Avni}, Y., {Branduardi}, G., {Elvis}, M., {Fabbiano}, G.,
  {Feigelson}, E., {Giacconi}, R., {Henry}, J.~P., {Pye}, J.~P., {Soltan}, A.,
  \& {Zamorani}, G. 1979, \apjl, 234, L9

\bibitem[{{Terlevich} \& {Melnick}(1985)}]{terlevich1985}
{Terlevich}, R., \& {Melnick}, J. 1985, \mnras, 213, 841

\bibitem[{{Tzanavaris} {et~al.}(2010){Tzanavaris}, {Hornschemeier},
  {Gallagher}, {Johnson}, {Gronwall}, {Immler}, {Reines}, {Hoversten}, \&
  {Charlton}}]{tzanavaris2010}
{Tzanavaris}, P., {Hornschemeier}, A.~E., {Gallagher}, S.~C., {Johnson}, K.~E.,
  {Gronwall}, C., {Immler}, S., {Reines}, A.~E., {Hoversten}, E., \&
  {Charlton}, J.~C. 2010, \apj, 716, 556

\bibitem[{{Veilleux} \& {Osterbrock}(1987)}]{veilleux1987}
{Veilleux}, S., \& {Osterbrock}, D.~E. 1987, \apjs, 63, 295

\bibitem[{{Verdes-Montenegro} {et~al.}(2001){Verdes-Montenegro}, {Yun},
  {Williams}, {Huchtmeier}, {Del Olmo}, \& {Perea}}]{verdes2001}
{Verdes-Montenegro}, L., {Yun}, M.~S., {Williams}, B.~A., {Huchtmeier}, W.~K.,
  {Del Olmo}, A., \& {Perea}, J. 2001, \aap, 377, 812

\bibitem[{{Walker} {et~al.}(2012){Walker}, {Johnson}, {Gallagher}, {Charlton},
  {Hornschemeier}, \& {Hibbard}}]{walker2012}
{Walker}, L.~M., {Johnson}, K.~E., {Gallagher}, S.~C., {Charlton}, J.~C.,
  {Hornschemeier}, A.~E., \& {Hibbard}, J.~E. 2012, \aj, 143, 69

\bibitem[{{Walker} {et~al.}(2010){Walker}, {Johnson}, {Gallagher}, {Hibbard},
  {Hornschemeier}, {Tzanavaris}, {Charlton}, \& {Jarrett}}]{walker2010}
{Walker}, L.~M., {Johnson}, K.~E., {Gallagher}, S.~C., {Hibbard}, J.~E.,
  {Hornschemeier}, A.~E., {Tzanavaris}, P., {Charlton}, J.~C., \& {Jarrett},
  T.~H. 2010, \aj, 140, 1254

\bibitem[{{Xu} {et~al.}(2005){Xu}, {Iglesias-P{\'a}ramo}, {Burgarella}, {Rich},
  {Neff}, {Lauger}, {Barlow}, {Bianchi}, {Byun}, {Forster}, {Friedman},
  {Heckman}, {Jelinsky}, {Lee}, {Madore}, {Malina}, {Martin}, {Milliard},
  {Morrissey}, {Schiminovich}, {Siegmund}, {Small}, {Szalay}, {Welsh}, \&
  {Wyder}}]{xu2005}
{Xu}, C.~K., {Iglesias-P{\'a}ramo}, J., {Burgarella}, D., {Rich}, R.~M.,
  {Neff}, S.~G., {Lauger}, S., {Barlow}, T.~A., {Bianchi}, L., {Byun}, Y.-I.,
  {Forster}, K., {Friedman}, P.~G., {Heckman}, T.~M., {Jelinsky}, P.~N., {Lee},
  Y.-W., {Madore}, B.~F., {Malina}, R.~F., {Martin}, D.~C., {Milliard}, B.,
  {Morrissey}, P., {Schiminovich}, D., {Siegmund}, O.~H.~W., {Small}, T.,
  {Szalay}, A.~S., {Welsh}, B.~Y., \& {Wyder}, T.~K. 2005, \apjl, 619, L95

\bibitem[{{Xue} {et~al.}(2011){Xue}, {Luo}, {Brandt}, {Bauer}, {Lehmer},
  {Broos}, {Schneider}, {Alexander}, {Brusa}, {Comastri}, {Fabian}, {Gilli},
  {Hasinger}, {Hornschemeier}, {Koekemoer}, {Liu}, {Mainieri}, {Paolillo},
  {Rafferty}, {Rosati}, {Shemmer}, {Silverman}, {Smail}, {Tozzi}, \&
  {Vignali}}]{xue2011}
{Xue}, Y.~Q., {Luo}, B., {Brandt}, W.~N., {Bauer}, F.~E., {Lehmer}, B.~D.,
  {Broos}, P.~S., {Schneider}, D.~P., {Alexander}, D.~M., {Brusa}, M.,
  {Comastri}, A., {Fabian}, A.~C., {Gilli}, R., {Hasinger}, G.,
  {Hornschemeier}, A.~E., {Koekemoer}, A., {Liu}, T., {Mainieri}, V.,
  {Paolillo}, M., {Rafferty}, D.~A., {Rosati}, P., {Shemmer}, O., {Silverman},
  J.~D., {Smail}, I., {Tozzi}, P., \& {Vignali}, C. 2011, \apjs, 195, 10

\end{thebibliography}

\clearpage
\clearpage

\clearpage
\LongTables
\begin{landscape}
\begin{deluxetable*}{c ccc ccc cc c    c  c  c  c  c  c    c  c  c     c  c  c}
\small
\tablecolumns{22}
\tablecaption{\small Derived quantities and detection codes for 
\x\ point sources in HCG~7 (online only). \label{tab-ae4-7}}
\tablehead{
\colhead{}
& \colhead{}
& \colhead{}
& \colhead{}
& \colhead{}
& \colhead{}
& \colhead{}
& \colhead{}
& \colhead{}
& \colhead{}
& \multicolumn{3}{c}{$L_X$}
& \colhead{}
& \colhead{}
& \multicolumn{4}{c}{lim}
& \colhead{}
& \colhead{}
& \colhead{}
\\
\colhead{}
& \colhead{}
& \colhead{}
& \colhead{}
& \colhead{}
& \colhead{}
& \colhead{}
& \colhead{}
& \colhead{}
& \colhead{}
& \multicolumn{3}{c}{\rule{2cm}{.02cm}}
& \colhead{}
& \colhead{}
& \multicolumn{4}{c}{\rule{2.5cm}{.02cm}}
& \colhead{}
& \colhead{}
& \colhead{}
\\
\colhead{ID}
& \colhead{$\Gamma$}
& \colhead{$+$}
& \colhead{$-$}
& \colhead{$f_F$}
& \colhead{$\pm$}
& \colhead{$f_S$}
& \colhead{$\pm$}
& \colhead{$f_H$}
& \colhead{$\pm$}
& \colhead{$F$}
& \colhead{$S$}
& \colhead{$H$}
& \colhead{$f_{\nu,2{\rm keV}}$}
& \colhead{$\pm$}
& \colhead{\tgt HR}
& \colhead{\tgt$F$}
& \colhead{\tgt$S$}
& \colhead{\tgt$H$}
& \colhead{EEF}
& \colhead{RA}
& \colhead{DEC}
\\
\colhead{(1)}
& \colhead{(2)}
& \colhead{(3)}
& \colhead{(4)}
& \colhead{(5)}
& \colhead{(6)}
& \colhead{(7)}
& \colhead{(8)}
& \colhead{(9)}
& \colhead{(10)}
& \colhead{(11)}
& \colhead{(12)}
& \colhead{(13)}
& \colhead{(14)}
& \colhead{(15)}
& \colhead{\tgt(16)}
& \colhead{\tgt(17)}
& \colhead{\tgt(18)}
& \colhead{\tgt(19)}
& \colhead{(20)}
& \colhead{(21)}
& \colhead{(22)}
\\
}
\startdata
    1 & $1.0$ & $0.0$ & $-0.0$ & $-15.0$ & $0.0$ & $-15.7$ & $0.0$ & $-15.1$ & $0.0$ & $38.6$ & $37.9$ & $38.5$ & $-33.2$ & $0.0$ & \tgt$   -2$ & \tgt$    5$ & \tgt$    1$ & \tgt$    5$ & $0.9$ & $9.7884$ & $0.8779$ \\
    2A &  $0.7$  &  $0.0$  &  $-0.0$  &  $-14.6$  &  $0.2$  &  $-15.4$  &  $0.0$  &  $-14.7$  &  $0.0$  &  $39.0$  &  $38.1$  &  $38.9$  &  $-32.9$  &  $0.2$  &  \tgt$   -2$  &  \tgt$    0$  &  \tgt$    5$  &  \tgt$    5$  &  $0.9$  &  $9.7965$  &  $0.8677$ \\
    3 & $1.0$ & $0.0$ & $-0.0$ & $-15.0$ & $0.0$ & $-15.7$ & $0.0$ & $-15.1$ & $0.0$ & $38.6$ & $37.9$ & $38.5$ & $-33.3$ & $0.0$ & \tgt$   -2$ & \tgt$    5$ & \tgt$    5$ & \tgt$    1$ & $0.9$ & $9.8001$ & $0.8911$ \\
    4A &  $0.7$  &  $0.0$  &  $-0.0$  &  $-14.5$  &  $0.2$  &  $-15.4$  &  $0.0$  &  $-14.6$  &  $0.0$  &  $39.0$  &  $38.2$  &  $39.0$  &  $-32.9$  &  $0.2$  &  \tgt$   -2$  &  \tgt$    0$  &  \tgt$    5$  &  \tgt$    5$  &  $0.9$  &  $9.8005$  &  $0.8649$ \\
    5A &  $0.9$  &  $0.1$  &  $-0.1$  &  $-13.1$  &  $0.0$  &  $-13.9$  &  $0.0$  &  $-13.2$  &  $0.0$  &  $40.4$  &  $39.7$  &  $40.3$  &  $-31.4$  &  $0.0$  &  \tgt$    0$  &  \tgt$    0$  &  \tgt$    0$  &  \tgt$    0$  &  $0.9$  &  $9.8050$  &  $0.8617$ \\
    6A\na &  $2.0$  &  $0.2$  &  $-0.2$  &  $-13.6$  &  $0.0$  &  $-13.9$  &  $0.0$  &  $-13.9$  &  $0.0$  &  $40.0$  &  $39.7$  &  $39.7$  &  $-31.7$  &  $0.0$  &  \tgt$    0$  &  \tgt$    0$  &  \tgt$    0$  &  \tgt$    0$  &  $0.9$  &  $9.8059$  &  $0.8643$ \\
    7A &  $0.7$  &  $0.0$  &  $-0.0$  &  $-14.7$  &  $0.2$  &  $-15.5$  &  $0.0$  &  $-14.7$  &  $0.0$  &  $38.9$  &  $38.1$  &  $38.9$  &  $-33.0$  &  $0.2$  &  \tgt$   -2$  &  \tgt$    0$  &  \tgt$    5$  &  \tgt$    5$  &  $0.9$  &  $9.8081$  &  $0.8685$ \\
    8 & $1.6$ & $0.4$ & $-0.4$ & $-14.0$ & $0.1$ & $-14.4$ & $0.1$ & $-14.2$ & $0.1$ & $39.6$ & $39.1$ & $39.4$ & $-32.2$ & $0.1$ & \tgt$    0$ & \tgt$    0$ & \tgt$    0$ & \tgt$    0$ & $0.9$ & $9.8082$ & $0.8986$ \\
\enddata
\tablecomments{\small
Columns are: (1) source ID (for sources inside
galaxy regions, the galaxy is indicated by an upper case letter); 
(2) $\Gamma$ power law index; 
(3) upper error in $\Gamma$; (4) lower error in $\Gamma$;
(5) log of full band flux (\funits);
(6) error;
(7) log of soft band flux (\funits);
(8) error;
(9) log of hard band flux (\funits);
(10) error;
(11) \lx (\lunits) (full band);
(12) \lx (\lunits) (soft band);
(13) \lx (\lunits) (hard band);
(14) log of flux density at 2 keV (\funits~Hz$^{-1}$);
(15) error;
(16) detection code for hardness ratio: 0 (detection
in both hard and soft bands); $-1$ (detection in hard but
not soft band; HR is a lower limit);
$1$ (detection in soft but not hard band;
HR is an upper limit); $-2$ (no detection in either hard or
soft band);
(17) detection code for full band: 0 (detection in
band); 1 (no detection in band because net counts $<0$);
5 (no detection in band because net counts $-$ Gehrels
$2\sigma$ error $<0$);
(18) as previous column but for the soft band;
(19) as previous column but for the hard band;
(20) encircled energy fraction by \ace\ PSF at 1.5 keV;
(21) right ascension;
(22) declination;
}
\end{deluxetable*}
\clearpage
\end{landscape}

\clearpage
\begin{landscape}
\begin{deluxetable*}{c ccc ccc cc c    c  c  c  c  c  c    c  c  c     c  c  c}
\small
\tablecolumns{22}
\tablecaption{\small Derived quantities and detection codes for 
X-ray point sources in HCG~16 (online only). \label{tab-ae4-16}}
\tablehead{
\colhead{}
& \colhead{}
& \colhead{}
& \colhead{}
& \colhead{}
& \colhead{}
& \colhead{}
& \colhead{}
& \colhead{}
& \colhead{}
& \multicolumn{3}{c}{$L_X$}
& \colhead{}
& \colhead{}
& \multicolumn{4}{c}{lim}
& \colhead{}
& \colhead{}
& \colhead{}
\\
\colhead{}
& \colhead{}
& \colhead{}
& \colhead{}
& \colhead{}
& \colhead{}
& \colhead{}
& \colhead{}
& \colhead{}
& \colhead{}
& \multicolumn{3}{c}{\rule{2cm}{.02cm}}
& \colhead{}
& \colhead{}
& \multicolumn{4}{c}{\rule{2.5cm}{.02cm}}
& \colhead{}
& \colhead{}
& \colhead{}
\\
\colhead{ID}
& \colhead{$\Gamma$}
& \colhead{$+$}
& \colhead{$-$}
& \colhead{$f_F$}
& \colhead{$\pm$}
& \colhead{$f_S$}
& \colhead{$\pm$}
& \colhead{$f_H$}
& \colhead{$\pm$}
& \colhead{$F$}
& \colhead{$S$}
& \colhead{$H$}
& \colhead{$f_{\nu,2{\rm keV}}$}
& \colhead{$\pm$}
& \colhead{\tgt HR}
& \colhead{\tgt$F$}
& \colhead{\tgt$S$}
& \colhead{\tgt$H$}
& \colhead{EEF}
& \colhead{RA}
& \colhead{DEC}
\\
\colhead{(1)}
& \colhead{(2)}
& \colhead{(3)}
& \colhead{(4)}
& \colhead{(5)}
& \colhead{(6)}
& \colhead{(7)}
& \colhead{(8)}
& \colhead{(9)}
& \colhead{(10)}
& \colhead{(11)}
& \colhead{(12)}
& \colhead{(13)}
& \colhead{(14)}
& \colhead{(15)}
& \colhead{\tgt(16)}
& \colhead{\tgt(17)}
& \colhead{\tgt(18)}
& \colhead{\tgt(19)}
& \colhead{(20)}
& \colhead{(21)}
& \colhead{(22)}
\\
}
\startdata
    1 & $0.3$ & $0.5$ & $-0.6$ & $-13.5$ & $0.1$ & $-14.6$ & $0.1$ & $-13.6$ & $0.1$ & $40.0$ & $39.0$ & $40.0$ & $-32.0$ & $0.1$ & \tgt$    0$ & \tgt$    0$ & \tgt$    0$ & \tgt$    0$ & $0.9$ & $32.3125$ & $-10.2028$ \\
    2 & $0.4$ & $0.8$ & $-0.8$ & $-13.8$ & $0.2$ & $-14.8$ & $0.2$ & $-13.8$ & $0.2$ & $39.7$ & $38.7$ & $39.7$ & $-32.3$ & $0.2$ & \tgt$    0$ & \tgt$    0$ & \tgt$    0$ & \tgt$    0$ & $0.9$ & $32.3261$ & $-10.2070$ \\
    3 & $1.0$ & $0.0$ & $-0.0$ & $-14.6$ & $0.0$ & $-15.3$ & $0.0$ & $-14.7$ & $0.0$ & $38.9$ & $38.2$ & $38.8$ & $-32.9$ & $0.0$ & \tgt$   -2$ & \tgt$    5$ & \tgt$    1$ & \tgt$    5$ & $0.9$ & $32.3344$ & $-10.1731$ \\
    4B\na &  $-0.7$  &  $0.2$  &  $-0.2$  &  $-12.2$  &  $0.0$  &  $-13.8$  &  $0.0$  &  $-12.2$  &  $0.0$  &  $41.3$  &  $39.7$  &  $41.3$  &  $-31.1$  &  $0.0$  &  \tgt$    0$  &  \tgt$    0$  &  \tgt$    0$  &  \tgt$    0$  &  $0.9$  &  $32.3370$  &  $-10.1332$ \\
    5 & $1.5$ & $2.8$ & $-0.0$ & $-14.0$ & $0.1$ & $-14.5$ & $0.1$ & $-14.2$ & $0.0$ & $39.5$ & $39.0$ & $39.3$ & $-32.2$ & $0.1$ & \tgt$    1$ & \tgt$    0$ & \tgt$    0$ & \tgt$    5$ & $0.9$ & $32.3422$ & $-10.2364$ \\
    6 & $1.6$ & $2.8$ & $-0.0$ & $-14.1$ & $0.2$ & $-14.5$ & $0.2$ & $-14.3$ & $0.0$ & $39.5$ & $39.0$ & $39.3$ & $-32.2$ & $0.2$ & \tgt$    1$ & \tgt$    0$ & \tgt$    0$ & \tgt$    5$ & $0.9$ & $32.3454$ & $-10.1402$ \\
    7 & $0.6$ & $0.0$ & $-0.0$ & $-14.2$ & $0.3$ & $-15.1$ & $0.0$ & $-14.3$ & $0.0$ & $39.3$ & $38.4$ & $39.3$ & $-32.6$ & $0.3$ & \tgt$   -2$ & \tgt$    0$ & \tgt$    5$ & \tgt$    5$ & $0.9$ & $32.3472$ & $-10.1683$ \\
    8 & $1.0$ & $0.0$ & $-0.0$ & $-14.6$ & $0.0$ & $-15.3$ & $0.0$ & $-14.7$ & $0.0$ & $39.0$ & $38.3$ & $38.9$ & $-32.8$ & $0.0$ & \tgt$   -2$ & \tgt$    5$ & \tgt$    5$ & \tgt$    1$ & $0.9$ & $32.3518$ & $-10.1864$ \\
\enddata
\end{deluxetable*}
\clearpage
\end{landscape}

\clearpage
\begin{landscape}
\begin{deluxetable*}{c ccc ccc cc c    c  c  c  c  c  c    c  c  c     c  c  c}
\small
\tablecolumns{22}
\tablecaption{\small Derived quantities and detection codes for 
X-ray point sources in HCG~22 (online only). \label{tab-ae4-22}}
\tablehead{
\colhead{}
& \colhead{}
& \colhead{}
& \colhead{}
& \colhead{}
& \colhead{}
& \colhead{}
& \colhead{}
& \colhead{}
& \colhead{}
& \multicolumn{3}{c}{$L_X$}
& \colhead{}
& \colhead{}
& \multicolumn{4}{c}{lim}
& \colhead{}
& \colhead{}
& \colhead{}
\\
\colhead{}
& \colhead{}
& \colhead{}
& \colhead{}
& \colhead{}
& \colhead{}
& \colhead{}
& \colhead{}
& \colhead{}
& \colhead{}
& \multicolumn{3}{c}{\rule{2cm}{.02cm}}
& \colhead{}
& \colhead{}
& \multicolumn{4}{c}{\rule{2.5cm}{.02cm}}
& \colhead{}
& \colhead{}
& \colhead{}
\\
\colhead{ID}
& \colhead{$\Gamma$}
& \colhead{$+$}
& \colhead{$-$}
& \colhead{$f_F$}
& \colhead{$\pm$}
& \colhead{$f_S$}
& \colhead{$\pm$}
& \colhead{$f_H$}
& \colhead{$\pm$}
& \colhead{$F$}
& \colhead{$S$}
& \colhead{$H$}
& \colhead{$f_{\nu,2{\rm keV}}$}
& \colhead{$\pm$}
& \colhead{\tgt HR}
& \colhead{\tgt$F$}
& \colhead{\tgt$S$}
& \colhead{\tgt$H$}
& \colhead{EEF}
& \colhead{RA}
& \colhead{DEC}
\\
\colhead{(1)}
& \colhead{(2)}
& \colhead{(3)}
& \colhead{(4)}
& \colhead{(5)}
& \colhead{(6)}
& \colhead{(7)}
& \colhead{(8)}
& \colhead{(9)}
& \colhead{(10)}
& \colhead{(11)}
& \colhead{(12)}
& \colhead{(13)}
& \colhead{(14)}
& \colhead{(15)}
& \colhead{\tgt(16)}
& \colhead{\tgt(17)}
& \colhead{\tgt(18)}
& \colhead{\tgt(19)}
& \colhead{(20)}
& \colhead{(21)}
& \colhead{(22)}
\\
}
\startdata
    1 & $1.0$ & $0.0$ & $-0.0$ & $-15.0$ & $0.0$ & $-15.7$ & $0.0$ & $-15.1$ & $0.0$ & $38.2$ & $37.5$ & $38.1$ & $-33.2$ & $0.0$ & \tgt$   -2$ & \tgt$    5$ & \tgt$    5$ & \tgt$    5$ & $0.9$ & $45.8077$ & $-15.5924$ \\
    2 & $1.8$ & $2.0$ & $-0.0$ & $-14.5$ & $0.2$ & $-14.9$ & $0.2$ & $-14.8$ & $0.0$ & $38.7$ & $38.3$ & $38.4$ & $-32.7$ & $0.2$ & \tgt$    1$ & \tgt$    0$ & \tgt$    0$ & \tgt$    5$ & $0.9$ & $45.8096$ & $-15.6824$ \\
    3 & $1.0$ & $0.0$ & $-0.0$ & $-15.0$ & $0.0$ & $-15.7$ & $0.0$ & $-15.1$ & $0.0$ & $38.2$ & $37.5$ & $38.1$ & $-33.2$ & $0.0$ & \tgt$   -2$ & \tgt$    5$ & \tgt$    5$ & \tgt$    5$ & $0.9$ & $45.8175$ & $-15.6480$ \\
    4 & $2.1$ & $0.6$ & $-0.6$ & $-14.2$ & $0.1$ & $-14.5$ & $0.1$ & $-14.5$ & $0.1$ & $39.0$ & $38.7$ & $38.7$ & $-32.3$ & $0.1$ & \tgt$    0$ & \tgt$    0$ & \tgt$    0$ & \tgt$    0$ & $0.9$ & $45.8261$ & $-15.5570$ \\
    5 & $1.2$ & $0.8$ & $-0.7$ & $-14.3$ & $0.1$ & $-14.9$ & $0.1$ & $-14.5$ & $0.1$ & $38.9$ & $38.3$ & $38.7$ & $-32.5$ & $0.1$ & \tgt$    0$ & \tgt$    0$ & \tgt$    0$ & \tgt$    0$ & $0.9$ & $45.8268$ & $-15.5977$ \\
    6 & $1.0$ & $0.0$ & $-0.0$ & $-14.9$ & $0.0$ & $-15.6$ & $0.0$ & $-15.0$ & $0.0$ & $38.3$ & $37.6$ & $38.2$ & $-33.2$ & $0.0$ & \tgt$   -2$ & \tgt$    5$ & \tgt$    5$ & \tgt$    5$ & $0.9$ & $45.8312$ & $-15.6053$ \\
    7 & $0.1$ & $0.0$ & $-1.2$ & $-14.5$ & $0.3$ & $-15.7$ & $0.0$ & $-14.5$ & $0.3$ & $38.7$ & $37.5$ & $38.7$ & $-33.0$ & $0.3$ & \tgt$   -1$ & \tgt$    0$ & \tgt$    1$ & \tgt$    0$ & $0.9$ & $45.8338$ & $-15.6422$ \\
    8 & $1.0$ & $0.0$ & $-0.0$ & $-15.0$ & $0.0$ & $-15.7$ & $0.0$ & $-15.1$ & $0.0$ & $38.2$ & $37.5$ & $38.1$ & $-33.2$ & $0.0$ & \tgt$   -2$ & \tgt$    5$ & \tgt$    5$ & \tgt$    5$ & $0.9$ & $45.8340$ & $-15.5516$ \\
\enddata
\end{deluxetable*}
\clearpage
\end{landscape}

\clearpage
\begin{landscape}
\begin{deluxetable*}{c ccc ccc cc c    c  c  c  c  c  c    c  c  c     c  c  c}
\small
\tablecolumns{22}
\tablecaption{\small Derived quantities and detection codes for 
X-ray point sources in HCG~31 (online only). \label{tab-ae4-31}}
\tablehead{
\colhead{}
& \colhead{}
& \colhead{}
& \colhead{}
& \colhead{}
& \colhead{}
& \colhead{}
& \colhead{}
& \colhead{}
& \colhead{}
& \multicolumn{3}{c}{$L_X$}
& \colhead{}
& \colhead{}
& \multicolumn{4}{c}{lim}
& \colhead{}
& \colhead{}
& \colhead{}
\\
\colhead{}
& \colhead{}
& \colhead{}
& \colhead{}
& \colhead{}
& \colhead{}
& \colhead{}
& \colhead{}
& \colhead{}
& \colhead{}
& \multicolumn{3}{c}{\rule{2cm}{.02cm}}
& \colhead{}
& \colhead{}
& \multicolumn{4}{c}{\rule{2.5cm}{.02cm}}
& \colhead{}
& \colhead{}
& \colhead{}
\\
\colhead{ID}
& \colhead{$\Gamma$}
& \colhead{$+$}
& \colhead{$-$}
& \colhead{$f_F$}
& \colhead{$\pm$}
& \colhead{$f_S$}
& \colhead{$\pm$}
& \colhead{$f_H$}
& \colhead{$\pm$}
& \colhead{$F$}
& \colhead{$S$}
& \colhead{$H$}
& \colhead{$f_{\nu,2{\rm keV}}$}
& \colhead{$\pm$}
& \colhead{\tgt HR}
& \colhead{\tgt$F$}
& \colhead{\tgt$S$}
& \colhead{\tgt$H$}
& \colhead{EEF}
& \colhead{RA}
& \colhead{DEC}
\\
\colhead{(1)}
& \colhead{(2)}
& \colhead{(3)}
& \colhead{(4)}
& \colhead{(5)}
& \colhead{(6)}
& \colhead{(7)}
& \colhead{(8)}
& \colhead{(9)}
& \colhead{(10)}
& \colhead{(11)}
& \colhead{(12)}
& \colhead{(13)}
& \colhead{(14)}
& \colhead{(15)}
& \colhead{\tgt(16)}
& \colhead{\tgt(17)}
& \colhead{\tgt(18)}
& \colhead{\tgt(19)}
& \colhead{(20)}
& \colhead{(21)}
& \colhead{(22)}
\\
}
\startdata
    1 & $1.8$ & $0.6$ & $-0.5$ & $-14.2$ & $0.1$ & $-14.6$ & $0.1$ & $-14.4$ & $0.1$ & $39.4$ & $39.0$ & $39.2$ & $-32.3$ & $0.1$ & \tgt$    0$ & \tgt$    0$ & \tgt$    0$ & \tgt$    0$ & $0.9$ & $75.3102$ & $-4.2743$ \\
    2 & $0.7$ & $0.8$ & $-0.8$ & $-14.3$ & $0.2$ & $-15.1$ & $0.2$ & $-14.4$ & $0.2$ & $39.3$ & $38.5$ & $39.2$ & $-32.6$ & $0.2$ & \tgt$    0$ & \tgt$    0$ & \tgt$    0$ & \tgt$    0$ & $0.9$ & $75.3159$ & $-4.2936$ \\
    3 & $0.9$ & $0.8$ & $-0.7$ & $-14.3$ & $0.1$ & $-15.0$ & $0.1$ & $-14.4$ & $0.1$ & $39.3$ & $38.6$ & $39.2$ & $-32.6$ & $0.1$ & \tgt$    0$ & \tgt$    0$ & \tgt$    0$ & \tgt$    0$ & $0.4$ & $75.3268$ & $-4.2817$ \\
    4 & $1.0$ & $0.0$ & $-0.0$ & $-14.6$ & $0.0$ & $-15.3$ & $0.0$ & $-14.7$ & $0.0$ & $39.0$ & $38.3$ & $38.9$ & $-32.9$ & $0.0$ & \tgt$   -2$ & \tgt$    5$ & \tgt$    0$ & \tgt$    1$ & $0.4$ & $75.3269$ & $-4.2817$ \\
    5 & $-0.1$ & $0.0$ & $-1.0$ & $-14.3$ & $0.2$ & $-15.5$ & $0.0$ & $-14.3$ & $0.2$ & $39.3$ & $38.1$ & $39.3$ & $-32.9$ & $0.2$ & \tgt$   -1$ & \tgt$    0$ & \tgt$    5$ & \tgt$    0$ & $0.9$ & $75.3348$ & $-4.3111$ \\
    6 & $1.6$ & $2.2$ & $-0.0$ & $-14.6$ & $0.2$ & $-15.1$ & $0.2$ & $-14.8$ & $0.0$ & $39.0$ & $38.5$ & $38.8$ & $-32.8$ & $0.2$ & \tgt$    1$ & \tgt$    0$ & \tgt$    0$ & \tgt$    5$ & $0.9$ & $75.3411$ & $-4.2934$ \\
    7 & $0.7$ & $0.0$ & $-0.0$ & $-14.6$ & $0.2$ & $-15.5$ & $0.0$ & $-14.7$ & $0.0$ & $39.0$ & $38.1$ & $38.9$ & $-33.0$ & $0.2$ & \tgt$   -2$ & \tgt$    0$ & \tgt$    5$ & \tgt$    5$ & $0.9$ & $75.3468$ & $-4.2624$ \\
    8 & $1.0$ & $0.0$ & $-0.0$ & $-15.0$ & $0.0$ & $-15.7$ & $0.0$ & $-15.1$ & $0.0$ & $38.6$ & $37.9$ & $38.5$ & $-33.2$ & $0.0$ & \tgt$   -2$ & \tgt$    5$ & \tgt$    1$ & \tgt$    5$ & $0.9$ & $75.3514$ & $-4.2629$ \\
\enddata
\end{deluxetable*}
\clearpage
\end{landscape}

\clearpage
\begin{landscape}
\begin{deluxetable*}{c ccc ccc cc c    c  c  c  c  c  c    c  c  c     c  c  c}
\small
\tablecolumns{22}
\tablecaption{\small Derived quantities and detection codes for 
X-ray point sources in HCG~42 (online only). \label{tab-ae4-42}}
\tablehead{
\colhead{}
& \colhead{}
& \colhead{}
& \colhead{}
& \colhead{}
& \colhead{}
& \colhead{}
& \colhead{}
& \colhead{}
& \colhead{}
& \multicolumn{3}{c}{$L_X$}
& \colhead{}
& \colhead{}
& \multicolumn{4}{c}{lim}
& \colhead{}
& \colhead{}
& \colhead{}
\\
\colhead{}
& \colhead{}
& \colhead{}
& \colhead{}
& \colhead{}
& \colhead{}
& \colhead{}
& \colhead{}
& \colhead{}
& \colhead{}
& \multicolumn{3}{c}{\rule{2cm}{.02cm}}
& \colhead{}
& \colhead{}
& \multicolumn{4}{c}{\rule{2.5cm}{.02cm}}
& \colhead{}
& \colhead{}
& \colhead{}
\\
\colhead{ID}
& \colhead{$\Gamma$}
& \colhead{$+$}
& \colhead{$-$}
& \colhead{$f_F$}
& \colhead{$\pm$}
& \colhead{$f_S$}
& \colhead{$\pm$}
& \colhead{$f_H$}
& \colhead{$\pm$}
& \colhead{$F$}
& \colhead{$S$}
& \colhead{$H$}
& \colhead{$f_{\nu,2{\rm keV}}$}
& \colhead{$\pm$}
& \colhead{\tgt HR}
& \colhead{\tgt$F$}
& \colhead{\tgt$S$}
& \colhead{\tgt$H$}
& \colhead{EEF}
& \colhead{RA}
& \colhead{DEC}
\\
\colhead{(1)}
& \colhead{(2)}
& \colhead{(3)}
& \colhead{(4)}
& \colhead{(5)}
& \colhead{(6)}
& \colhead{(7)}
& \colhead{(8)}
& \colhead{(9)}
& \colhead{(10)}
& \colhead{(11)}
& \colhead{(12)}
& \colhead{(13)}
& \colhead{(14)}
& \colhead{(15)}
& \colhead{\tgt(16)}
& \colhead{\tgt(17)}
& \colhead{\tgt(18)}
& \colhead{\tgt(19)}
& \colhead{(20)}
& \colhead{(21)}
& \colhead{(22)}
\\
}
\startdata
    1 & $1.2$ & $0.8$ & $-0.7$ & $-14.4$ & $0.2$ & $-15.0$ & $0.2$ & $-14.5$ & $0.2$ & $39.3$ & $38.7$ & $39.2$ & $-32.6$ & $0.2$ & \tgt$    0$ & \tgt$    0$ & \tgt$    0$ & \tgt$    0$ & $0.9$ & $150.0048$ & $-19.6321$ \\
    2 & $1.0$ & $0.0$ & $-0.0$ & $-14.9$ & $0.0$ & $-15.6$ & $0.0$ & $-15.0$ & $0.0$ & $38.8$ & $38.1$ & $38.7$ & $-33.1$ & $0.0$ & \tgt$   -2$ & \tgt$    5$ & \tgt$    5$ & \tgt$    5$ & $0.9$ & $150.0286$ & $-19.6759$ \\
    3C &  $1.4$  &  $2.4$  &  $-0.0$  &  $-14.7$  &  $0.2$  &  $-15.2$  &  $0.2$  &  $-14.9$  &  $0.0$  &  $39.0$  &  $38.5$  &  $38.8$  &  $-32.9$  &  $0.2$  &  \tgt$    1$  &  \tgt$    0$  &  \tgt$    0$  &  \tgt$    1$  &  $0.9$  &  $150.0355$  &  $-19.6309$ \\
    4 & $1.0$ & $0.0$ & $-0.0$ & $-15.0$ & $0.0$ & $-15.7$ & $0.0$ & $-15.1$ & $0.0$ & $38.7$ & $38.0$ & $38.6$ & $-33.3$ & $0.0$ & \tgt$   -2$ & \tgt$    5$ & \tgt$    1$ & \tgt$    5$ & $0.9$ & $150.0412$ & $-19.6145$ \\
    5 & $1.0$ & $0.0$ & $-0.0$ & $-15.0$ & $0.0$ & $-15.7$ & $0.0$ & $-15.1$ & $0.0$ & $38.7$ & $38.0$ & $38.6$ & $-33.2$ & $0.0$ & \tgt$   -2$ & \tgt$    5$ & \tgt$    5$ & \tgt$    1$ & $0.9$ & $150.0416$ & $-19.6763$ \\
    6C\na &  $2.4$  &  $1.4$  &  $-0.0$  &  $-14.3$  &  $0.1$  &  $-14.5$  &  $0.1$  &  $-14.8$  &  $0.0$  &  $39.3$  &  $39.1$  &  $38.9$  &  $-32.5$  &  $0.1$  &  \tgt$    1$  &  \tgt$    0$  &  \tgt$    0$  &  \tgt$    5$  &  $0.9$  &  $150.0430$  &  $-19.6221$ \\
    7 & $0.7$ & $0.0$ & $-0.0$ & $-14.5$ & $0.2$ & $-15.4$ & $0.0$ & $-14.6$ & $0.0$ & $39.2$ & $38.3$ & $39.1$ & $-32.9$ & $0.2$ & \tgt$   -2$ & \tgt$    0$ & \tgt$    5$ & \tgt$    5$ & $0.9$ & $150.0430$ & $-19.7098$ \\
    8C &  $1.6$  &  $2.2$  &  $-0.0$  &  $-14.7$  &  $0.2$  &  $-15.1$  &  $0.2$  &  $-14.9$  &  $0.0$  &  $39.0$  &  $38.5$  &  $38.8$  &  $-32.8$  &  $0.2$  &  \tgt$    1$  &  \tgt$    0$  &  \tgt$    0$  &  \tgt$    1$  &  $0.9$  &  $150.0445$  &  $-19.6279$ \\
\enddata
\end{deluxetable*}
\clearpage
\end{landscape}

\clearpage
\begin{landscape}
\begin{deluxetable*}{c ccc ccc cc c    c  c  c  c  c  c    c  c  c     c  c  c}
\small
\tablecolumns{22}
\tablecaption{\small Derived quantities and detection codes for 
X-ray point sources in HCG~59 (online only). \label{tab-ae4-59}}
\tablehead{
\colhead{}
& \colhead{}
& \colhead{}
& \colhead{}
& \colhead{}
& \colhead{}
& \colhead{}
& \colhead{}
& \colhead{}
& \colhead{}
& \multicolumn{3}{c}{$L_X$}
& \colhead{}
& \colhead{}
& \multicolumn{4}{c}{lim}
& \colhead{}
& \colhead{}
& \colhead{}
\\
\colhead{}
& \colhead{}
& \colhead{}
& \colhead{}
& \colhead{}
& \colhead{}
& \colhead{}
& \colhead{}
& \colhead{}
& \colhead{}
& \multicolumn{3}{c}{\rule{2cm}{.02cm}}
& \colhead{}
& \colhead{}
& \multicolumn{4}{c}{\rule{2.5cm}{.02cm}}
& \colhead{}
& \colhead{}
& \colhead{}
\\
\colhead{ID}
& \colhead{$\Gamma$}
& \colhead{$+$}
& \colhead{$-$}
& \colhead{$f_F$}
& \colhead{$\pm$}
& \colhead{$f_S$}
& \colhead{$\pm$}
& \colhead{$f_H$}
& \colhead{$\pm$}
& \colhead{$F$}
& \colhead{$S$}
& \colhead{$H$}
& \colhead{$f_{\nu,2{\rm keV}}$}
& \colhead{$\pm$}
& \colhead{\tgt HR}
& \colhead{\tgt$F$}
& \colhead{\tgt$S$}
& \colhead{\tgt$H$}
& \colhead{EEF}
& \colhead{RA}
& \colhead{DEC}
\\
\colhead{(1)}
& \colhead{(2)}
& \colhead{(3)}
& \colhead{(4)}
& \colhead{(5)}
& \colhead{(6)}
& \colhead{(7)}
& \colhead{(8)}
& \colhead{(9)}
& \colhead{(10)}
& \colhead{(11)}
& \colhead{(12)}
& \colhead{(13)}
& \colhead{(14)}
& \colhead{(15)}
& \colhead{\tgt(16)}
& \colhead{\tgt(17)}
& \colhead{\tgt(18)}
& \colhead{\tgt(19)}
& \colhead{(20)}
& \colhead{(21)}
& \colhead{(22)}
\\
}
\startdata
    1 & $1.0$ & $0.0$ & $-0.0$ & $-14.9$ & $0.0$ & $-15.6$ & $0.0$ & $-15.0$ & $0.0$ & $38.8$ & $38.1$ & $38.7$ & $-33.1$ & $0.0$ & \tgt$   -2$ & \tgt$    5$ & \tgt$    1$ & \tgt$    5$ & $0.9$ & $177.0392$ & $12.7376$ \\
    2 & $1.0$ & $0.0$ & $-0.0$ & $-14.7$ & $0.0$ & $-15.4$ & $0.0$ & $-14.8$ & $0.0$ & $39.0$ & $38.3$ & $38.9$ & $-32.9$ & $0.0$ & \tgt$   -2$ & \tgt$    5$ & \tgt$    5$ & \tgt$    5$ & $0.9$ & $177.0435$ & $12.7450$ \\
    3 & $1.0$ & $0.0$ & $-0.0$ & $-15.0$ & $0.0$ & $-15.7$ & $0.0$ & $-15.1$ & $0.0$ & $38.7$ & $38.0$ & $38.6$ & $-33.2$ & $0.0$ & \tgt$   -2$ & \tgt$    5$ & \tgt$    5$ & \tgt$    1$ & $0.9$ & $177.0456$ & $12.7168$ \\
    4 & $1.2$ & $3.2$ & $-0.0$ & $-14.5$ & $0.2$ & $-15.1$ & $0.2$ & $-14.6$ & $0.0$ & $39.2$ & $38.6$ & $39.1$ & $-32.7$ & $0.2$ & \tgt$    1$ & \tgt$    0$ & \tgt$    0$ & \tgt$    5$ & $0.9$ & $177.0500$ & $12.7112$ \\
    5 & $1.0$ & $0.0$ & $-0.0$ & $-14.9$ & $0.0$ & $-15.6$ & $0.0$ & $-15.0$ & $0.0$ & $38.8$ & $38.1$ & $38.7$ & $-33.2$ & $0.0$ & \tgt$   -2$ & \tgt$    5$ & \tgt$    1$ & \tgt$    5$ & $0.9$ & $177.0528$ & $12.7024$ \\
    6 & $-0.9$ & $0.0$ & $-0.2$ & $-13.7$ & $0.1$ & $-15.4$ & $0.0$ & $-13.7$ & $0.1$ & $40.0$ & $38.2$ & $40.0$ & $-32.7$ & $0.1$ & \tgt$   -1$ & \tgt$    0$ & \tgt$    1$ & \tgt$    0$ & $0.9$ & $177.0570$ & $12.7208$ \\
    7 & $0.7$ & $0.0$ & $-0.0$ & $-14.4$ & $0.2$ & $-15.2$ & $0.0$ & $-14.4$ & $0.0$ & $39.3$ & $38.5$ & $39.2$ & $-32.7$ & $0.2$ & \tgt$   -2$ & \tgt$    0$ & \tgt$    5$ & \tgt$    5$ & $0.9$ & $177.0631$ & $12.7514$ \\
    8 & $1.0$ & $0.0$ & $-0.0$ & $-15.0$ & $0.0$ & $-15.7$ & $0.0$ & $-15.1$ & $0.0$ & $38.7$ & $38.0$ & $38.6$ & $-33.3$ & $0.0$ & \tgt$   -2$ & \tgt$    5$ & \tgt$    5$ & \tgt$    1$ & $0.9$ & $177.0669$ & $12.7091$ \\
\enddata
\end{deluxetable*}
\clearpage
\end{landscape}

\clearpage
\begin{landscape}
\begin{deluxetable*}{c ccc ccc cc c    c  c  c  c  c  c    c  c  c     c  c  c}
\small
\tablecolumns{22}
\tablecaption{\small Derived quantities and detection codes for 
X-ray point sources in HCG~62 (online only). \label{tab-ae4-62}}
\tablehead{
\colhead{}
& \colhead{}
& \colhead{}
& \colhead{}
& \colhead{}
& \colhead{}
& \colhead{}
& \colhead{}
& \colhead{}
& \colhead{}
& \multicolumn{3}{c}{$L_X$}
& \colhead{}
& \colhead{}
& \multicolumn{4}{c}{lim}
& \colhead{}
& \colhead{}
& \colhead{}
\\
\colhead{}
& \colhead{}
& \colhead{}
& \colhead{}
& \colhead{}
& \colhead{}
& \colhead{}
& \colhead{}
& \colhead{}
& \colhead{}
& \multicolumn{3}{c}{\rule{2cm}{.02cm}}
& \colhead{}
& \colhead{}
& \multicolumn{4}{c}{\rule{2.5cm}{.02cm}}
& \colhead{}
& \colhead{}
& \colhead{}
\\
\colhead{ID}
& \colhead{$\Gamma$}
& \colhead{$+$}
& \colhead{$-$}
& \colhead{$f_F$}
& \colhead{$\pm$}
& \colhead{$f_S$}
& \colhead{$\pm$}
& \colhead{$f_H$}
& \colhead{$\pm$}
& \colhead{$F$}
& \colhead{$S$}
& \colhead{$H$}
& \colhead{$f_{\nu,2{\rm keV}}$}
& \colhead{$\pm$}
& \colhead{\tgt HR}
& \colhead{\tgt$F$}
& \colhead{\tgt$S$}
& \colhead{\tgt$H$}
& \colhead{EEF}
& \colhead{RA}
& \colhead{DEC}
\\
\colhead{(1)}
& \colhead{(2)}
& \colhead{(3)}
& \colhead{(4)}
& \colhead{(5)}
& \colhead{(6)}
& \colhead{(7)}
& \colhead{(8)}
& \colhead{(9)}
& \colhead{(10)}
& \colhead{(11)}
& \colhead{(12)}
& \colhead{(13)}
& \colhead{(14)}
& \colhead{(15)}
& \colhead{\tgt(16)}
& \colhead{\tgt(17)}
& \colhead{\tgt(18)}
& \colhead{\tgt(19)}
& \colhead{(20)}
& \colhead{(21)}
& \colhead{(22)}
\\
}
\startdata
    1 & $0.7$ & $0.4$ & $-0.4$ & $-14.5$ & $0.1$ & $-15.3$ & $0.1$ & $-14.5$ & $0.1$ & $39.2$ & $38.4$ & $39.2$ & $-32.8$ & $0.1$ & \tgt$    0$ & \tgt$    0$ & \tgt$    0$ & \tgt$    0$ & $0.9$ & $193.1794$ & $-9.2492$ \\
    2 & $0.5$ & $0.6$ & $-0.7$ & $-14.8$ & $0.1$ & $-15.7$ & $0.1$ & $-14.8$ & $0.1$ & $38.9$ & $38.0$ & $38.9$ & $-33.2$ & $0.1$ & \tgt$    0$ & \tgt$    0$ & \tgt$    0$ & \tgt$    0$ & $0.9$ & $193.1884$ & $-9.2188$ \\
    3 & $0.4$ & $0.4$ & $-0.3$ & $-14.3$ & $0.1$ & $-15.3$ & $0.1$ & $-14.4$ & $0.1$ & $39.4$ & $38.4$ & $39.3$ & $-32.8$ & $0.1$ & \tgt$    0$ & \tgt$    0$ & \tgt$    0$ & \tgt$    0$ & $0.9$ & $193.1889$ & $-9.2643$ \\
    4 & $1.0$ & $0.0$ & $-0.0$ & $-15.4$ & $0.0$ & $-16.1$ & $0.0$ & $-15.5$ & $0.0$ & $38.3$ & $37.6$ & $38.2$ & $-33.7$ & $0.0$ & \tgt$   -2$ & \tgt$    5$ & \tgt$    5$ & \tgt$    1$ & $0.9$ & $193.1988$ & $-9.1934$ \\
    5 & $0.8$ & $0.3$ & $-0.3$ & $-14.2$ & $0.1$ & $-15.0$ & $0.1$ & $-14.3$ & $0.1$ & $39.5$ & $38.7$ & $39.4$ & $-32.5$ & $0.1$ & \tgt$    0$ & \tgt$    0$ & \tgt$    0$ & \tgt$    0$ & $0.9$ & $193.2010$ & $-9.2535$ \\
    6 & $1.1$ & $0.4$ & $-0.4$ & $-14.5$ & $0.1$ & $-15.2$ & $0.1$ & $-14.6$ & $0.1$ & $39.2$ & $38.5$ & $39.1$ & $-32.7$ & $0.1$ & \tgt$    0$ & \tgt$    0$ & \tgt$    0$ & \tgt$    0$ & $0.9$ & $193.2026$ & $-9.2236$ \\
    7 & $1.7$ & $0.2$ & $-0.2$ & $-14.1$ & $0.0$ & $-14.5$ & $0.0$ & $-14.3$ & $0.0$ & $39.6$ & $39.2$ & $39.3$ & $-32.3$ & $0.0$ & \tgt$    0$ & \tgt$    0$ & \tgt$    0$ & \tgt$    0$ & $0.9$ & $193.2058$ & $-9.2073$ \\
    8 & $1.9$ & $0.3$ & $-0.3$ & $-14.4$ & $0.1$ & $-14.7$ & $0.1$ & $-14.7$ & $0.1$ & $39.3$ & $39.0$ & $39.0$ & $-32.5$ & $0.1$ & \tgt$    0$ & \tgt$    0$ & \tgt$    0$ & \tgt$    0$ & $0.9$ & $193.2071$ & $-9.1961$ \\
\enddata
\end{deluxetable*}
\clearpage
\end{landscape}

\clearpage
\begin{landscape}
\begin{deluxetable*}{c ccc ccc cc c    c  c  c  c  c  c    c  c  c     c  c  c}
\small
\tablecolumns{22}
\tablecaption{\small Derived quantities and detection codes for 
X-ray point sources in HCG~90 (online only). \label{tab-ae4-90}}
\tablehead{
\colhead{}
& \colhead{}
& \colhead{}
& \colhead{}
& \colhead{}
& \colhead{}
& \colhead{}
& \colhead{}
& \colhead{}
& \colhead{}
& \multicolumn{3}{c}{$L_X$}
& \colhead{}
& \colhead{}
& \multicolumn{4}{c}{lim}
& \colhead{}
& \colhead{}
& \colhead{}
\\
\colhead{}
& \colhead{}
& \colhead{}
& \colhead{}
& \colhead{}
& \colhead{}
& \colhead{}
& \colhead{}
& \colhead{}
& \colhead{}
& \multicolumn{3}{c}{\rule{2cm}{.02cm}}
& \colhead{}
& \colhead{}
& \multicolumn{4}{c}{\rule{2.5cm}{.02cm}}
& \colhead{}
& \colhead{}
& \colhead{}
\\
\colhead{ID}
& \colhead{$\Gamma$}
& \colhead{$+$}
& \colhead{$-$}
& \colhead{$f_F$}
& \colhead{$\pm$}
& \colhead{$f_S$}
& \colhead{$\pm$}
& \colhead{$f_H$}
& \colhead{$\pm$}
& \colhead{$F$}
& \colhead{$S$}
& \colhead{$H$}
& \colhead{$f_{\nu,2{\rm keV}}$}
& \colhead{$\pm$}
& \colhead{\tgt HR}
& \colhead{\tgt$F$}
& \colhead{\tgt$S$}
& \colhead{\tgt$H$}
& \colhead{EEF}
& \colhead{RA}
& \colhead{DEC}
\\
\colhead{(1)}
& \colhead{(2)}
& \colhead{(3)}
& \colhead{(4)}
& \colhead{(5)}
& \colhead{(6)}
& \colhead{(7)}
& \colhead{(8)}
& \colhead{(9)}
& \colhead{(10)}
& \colhead{(11)}
& \colhead{(12)}
& \colhead{(13)}
& \colhead{(14)}
& \colhead{(15)}
& \colhead{\tgt(16)}
& \colhead{\tgt(17)}
& \colhead{\tgt(18)}
& \colhead{\tgt(19)}
& \colhead{(20)}
& \colhead{(21)}
& \colhead{(22)}
\\
}
\startdata
    1 & $2.0$ & $0.5$ & $-0.5$ & $-14.1$ & $0.1$ & $-14.4$ & $0.1$ & $-14.3$ & $0.1$ & $39.1$ & $38.8$ & $38.8$ & $-32.2$ & $0.1$ & \tgt$    0$ & \tgt$    0$ & \tgt$    0$ & \tgt$    0$ & $0.9$ & $330.3675$ & $-32.0938$ \\
    2 & $2.1$ & $0.4$ & $-0.3$ & $-13.7$ & $0.1$ & $-14.0$ & $0.1$ & $-14.0$ & $0.1$ & $39.4$ & $39.1$ & $39.1$ & $-31.8$ & $0.1$ & \tgt$    0$ & \tgt$    0$ & \tgt$    0$ & \tgt$    0$ & $0.9$ & $330.3732$ & $-31.9535$ \\
    3 & $-0.1$ & $0.3$ & $-0.3$ & $-13.4$ & $0.1$ & $-14.7$ & $0.1$ & $-13.4$ & $0.1$ & $39.8$ & $38.5$ & $39.7$ & $-32.0$ & $0.1$ & \tgt$    0$ & \tgt$    0$ & \tgt$    0$ & \tgt$    0$ & $0.9$ & $330.3733$ & $-31.8821$ \\
    4 & $1.9$ & $0.2$ & $-0.2$ & $-13.5$ & $0.0$ & $-13.8$ & $0.0$ & $-13.8$ & $0.0$ & $39.6$ & $39.3$ & $39.3$ & $-31.6$ & $0.0$ & \tgt$    0$ & \tgt$    0$ & \tgt$    0$ & \tgt$    0$ & $0.9$ & $330.3776$ & $-31.9212$ \\
    5 & $1.8$ & $0.5$ & $-0.5$ & $-14.1$ & $0.1$ & $-14.4$ & $0.1$ & $-14.3$ & $0.1$ & $39.1$ & $38.7$ & $38.8$ & $-32.2$ & $0.1$ & \tgt$    0$ & \tgt$    0$ & \tgt$    0$ & \tgt$    0$ & $0.9$ & $330.3796$ & $-32.0583$ \\
    6 & $1.8$ & $0.6$ & $-0.5$ & $-14.1$ & $0.1$ & $-14.5$ & $0.1$ & $-14.4$ & $0.1$ & $39.0$ & $38.7$ & $38.8$ & $-32.2$ & $0.1$ & \tgt$    0$ & \tgt$    0$ & \tgt$    0$ & \tgt$    0$ & $0.9$ & $330.3834$ & $-31.8770$ \\
    7 & $1.8$ & $0.4$ & $-0.4$ & $-14.0$ & $0.1$ & $-14.3$ & $0.1$ & $-14.2$ & $0.1$ & $39.2$ & $38.8$ & $38.9$ & $-32.1$ & $0.1$ & \tgt$    0$ & \tgt$    0$ & \tgt$    0$ & \tgt$    0$ & $0.9$ & $330.3838$ & $-31.9160$ \\
    8 & $0.7$ & $0.5$ & $-0.5$ & $-14.0$ & $0.1$ & $-14.8$ & $0.1$ & $-14.0$ & $0.1$ & $39.2$ & $38.3$ & $39.1$ & $-32.3$ & $0.1$ & \tgt$    0$ & \tgt$    0$ & \tgt$    0$ & \tgt$    0$ & $0.9$ & $330.3868$ & $-32.0294$ \\
\enddata
\end{deluxetable*}
\clearpage
\end{landscape}

\clearpage
\begin{landscape}
\begin{deluxetable*}{c ccc ccc cc c    c  c  c  c  c  c    c  c  c     c  c  c}
\small
\tablecolumns{22}
\tablecaption{\small Derived quantities and detection codes for 
X-ray point sources in HCG~92 (online only). \label{tab-ae4-92}}
\tablehead{
\colhead{}
& \colhead{}
& \colhead{}
& \colhead{}
& \colhead{}
& \colhead{}
& \colhead{}
& \colhead{}
& \colhead{}
& \colhead{}
& \multicolumn{3}{c}{$L_X$}
& \colhead{}
& \colhead{}
& \multicolumn{4}{c}{lim}
& \colhead{}
& \colhead{}
& \colhead{}
\\
\colhead{}
& \colhead{}
& \colhead{}
& \colhead{}
& \colhead{}
& \colhead{}
& \colhead{}
& \colhead{}
& \colhead{}
& \colhead{}
& \multicolumn{3}{c}{\rule{2cm}{.02cm}}
& \colhead{}
& \colhead{}
& \multicolumn{4}{c}{\rule{2.5cm}{.02cm}}
& \colhead{}
& \colhead{}
& \colhead{}
\\
\colhead{ID}
& \colhead{$\Gamma$}
& \colhead{$+$}
& \colhead{$-$}
& \colhead{$f_F$}
& \colhead{$\pm$}
& \colhead{$f_S$}
& \colhead{$\pm$}
& \colhead{$f_H$}
& \colhead{$\pm$}
& \colhead{$F$}
& \colhead{$S$}
& \colhead{$H$}
& \colhead{$f_{\nu,2{\rm keV}}$}
& \colhead{$\pm$}
& \colhead{\tgt HR}
& \colhead{\tgt$F$}
& \colhead{\tgt$S$}
& \colhead{\tgt$H$}
& \colhead{EEF}
& \colhead{RA}
& \colhead{DEC}
\\
\colhead{(1)}
& \colhead{(2)}
& \colhead{(3)}
& \colhead{(4)}
& \colhead{(5)}
& \colhead{(6)}
& \colhead{(7)}
& \colhead{(8)}
& \colhead{(9)}
& \colhead{(10)}
& \colhead{(11)}
& \colhead{(12)}
& \colhead{(13)}
& \colhead{(14)}
& \colhead{(15)}
& \colhead{\tgt(16)}
& \colhead{\tgt(17)}
& \colhead{\tgt(18)}
& \colhead{\tgt(19)}
& \colhead{(20)}
& \colhead{(21)}
& \colhead{(22)}
\\
}
\startdata
    1 & $1.1$ & $3.3$ & $-0.0$ & $-15.1$ & $0.2$ & $-15.7$ & $0.2$ & $-15.2$ & $0.0$ & $38.9$ & $38.3$ & $38.8$ & $-33.3$ & $0.2$ & \tgt$    1$ & \tgt$    0$ & \tgt$    0$ & \tgt$    5$ & $0.9$ & $338.9142$ & $33.9697$ \\
    2 & $1.0$ & $0.0$ & $-0.0$ & $-15.5$ & $0.0$ & $-16.2$ & $0.0$ & $-15.6$ & $0.0$ & $38.4$ & $37.7$ & $38.3$ & $-33.8$ & $0.0$ & \tgt$   -2$ & \tgt$    5$ & \tgt$    0$ & \tgt$    1$ & $0.9$ & $338.9196$ & $33.9596$ \\
    3 & $1.1$ & $3.3$ & $-0.0$ & $-14.9$ & $0.2$ & $-15.5$ & $0.2$ & $-15.0$ & $0.0$ & $39.1$ & $38.4$ & $39.0$ & $-33.1$ & $0.2$ & \tgt$    1$ & \tgt$    0$ & \tgt$    0$ & \tgt$    5$ & $0.9$ & $338.9382$ & $33.9753$ \\
    4 & $-0.7$ & $0.0$ & $-0.4$ & $-14.2$ & $0.1$ & $-15.8$ & $0.0$ & $-14.2$ & $0.1$ & $39.8$ & $38.2$ & $39.8$ & $-33.1$ & $0.1$ & \tgt$   -1$ & \tgt$    0$ & \tgt$    5$ & \tgt$    0$ & $0.9$ & $338.9413$ & $33.9651$ \\
    5 & $1.1$ & $3.3$ & $-0.0$ & $-15.1$ & $0.2$ & $-15.7$ & $0.2$ & $-15.2$ & $0.0$ & $38.9$ & $38.3$ & $38.8$ & $-33.3$ & $0.2$ & \tgt$    1$ & \tgt$    0$ & \tgt$    0$ & \tgt$    5$ & $0.9$ & $338.9429$ & $33.9902$ \\
    6 & $0.5$ & $0.8$ & $-0.9$ & $-14.7$ & $0.2$ & $-15.6$ & $0.2$ & $-14.7$ & $0.2$ & $39.3$ & $38.4$ & $39.3$ & $-33.1$ & $0.2$ & \tgt$    0$ & \tgt$    0$ & \tgt$    0$ & \tgt$    0$ & $0.9$ & $338.9438$ & $33.9632$ \\
    7 & $0.2$ & $0.0$ & $-1.3$ & $-14.6$ & $0.2$ & $-15.8$ & $0.0$ & $-14.7$ & $0.2$ & $39.3$ & $38.2$ & $39.3$ & $-33.2$ & $0.2$ & \tgt$   -1$ & \tgt$    0$ & \tgt$    5$ & \tgt$    0$ & $0.9$ & $338.9442$ & $33.9478$ \\
    8 & $1.0$ & $0.0$ & $-0.0$ & $-15.3$ & $0.0$ & $-16.0$ & $0.0$ & $-15.4$ & $0.0$ & $38.6$ & $37.9$ & $38.5$ & $-33.6$ & $0.0$ & \tgt$   -2$ & \tgt$    5$ & \tgt$    5$ & \tgt$    1$ & $0.9$ & $338.9473$ & $34.0095$ \\
\enddata
\end{deluxetable*}
\clearpage
\end{landscape}

\clearpage
\begin{landscape}
\begin{deluxetable*}{cccc cccc cccc cc}
\small
\tablecolumns{14}
\tablecaption{\small Counts, count rates and \ace\ detection significance and
no-source probabilities for X-ray point sources in HCG~7 (online only).\label{tab-count-7}}
\tablehead{
\colhead{ID}
& \colhead{c(FB)}
& \colhead{lim$_F$}
& \colhead{c(SB)}
& \colhead{lim$_S$}
& \colhead{c(HB)}
& \colhead{lim$_S$}
& \colhead{HR}
& \colhead{lim$_{\rm HR}$}
& \colhead{cr(FB)}
& \colhead{cr(SB)}
& \colhead{cr(HB)}
& \colhead{\ace(sig)}
& \colhead{\ace($P$)}
\\
\colhead{(1)}
& \colhead{(2)}
& \colhead{(3)}
& \colhead{(4)}
& \colhead{(5)}
& \colhead{(6)}
& \colhead{(7)}
& \colhead{(8)}
& \colhead{(9)}
& \colhead{(10)}
& \colhead{(11)}
& \colhead{(12)}
& \colhead{(13)}
& \colhead{(14)}
\\
}
\startdata
    1 & \aer{    2}{+    0}{   -2} & $    5$ & \aer{    2}{+    0}{   -2} & $    1$ & \aer{    2}{+    0}{   -2} & $    5$ & \aer{0.00}{+0.00}{-0.00} & $-2$ & \aer{0.07}{+0.00}{-0.07} & \aer{0.06}{+0.00}{-0.06} & \aer{0.07}{+0.00}{-0.07} & 8E-01 & 3E-02 \\
    2A &  \aer{    5}{+    3}{   -2}  &  $    0$  &  \aer{    2}{+    0}{   -2}  &  $    5$  &  \aer{    2}{+    0}{   -2}  &  $    5$  &  \aer{0.00}{+0.00}{-0.00}  &  $-2$  &  \aer{0.14}{+0.09}{-0.06}  &  \aer{0.07}{+0.00}{-0.07}  &  \aer{0.07}{+0.00}{-0.07}  &  8E-01  &  4E-02 \\
    3 & \aer{    2}{+    0}{   -2} & $    5$ & \aer{    2}{+    0}{   -2} & $    5$ & \aer{    2}{+    0}{   -2} & $    1$ & \aer{0.00}{+0.00}{-0.00} & $-2$ & \aer{0.07}{+0.00}{-0.07} & \aer{0.07}{+0.00}{-0.07} & \aer{0.06}{+0.00}{-0.06} & -2E-01 & 1E+00 \\
    4A &  \aer{    5}{+    3}{   -2}  &  $    0$  &  \aer{    2}{+    0}{   -2}  &  $    5$  &  \aer{    2}{+    0}{   -2}  &  $    5$  &  \aer{0.00}{+0.00}{-0.00}  &  $-2$  &  \aer{0.16}{+0.10}{-0.06}  &  \aer{0.07}{+0.00}{-0.07}  &  \aer{0.07}{+0.00}{-0.07}  &  8E-01  &  2E-02 \\
    5A &  \aer{  160}{+   13}{  -12}  &  $    0$  &  \aer{   88}{+   10}{   -9}  &  $    0$  &  \aer{   71}{+    9}{   -8}  &  $    0$  &  \aer{-0.11}{+0.08}{-0.08}  &  $ 0$  &  \aer{4.46}{+0.38}{-0.35}  &  \aer{2.47}{+0.29}{-0.26}  &  \aer{1.99}{+0.26}{-0.23}  &  7E+00  &  0E+00 \\
    6A &  \aer{  102}{+   11}{  -10}  &  $    0$  &  \aer{   82}{+   10}{   -9}  &  $    0$  &  \aer{   19}{+    5}{   -4}  &  $    0$  &  \aer{-0.62}{+0.09}{-0.08}  &  $ 0$  &  \aer{2.85}{+0.31}{-0.28}  &  \aer{2.31}{+0.28}{-0.25}  &  \aer{0.54}{+0.15}{-0.12}  &  3E+00  &  3E-22 \\
    7A &  \aer{    4}{+    3}{   -1}  &  $    0$  &  \aer{    2}{+    0}{   -2}  &  $    5$  &  \aer{    2}{+    0}{   -2}  &  $    5$  &  \aer{0.00}{+0.00}{-0.00}  &  $-2$  &  \aer{0.12}{+0.09}{-0.05}  &  \aer{0.07}{+0.00}{-0.07}  &  \aer{0.07}{+0.00}{-0.07}  &  6E-01  &  5E-02 \\
\enddata
\end{deluxetable*}
\clearpage
\end{landscape}

\clearpage
\begin{landscape}
\begin{deluxetable*}{cccc cccc cccc cc}
\small
\tablecolumns{14}
\tablecaption{\small Counts, count rates and \ace\ detection significance and
no-source probabilities for X-ray point sources in HCG~16 (online only).\label{tab-count-16}}
\tablehead{
\colhead{ID}
& \colhead{c(FB)}
& \colhead{lim$_F$}
& \colhead{c(SB)}
& \colhead{lim$_S$}
& \colhead{c(HB)}
& \colhead{lim$_S$}
& \colhead{HR}
& \colhead{lim$_{\rm HR}$}
& \colhead{cr(FB)}
& \colhead{cr(SB)}
& \colhead{cr(HB)}
& \colhead{\ace(sig)}
& \colhead{\ace($P$)}
\\
\colhead{(1)}
& \colhead{(2)}
& \colhead{(3)}
& \colhead{(4)}
& \colhead{(5)}
& \colhead{(6)}
& \colhead{(7)}
& \colhead{(8)}
& \colhead{(9)}
& \colhead{(10)}
& \colhead{(11)}
& \colhead{(12)}
& \colhead{(13)}
& \colhead{(14)}
\\
}
\startdata
    1 & \aer{   15}{+    5}{   -3} & $    0$ & \aer{    6}{+    3}{   -2} & $    0$ & \aer{    8}{+    4}{   -2} & $    0$ & \aer{0.12}{+0.29}{-0.30} & $ 0$ & \aer{1.22}{+0.40}{-0.31} & \aer{0.54}{+0.30}{-0.20} & \aer{0.68}{+0.32}{-0.23} & 2E+00 & 3E-09 \\
    2 & \aer{    8}{+    4}{   -2} & $    0$ & \aer{    3}{+    3}{   -1} & $    0$ & \aer{    4}{+    3}{   -2} & $    0$ & \aer{0.11}{+0.40}{-0.42} & $ 0$ & \aer{0.67}{+0.32}{-0.23} & \aer{0.30}{+0.25}{-0.15} & \aer{0.37}{+0.27}{-0.17} & 1E+00 & 4E-05 \\
    3 & \aer{    2}{+    0}{   -2} & $    5$ & \aer{    2}{+    0}{   -2} & $    1$ & \aer{    2}{+    0}{   -2} & $    5$ & \aer{0.00}{+0.00}{-0.00} & $-2$ & \aer{0.19}{+0.00}{-0.19} & \aer{0.18}{+0.00}{-0.18} & \aer{0.20}{+0.00}{-0.20} & 7E-01 & 3E-02 \\
    4B &  \aer{  189}{+   14}{  -13}  &  $    0$  &  \aer{   39}{+    7}{   -6}  &  $    0$  &  \aer{  150}{+   13}{  -12}  &  $    0$  &  \aer{0.58}{+0.06}{-0.07}  &  $ 0$  &  \aer{15.11}{+1.18}{-1.10}  &  \aer{3.17}{+0.59}{-0.50}  &  \aer{11.95}{+1.06}{-0.97}  &  1E+01  &  0E+00 \\
    5 & \aer{   10}{+    4}{   -3} & $    0$ & \aer{    7}{+    3}{   -2} & $    0$ & \aer{    2}{+    0}{   -2} & $    5$ & \aer{-0.47}{+0.00}{-0.53} & $ 1$ & \aer{0.85}{+0.35}{-0.26} & \aer{0.62}{+0.31}{-0.22} & \aer{0.22}{+0.00}{-0.22} & 1E+00 & 1E-03 \\
    6 & \aer{    9}{+    4}{   -3} & $    0$ & \aer{    7}{+    3}{   -2} & $    0$ & \aer{    2}{+    0}{   -2} & $    5$ & \aer{-0.51}{+0.00}{-0.49} & $ 1$ & \aer{0.79}{+0.34}{-0.25} & \aer{0.60}{+0.31}{-0.21} & \aer{0.20}{+0.00}{-0.20} & 8E-01 & 3E-02 \\
    7 & \aer{    3}{+    3}{   -1} & $    0$ & \aer{    3}{+    0}{   -3} & $    5$ & \aer{    2}{+    0}{   -2} & $    5$ & \aer{0.00}{+0.00}{-0.00} & $-2$ & \aer{0.30}{+0.25}{-0.15} & \aer{0.29}{+0.00}{-0.29} & \aer{0.20}{+0.00}{-0.20} & 4E-01 & 1E-01 \\
\enddata
\end{deluxetable*}
\clearpage
\end{landscape}

\clearpage
\begin{landscape}
\begin{deluxetable*}{cccc cccc cccc cc}
\small
\tablecolumns{14}
\tablecaption{\small Counts, count rates and \ace\ detection significance and
no-source probabilities for X-ray point sources in HCG~22 (online only).\label{tab-count-22}}
\tablehead{
\colhead{ID}
& \colhead{c(FB)}
& \colhead{lim$_F$}
& \colhead{c(SB)}
& \colhead{lim$_S$}
& \colhead{c(HB)}
& \colhead{lim$_S$}
& \colhead{HR}
& \colhead{lim$_{\rm HR}$}
& \colhead{cr(FB)}
& \colhead{cr(SB)}
& \colhead{cr(HB)}
& \colhead{\ace(sig)}
& \colhead{\ace($P$)}
\\
\colhead{(1)}
& \colhead{(2)}
& \colhead{(3)}
& \colhead{(4)}
& \colhead{(5)}
& \colhead{(6)}
& \colhead{(7)}
& \colhead{(8)}
& \colhead{(9)}
& \colhead{(10)}
& \colhead{(11)}
& \colhead{(12)}
& \colhead{(13)}
& \colhead{(14)}
\\
}
\startdata
    1 & \aer{    2}{+    0}{   -2} & $    5$ & \aer{    2}{+    0}{   -2} & $    5$ & \aer{    2}{+    0}{   -2} & $    5$ & \aer{0.00}{+0.00}{-0.00} & $-2$ & \aer{0.07}{+0.00}{-0.07} & \aer{0.07}{+0.00}{-0.07} & \aer{0.07}{+0.00}{-0.07} & 1E-01 & 5E-01 \\
    2 & \aer{   10}{+    4}{   -3} & $    0$ & \aer{    7}{+    3}{   -2} & $    0$ & \aer{    2}{+    0}{   -2} & $    5$ & \aer{-0.51}{+0.00}{-0.49} & $ 1$ & \aer{0.32}{+0.14}{-0.10} & \aer{0.23}{+0.12}{-0.08} & \aer{0.08}{+0.00}{-0.08} & 9E-01 & 3E-02 \\
    3 & \aer{    2}{+    0}{   -2} & $    5$ & \aer{    2}{+    0}{   -2} & $    5$ & \aer{    2}{+    0}{   -2} & $    5$ & \aer{0.00}{+0.00}{-0.00} & $-2$ & \aer{0.07}{+0.00}{-0.07} & \aer{0.08}{+0.00}{-0.08} & \aer{0.07}{+0.00}{-0.07} & 5E-01 & 2E-01 \\
    4 & \aer{   24}{+    6}{   -4} & $    0$ & \aer{   20}{+    5}{   -4} & $    0$ & \aer{    4}{+    3}{   -2} & $    0$ & \aer{-0.64}{+0.21}{-0.16} & $ 0$ & \aer{0.78}{+0.19}{-0.16} & \aer{0.64}{+0.18}{-0.14} & \aer{0.14}{+0.10}{-0.06} & 1E+00 & 3E-02 \\
    5 & \aer{   11}{+    4}{   -3} & $    0$ & \aer{    7}{+    3}{   -2} & $    0$ & \aer{    4}{+    3}{   -1} & $    0$ & \aer{-0.29}{+0.35}{-0.32} & $ 0$ & \aer{0.36}{+0.14}{-0.11} & \aer{0.23}{+0.12}{-0.08} & \aer{0.13}{+0.10}{-0.06} & 1E+00 & 1E-02 \\
    6 & \aer{    2}{+    0}{   -2} & $    5$ & \aer{    2}{+    0}{   -2} & $    5$ & \aer{    2}{+    0}{   -2} & $    5$ & \aer{0.00}{+0.00}{-0.00} & $-2$ & \aer{0.08}{+0.00}{-0.08} & \aer{0.07}{+0.00}{-0.07} & \aer{0.08}{+0.00}{-0.08} & 8E-01 & 6E-02 \\
    7 & \aer{    3}{+    3}{   -1} & $    0$ & \aer{    2}{+    0}{   -2} & $    1$ & \aer{    4}{+    3}{   -1} & $    0$ & \aer{0.28}{+0.72}{-0.00} & $-1$ & \aer{0.12}{+0.10}{-0.06} & \aer{0.07}{+0.00}{-0.07} & \aer{0.13}{+0.10}{-0.06} & 1E+00 & 2E-03 \\
\enddata
\end{deluxetable*}
\clearpage
\end{landscape}

\clearpage
\begin{landscape}
\begin{deluxetable*}{cccc cccc cccc cc}
\small
\tablecolumns{14}
\tablecaption{\small Counts, count rates and \ace\ detection significance and
no-source probabilities for X-ray point sources in HCG~31 (online only).\label{tab-count-31}}
\tablehead{
\colhead{ID}
& \colhead{c(FB)}
& \colhead{lim$_F$}
& \colhead{c(SB)}
& \colhead{lim$_S$}
& \colhead{c(HB)}
& \colhead{lim$_S$}
& \colhead{HR}
& \colhead{lim$_{\rm HR}$}
& \colhead{cr(FB)}
& \colhead{cr(SB)}
& \colhead{cr(HB)}
& \colhead{\ace(sig)}
& \colhead{\ace($P$)}
\\
\colhead{(1)}
& \colhead{(2)}
& \colhead{(3)}
& \colhead{(4)}
& \colhead{(5)}
& \colhead{(6)}
& \colhead{(7)}
& \colhead{(8)}
& \colhead{(9)}
& \colhead{(10)}
& \colhead{(11)}
& \colhead{(12)}
& \colhead{(13)}
& \colhead{(14)}
\\
}
\startdata
    1 & \aer{   21}{+    5}{   -4} & $    0$ & \aer{   15}{+    5}{   -3} & $    0$ & \aer{    5}{+    3}{   -2} & $    0$ & \aer{-0.48}{+0.24}{-0.20} & $ 0$ & \aer{0.60}{+0.16}{-0.13} & \aer{0.45}{+0.14}{-0.11} & \aer{0.16}{+0.10}{-0.06} & 1E+00 & 2E-02 \\
    2 & \aer{    9}{+    4}{   -3} & $    0$ & \aer{    4}{+    3}{   -2} & $    0$ & \aer{    4}{+    3}{   -2} & $    0$ & \aer{-0.02}{+0.39}{-0.38} & $ 0$ & \aer{0.27}{+0.12}{-0.09} & \aer{0.14}{+0.09}{-0.06} & \aer{0.13}{+0.09}{-0.06} & 1E+00 & 3E-02 \\
    3 & \aer{   10}{+    4}{   -3} & $    0$ & \aer{    5}{+    3}{   -2} & $    0$ & \aer{    4}{+    3}{   -2} & $    0$ & \aer{-0.13}{+0.38}{-0.36} & $ 0$ & \aer{0.29}{+0.12}{-0.09} & \aer{0.16}{+0.10}{-0.07} & \aer{0.12}{+0.09}{-0.06} & 1E+00 & 4E-04 \\
    4 & \aer{    2}{+    0}{   -2} & $    5$ & \aer{    3}{+    3}{   -1} & $    0$ & \aer{    2}{+    0}{   -2} & $    1$ & \aer{0.00}{+0.00}{-0.00} & $-2$ & \aer{0.07}{+0.00}{-0.07} & \aer{0.10}{+0.09}{-0.05} & \aer{0.07}{+0.00}{-0.07} & -4E-01 & 8E-01 \\
    5 & \aer{    6}{+    3}{   -2} & $    0$ & \aer{    2}{+    0}{   -2} & $    5$ & \aer{    5}{+    3}{   -2} & $    0$ & \aer{0.39}{+0.61}{-0.00} & $-1$ & \aer{0.18}{+0.10}{-0.07} & \aer{0.07}{+0.00}{-0.07} & \aer{0.15}{+0.10}{-0.06} & 1E+00 & 2E-02 \\
    6 & \aer{    7}{+    3}{   -2} & $    0$ & \aer{    5}{+    3}{   -2} & $    0$ & \aer{    2}{+    0}{   -2} & $    5$ & \aer{-0.43}{+0.00}{-0.57} & $ 1$ & \aer{0.20}{+0.11}{-0.07} & \aer{0.17}{+0.10}{-0.07} & \aer{0.07}{+0.00}{-0.07} & 4E-01 & 3E-01 \\
    7 & \aer{    4}{+    3}{   -1} & $    0$ & \aer{    2}{+    0}{   -2} & $    5$ & \aer{    2}{+    0}{   -2} & $    5$ & \aer{0.00}{+0.00}{-0.00} & $-2$ & \aer{0.12}{+0.09}{-0.05} & \aer{0.07}{+0.00}{-0.07} & \aer{0.07}{+0.00}{-0.07} & 3E-01 & 4E-01 \\
\enddata
\end{deluxetable*}
\clearpage
\end{landscape}

\clearpage
\begin{landscape}
\begin{deluxetable*}{cccc cccc cccc cc}
\small
\tablecolumns{14}
\tablecaption{\small Counts, count rates and \ace\ detection significance and
no-source probabilities for X-ray point sources in HCG~42 (online only).\label{tab-count-42}}
\tablehead{
\colhead{ID}
& \colhead{c(FB)}
& \colhead{lim$_F$}
& \colhead{c(SB)}
& \colhead{lim$_S$}
& \colhead{c(HB)}
& \colhead{lim$_S$}
& \colhead{HR}
& \colhead{lim$_{\rm HR}$}
& \colhead{cr(FB)}
& \colhead{cr(SB)}
& \colhead{cr(HB)}
& \colhead{\ace(sig)}
& \colhead{\ace($P$)}
\\
\colhead{(1)}
& \colhead{(2)}
& \colhead{(3)}
& \colhead{(4)}
& \colhead{(5)}
& \colhead{(6)}
& \colhead{(7)}
& \colhead{(8)}
& \colhead{(9)}
& \colhead{(10)}
& \colhead{(11)}
& \colhead{(12)}
& \colhead{(13)}
& \colhead{(14)}
\\
}
\startdata
    1 & \aer{   10}{+    4}{   -3} & $    0$ & \aer{    6}{+    3}{   -2} & $    0$ & \aer{    3}{+    3}{   -1} & $    0$ & \aer{-0.26}{+0.38}{-0.34} & $ 0$ & \aer{0.32}{+0.14}{-0.10} & \aer{0.20}{+0.12}{-0.08} & \aer{0.12}{+0.10}{-0.06} & 1E+00 & 3E-02 \\
    2 & \aer{    2}{+    0}{   -2} & $    5$ & \aer{    2}{+    0}{   -2} & $    5$ & \aer{    2}{+    0}{   -2} & $    5$ & \aer{0.00}{+0.00}{-0.00} & $-2$ & \aer{0.08}{+0.00}{-0.08} & \aer{0.09}{+0.00}{-0.09} & \aer{0.08}{+0.00}{-0.08} & 3E-01 & 3E-01 \\
    3C &  \aer{    5}{+    3}{   -2}  &  $    0$  &  \aer{    5}{+    3}{   -2}  &  $    0$  &  \aer{    2}{+    0}{   -2}  &  $    1$  &  \aer{-0.40}{+0.00}{-0.60}  &  $ 1$  &  \aer{0.17}{+0.11}{-0.07}  &  \aer{0.17}{+0.11}{-0.07}  &  \aer{0.07}{+0.00}{-0.07}  &  -6E-03  &  6E-01 \\
    4 & \aer{    2}{+    0}{   -2} & $    5$ & \aer{    2}{+    0}{   -2} & $    1$ & \aer{    2}{+    0}{   -2} & $    5$ & \aer{0.00}{+0.00}{-0.00} & $-2$ & \aer{0.07}{+0.00}{-0.07} & \aer{0.07}{+0.00}{-0.07} & \aer{0.08}{+0.00}{-0.08} & 6E-01 & 2E-01 \\
    5 & \aer{    2}{+    0}{   -2} & $    5$ & \aer{    2}{+    0}{   -2} & $    5$ & \aer{    2}{+    0}{   -2} & $    1$ & \aer{0.00}{+0.00}{-0.00} & $-2$ & \aer{0.08}{+0.00}{-0.08} & \aer{0.08}{+0.00}{-0.08} & \aer{0.07}{+0.00}{-0.07} & -2E-01 & 1E+00 \\
    6C &  \aer{   19}{+    5}{   -4}  &  $    0$  &  \aer{   17}{+    5}{   -4}  &  $    0$  &  \aer{    2}{+    0}{   -2}  &  $    5$  &  \aer{-0.76}{+0.00}{-0.24}  &  $ 1$  &  \aer{0.60}{+0.17}{-0.14}  &  \aer{0.54}{+0.17}{-0.13}  &  \aer{0.08}{+0.00}{-0.08}  &  6E-01  &  1E-01 \\
    7 & \aer{    5}{+    3}{   -2} & $    0$ & \aer{    2}{+    0}{   -2} & $    5$ & \aer{    2}{+    0}{   -2} & $    5$ & \aer{0.00}{+0.00}{-0.00} & $-2$ & \aer{0.17}{+0.11}{-0.07} & \aer{0.08}{+0.00}{-0.08} & \aer{0.08}{+0.00}{-0.08} & 1E+00 & 1E-03 \\
\enddata
\end{deluxetable*}%
\clearpage
\end{landscape}

\end{document}